\documentclass[aps,nofootinbib,superscriptaddress,twocolumn,prd,10pt]{revtex4-1}
\usepackage[normalem]{ulem}

\usepackage{amsmath,amssymb}

\usepackage{graphicx}

\usepackage{lipsum}
\usepackage[dvipsnames]{xcolor}
\usepackage{booktabs}
\usepackage[T1]{fontenc}
\usepackage{dcolumn}
\usepackage{bm}

\usepackage{natbib}

\usepackage{hyperref}
\hypersetup{
    colorlinks=true,
    linkcolor=red,
    citecolor=blue,
    urlcolor=BlueViolet,
}

\newcount\tmpnum
\def\bionic{\futurelet\next\bionicA}
\def\bionicA{\ifx\next[\afterfi\bionicB \else \afterfi{\bionicB[50]}\fi}
\def\bionicB[#1]#2{{\def\wratio{#1}\bionicC#2 {} }}
\def\bionicC #1 {\ifx^#1^\else
    \spacebetweenwords
    \tmpnum=0 \lcount #1{}
    \multiply\tmpnum by\wratio \divide\tmpnum by100
    \testIm #1''\end 
    \bgroup \bf \wordpart #1%
    \expandafter\bionicC \fi
}
\def\spacebetweenwords{\def\spacebetweenwords{ }}
\def\lcount #1{\ifx^#1^\else \advance\tmpnum by1 \expandafter\lcount\fi}
\def\wordpart #1{%
   \ifnum\tmpnum=0 \egroup#1%
   \else #1\advance\tmpnum by-1 \expandafter \wordpart \fi}
\def\testIm #1'#2'#3\end{\ifx^#2^\else \tmpnum=0 \fi}
\def\afterfi#1#2\fi{\fi#1}

\definecolor{RL}{HTML}{258899}

\newcommand{\ETA}{\text{ETA}}
\newcommand{\EFA}{\text{EFA}}
\newcommand{\ct}{\tau}
\newcommand{\hL}{h}
\newcommand{\etaT}{\eta}
\newcommand{\shear}{\sigma}
\newcommand{\mt}{{m t}}

\newcommand{\ax}{{\rm ax}}

\newcommand{\ULA}{ULA}

\newcommand{\DM}{{\rm DM}}
\newcommand{\DE}{{\rm DE}}
\newcommand{\thetas}{\theta_\star}
\newcommand{\ini}{\text{ini}}
\newcommand{\ourcode}{\textsc{AxiECAMB}}

\usepackage{tcolorbox}

\begin{document}

\title{Accurate  method for ultralight axion CMB and matter power spectra}

\author{Rayne Liu}\email{rayneliu@uchicago.edu}
\affiliation{Kavli Institute for Cosmological Physics, Enrico Fermi Institute, and Department of Astronomy \& Astrophysics, University of Chicago, Chicago IL 60637
}

\author{Wayne Hu}
\affiliation{Kavli Institute for Cosmological Physics, Enrico Fermi Institute, and Department of Astronomy \& Astrophysics, University of Chicago, Chicago IL 60637
}

\author{Daniel Grin}
\affiliation{Department of Physics and Astronomy, Haverford College, 370 Lancaster Avenue, Haverford, PA 19041, United States}

\date{\today}

\begin{abstract}
Ultralight axions (ULAs) with masses $10^{-33} \lesssim m/{\rm eV} \lesssim 10^{-12}$ 
are well motivated in string-inspired models and can be part or all of the dark energy or the dark matter in this range. Since the \ULA\ field oscillates at a frequency $m$ that can be much larger than the expansion rate $H$, accurate and efficient calculation of cosmological observables requires an effective time averaged treatment. While these are well established for $m\gg 10 H_{\rm eq}$, the Hubble rate at matter radiation equality, here we extend and develop these techniques to cover the mass range $10^{-33}  \lesssim m/{\rm eV} \lesssim 10^{-18}$.  We implement this technique in a full cosmological Boltzmann code (\ourcode) with numerical precision sufficiently accurate for current and next-generation cosmic microwave background, as well as large-scale structure data analysis. New effects including the time averaging of metric perturbations and hydrostatic equilibrium of the effective fluid result in many orders of 
magnitude improvements for power spectra accuracy over some previous treatments such as \textsc{axionCAMB} in some extreme regions of parameter space 
and order unity changes of the \ULA\ effects near $\Lambda$CDM models. These improvements may impact the specific model parameters that have been suggested might resolve various tensions in $\Lambda$CDM at a
comparable level.
\end{abstract}

\maketitle

\section{\label{sec:Introduction}Introduction}

Cosmic microwave background (CMB) anisotropy measurements have facilitated the establishment of the $\Lambda$ cold dark matter ($\Lambda$CDM) model as the standard model for cosmology
\cite{Planck:2018vyg,ACT:2020gnv,SPT-3G:2024atg}.
Additional evidence for this paradigm comes from inferences of cosmic acceleration by Type Ia supernovae, as well as surveys probing the clustering of galaxies, baryon acoustic oscillations and weak gravitational lensing  (e.g.~\cite{Rubin:2023ovl,DES:2024jxu,DESI:2024mwx,DES:2021wwk,Heymans:2020gsg,Miyatake:2023njf}).

Despite these successes, the $\Lambda$CDM model is incomplete. The microphysical nature of the dark matter and dark energy remains unknown. Outstanding unresolved challenges to this paradigm also persist, such as the $\sim 5-6\sigma$ tension between CMB and Cepheid-supernovae inferences of the Hubble constant $H_0$ (e.g.~\cite{Breuval:2024lsv}; cf.~\cite{Freedman:2024eph}),
 the small discrepancy between values of the amplitude of structure parameter $S_8 = \sigma_8\sqrt{\Omega_{\rm m}/0.3}$ inferred from the CMB  and weak lensing measurements (e.g.~\cite{Heymans:2020gsg,DES:2021wwk,Miyatake:2023njf}), and differences in the relative distance measures between supernovae, baryon acoustic oscillations and the CMB \cite{DES:2024jxu,DESI:2024mwx}.
 These tensions may indicate a more complicated dark sector (see \cite{Schoneberg:2021qvd,DiValentino:2021izs} and references therein).

Some theoretical resolutions to these issues invoke new fundamental pseudo-scalar fields with extremely low mass scales ($10^{-33}\lesssim m/{\rm eV} \lesssim 10^{-12}$), where $m$ is the mass of the ultralight field and we set $c=\hbar=1$ throughout. The existence of such fields is motivated by considerations in string-inspired models, where new gauge fields on compactified extra dimensions acquire mass through nonperturbative interactions and have pseudoscalar couplings, like the axions which resolve the strong $\mathcal{CP}$ problem of the standard model of particle physics; these fields are thus usually referred to as ultralight axions (\ULA s). This has given rise to the idea of an \emph{axiverse} \cite{Svrcek:2006yi,Conlon:2006tq,Arvanitaki:2009fg,Cicoli:2012sz,Stott:2017hvl,Gendler:2023kjt}, in which many such particles exist with a log-flat distribution of masses.
Axion mass values may be even higher than $ m \gtrsim 10^{-12}$\,eV and one of these particles could be the QCD axion \cite{Arvanitaki:2009fg}, but we would then not classify them as ultralight. In these generalizations, \ULA s need not be pseudoscalar particles as they could equally well be scalars as long as they obey the same equations of motion, but we will continue to loosely refer to them as \ULA s. As another example,
similar dynamics occur for the longitudinal modes of ultralight vector dark matter or a dark photon (e.g.~\cite{Graham:2015rva,Amin:2022pzv}) and a complex scalar (e.g.~\cite{Yang:2025vcb,Suarez:2015fga}). 

Such fields are initially cosmologically misaligned from their vacuum, and begin to oscillate coherently once $m \sim H \equiv d\ln a/dt$, diluting thereafter as $a^{-3}$ in density, where $a$ is the usual cosmological scale factor, and behave as dark matter.
The Hubble rate at matter-radiation equality in $\Lambda$CDM 
($H_\text{eq}\sim 10^{-28}\,{\rm eV}$) therefore approximately partitions \ULA s into 
heavier ones that behave like dark matter on large scales and lighter ones that behave  as dark energy at least during the initial phase of structure formation
\cite{Frieman:1995pm, Choi:1999xn, Marsh:2010wq}.
\ULA s could thus contribute to some or all of the dark matter or energy. 

In the dark matter regime, \ULA s of differing masses could impact a number of observational challenges to the $\Lambda$CDM cosmological paradigm on small scales \cite{Marsh:2015xka}, such as the \emph{missing satellites} issue in the Local Group, the \emph{too big to fail} issue, and the \emph{cusp-core} issue with density profiles of low-mass galaxies \cite{Klypin:1999uc,deBlok:2009sp,Boylan-Kolchin:2011qkt,Bullock:2017xww,Lovell:2022vzx,2023MNRAS.520..461J,Homma:2023ppu,2024arXiv241209566M}, and play a significant role in changing their interpretation.

\ULA s have distinct cosmological signatures, altering the time evolution of the Hubble parameter, and affecting the evolution of density fluctuations in the universe. The macroscopically large de Broglie wavelengths of \ULA s would suppress the growth of density fluctuations on small scales, making them an example of fuzzy dark matter (FDM) \cite{Hu:2000ke,Hui:2016ltb}. As a component of the dark matter,
\ULA s can alter the acoustic peaks and the early integrated Sachs-Wolfe (ISW) effect in the CMB,  carry isocurvature perturbations \cite{Fox:2004kb}, contribute to an early dark energy component \cite{Poulin:2018dzj, Lin:2019qug}, and alter the matter power spectrum. \ULA s could thus alleviate the $S_8$ tension \cite{Rogers:2023ezo}.

Cosmological data thus offer a powerful probe of the \ULA\ possibility, with promising opportunities for experimental tests, e.g.\ ongoing and future CMB, galaxy clustering, and weak lensing data from the CMB-S4 experiment \cite{Abazajian:2019eic}, the Rubin Observatory \cite{LSST:2008ijt,2009arXiv0912.0201L,LSSTDarkEnergyScience:2018jkl}, the Euclid satellite \cite{Amendola:2016saw}, the DESI survey \cite{DESI:2016fyo}, and many others. To turn this data into robust \ULA\ constraints, one must sample the likelihood function, comparing \ULA\ model predictions to data. This requires rapid and accurate calculation of observables including CMB and matter power spectra.

For
$m\gg H_0$, the field oscillates on a time scale much faster than that which cosmological observables evolve, making it numerically challenging to efficiently and accurately make theoretical predictions for the impact of \ULA s. 
Boltzmann codes (such as \textsc{axionCAMB} \cite{Hlozek:2014lca} and \textsc{AxiCLASS} \cite{Poulin:2018dzj}), used to impose current state-of-the-art constraints to \ULA s, address this timescale problem using an effective fluid approximation (\EFA) that tracks the effective average of axion variables over oscillation cycles \cite{Ratra:1990me,Hu:2000ke}, essentially employing a WKB-type approach. 

Use of this approximation leads to numerical artefacts in mode evolution and observables \cite{Cookmeyer:2019rna}. The artefacts induced specifically by the \EFA\ 
were found to be below the precision required for \emph{Planck} data analysis but could introduce biases to parameter inferences in the future (e.g.\ from ongoing and upcoming experiments like the Simons Observatory or CMB-S4, respectively). \textsc{AxionCAMB} has some additional numerical artefacts, some of which were addressed in the update from version 1.0 to 2.0, and others of which are discussed here.

A new algorithm was developed in Ref.\ \cite{Passaglia:2022bcr}, which refines the corresponding theoretical predictions to sub-percent accuracy without loss of computational speed for $m\gg 10 H_{\rm eq}$. This algorithm works by following the exact Klein-Gordon (KG) equation at background and perturbation level for at least a few cycles, and then matching the EFA variables to their effective time averages (ETAs) at a switch point. The ETA matching  is  constructed to exactly obey conservation laws, yielding optimized functional forms for EFA equation-of-state parameter and  sound speed.

In Ref.\ \cite{Passaglia:2022bcr}, only
 $m\gg 10 H_{\rm eq}$ cases were considered, for which the \ULA\ time averages need not account for the back reaction on metric fluctuations. The resulting differences from $\Lambda$CDM do not 
change CMB observables. The new EFA approach was applied in Ref. \cite{Chen:2023unc} to compute matter power spectra and halo-model observables in \ULA\ models containing two light fields, inspired by the axiverse scenario. 
The method was also implemented in
the \textsc{CLASS} Boltzmann codebase  \cite{Blas:2011rf}  in Ref.~\cite{Baryakhtar:2024rky} again neglecting the metric effects when applied to the $m\ll 10 H_{\rm eq}$ regime.

An alternative approach was presented in Refs.~\cite{Urena-Lopez:2015gur,Cedeno:2017sou,Urena-Lopez:2023ngt}. There, fluctuations are factored into a smoothly evolving and oscillating component, with a specific phase-matching prescription applied at the onset of rapid oscillations, when numerical challenges emerge. Here again the techniques focused on the $m > 10 H_{\rm eq}$ regime.

Our work yields the first \ULA\ Boltzmann code that is accurate and efficient for the ULA background and perturbations over a broad range of masses $10^{-33}  \lesssim m/{\rm eV} \lesssim 10^{-18}$.  
We develop and extend the effective methods of
Ref.~\cite{Passaglia:2022bcr} to the $m \lesssim 10^{-25}$\,eV regime and implement them in a new code,
 \ourcode\footnote{\href{https://github.com/Ra-yne/AxiECAMB}{https://github.com/Ra-yne/AxiECAMB} } (Axion Effective-method in CAMB\footnote{\textsc{CAMB}: \href{http://camb.info}{http://camb.info}})   as a successor of \textsc{axionCAMB}.
 In this regime there are a number of additional effects due to the impact of the \ULA\ on metric fluctuations and their behavior well below their cosmologically large Jeans scale. 

We begin in Sec.~\ref{sec:EFA_methodology}, where we explain the elements of the effective method: Klein-Gordon field solutions until a given switch epoch $m/H\approx 10$, their effective time average at the switch, and their evolution as an effective fluid thereafter.   We discuss the impact of the \ULA\ on CMB and matter power spectra  in Sec.~\ref{sec:Performance} and show that \ourcode\ is sufficiently accurate for current and next-generation surveys.  

Technical issues are detailed in a series of Appendices: comparisons to previous work  (\ref{sec:constraint}) including \textsc{AxionCAMB} and fitting functions in the FDM regime;  Klein-Gordon field solutions  (\ref{app:KG}), including methods for predicting the oscillatory phase for $m\lesssim 10 H_{\rm eq}$; effective time averaging of variables (\ref{app:ETA_construction}), including metric perturbations; effective fluid construction  (\ref{app:EFAvars_and_quasistatic}), including the quasi-static limit for perturbations well under the Jeans scale; and CMB implementation (\ref{app:cmbsources_corr}), including their response to \ULA\ oscillations and the impact of effective sources at the switch.

We discuss all of these results in Sec.~\ref{sec:conclusions}.

\section{\label{sec:EFA_methodology}Effective Method for \ULA s}

Here we establish our effective method for \ULA s to linear order in perturbation theory.   
We first briefly review the behavior of the Klein-Gordon system when the mass time scale becomes much shorter than the Hubble time (KG; Sec.~\ref{sec:KG}).  
Extending the methods of Ref.~\cite{Passaglia:2022bcr}, we then construct an effective time average (\ETA; Sec.~\ref{sec:ETA}) of the \ULA\ and metric variables at some appropriate switch time (henceforth ``the switch'') and thereafter evolve this time average using an effective fluid approximation (\EFA; Sec.~\ref{sec:EFA}).  We address the choice of switch time and CMB sources in Sec.~\ref{sec:CMB_changes}, in particular for cases where $ 10^{-28} \lesssim m/{\rm eV} \lesssim 10^{-25}$.

For testing the accuracy of our effective method here and below, we take as our fiducial model a cosmology with \emph{Planck} baseline parameters $\Omega_{b}h^2=0.0224, \sum m_\nu = 0.06\,$eV,
 $H_0 = 67.36$\,km/s/Mpc \cite{Aghanim:2018eyx} in a spatially flat cosmology. 
In the presence of \ULA s with $m/H_0\geq 10$, we replace the cold dark matter density $\Omega_c h^2$ with the total dark matter $\Omega_{\DM}h^2=\Omega_c h^2+\Omega_\ax h^2=0.12$. When $m/H_0 < 10$, \ULA s behave as dark energy (DE) and can explain the current epoch of cosmic acceleration, hence we parametrize  its abundance instead with $f_\DE\equiv \Omega_\ax h^2/\Omega_\DE h^2$, where
 $\Omega_\DE h^2 = \Omega_\Lambda h^2+\Omega_\ax h^2$ and leave $\Omega_c h^2=0.12$.  In this section we will take 
 $m=10^{-27}$\,eV and $f_\DM=1$ as our fiducial example unless otherwise specified.

\subsection{Klein-Gordon Oscillations}\label{sec:KG}
The \ULA\ field $\phi$ evolves via the Klein-Gordon equation
$\Box \phi = -V'$. We adopt the common assumption that the field is sufficiently close to its minimum that the  quadratic approximation   $V(\phi)\approx m^2\phi^2/2$ applies, and the background evolution becomes
\begin{equation}
\ddot{\phi} + 2\frac{\dot{a}}{a}\dot{\phi} + a^2m^2\phi =0 ,
\label{eq:KG}
\end{equation}
where overdots denote conformal time derivatives $d/d\ct = a d/dt$, and $a$ is the usual cosmological scale factor. Once the Hubble rate $H=d\ln a/dt$ drops below the mass, the background field oscillates and $H$ can become much smaller than the frequency $m$ by the present for \ULA\ masses of interest.

For spatial perturbations $\delta\phi$, we study the Fourier space evolution in synchronous gauge:
\begin{equation}
\label{eq:perturbedKG}
      \ddot{\delta \phi} + 2\frac{\dot{a}}{a}\dot{\delta \phi} + (k^2 + a^2m^2)\delta \phi = -\frac{\dot{\hL}}{2}\dot{\phi},
\end{equation}
where $k$ is the comoving wavenumber and $\hL$ is the fluctuation in the trace of the spatial metric around the background 3-metric $g_{ij} = a^2(\gamma_{ij}+ h_{ij})$ 
\begin{eqnarray}
h_{ij} &=& \frac{h}{3} \gamma_{ij} + 
\left(\nabla_i \nabla_j \nabla^{-2}-\frac{1}{3}\gamma_{ij}\right)(\hL + 6\etaT),
\end{eqnarray}
where $\etaT$ is the space curvature perturbation and $\nabla$ is the covariant derivative constructed from $\gamma_{ij}$, the metric of constant spatial curvature. We also use  shear  of the local expansion
$\shear =(\dot \hL +6\dot\etaT)/2k$ below
(see e.g.~\cite{Ma:1995ey}).

The corresponding energy density $\rho_\ax$, pressure $p_\ax$, their perturbations, and the momentum density {$u_{\ax}\rho_{\ax}$} are given by
\begin{align}
\label{eq:axionstressenergy}
\rho_\ax &= \tfrac{1}{2}a^{-2}\dot{\phi}^2 + \tfrac{1}{2}m^2\phi^2,\nonumber\\
p_\ax &= \tfrac{1}{2}a^{-2}\dot{\phi}^2 - \tfrac{1}{2}m^2\phi^2,\nonumber\\
    \delta\rho_\ax &= a^{-2}\dot{\phi}{\delta \dot{\phi}} + m^2\phi \delta \phi,\nonumber\\
    \delta p_\ax&= a^{-2}\dot{\phi}{\delta \dot{\phi}} - m^2\phi \delta \phi,\nonumber\\
    u_\ax\rho_\ax &= {a^{-2}k\dot{\phi}\delta\phi},
\end{align}
and influence cosmological observables through their gravitational effects.
Once $m\gg H$, these quantities also exhibit rapid oscillations, making the joint system of linear perturbation equations difficult to solve directly.

To complete the \ULA\ system, we also require initial conditions for the \ULA\ field. In the limit $a\rightarrow 0$, i.e. $m/H\rightarrow 0$, the field is frozen. Only its initial background value $\phi_\ini$ is finite, and all other field quantities approach zero for adiabatic fluctuations. Numerically, however, one evolves from a finite initial time and at that time
the field has a finite time derivative and its perturbations are generated by fluctuations in the other matter species.

We improve on current state-of-the-art \ULA\ Boltzmann codes such as \textsc{axionCAMB}~\cite{Hlozek:2014lca} by keeping the leading order term in powers of $m/H$ as well as $k\tau$, i.e.\ the solution to the KG system in the radiation-dominated epoch becomes
\begin{equation}\label{eq:KG_IC}
\begin{aligned}
\phi & \approx \phi_\ini, & 
\dot{\phi}  &\approx -\frac{am^2}{5H}\phi_\text{ini}, \\ 
    \delta\phi  &\approx \frac{ m^2\dot{\hL}}{ 420 aH^3} \phi_\text{ini}, & 
        \dot{\delta \phi}
    &\approx \frac{m^2 \dot{\hL}}{70 H^2} \phi_\text{ini}.\\
\end{aligned}
\end{equation}
Thus $\phi_{\rm ini}$ specifies all of the initial conditions and also determines the \ULA\ abundance today, $\Omega_\ax h^2$.

\begin{figure}
    \includegraphics[width=1\linewidth]{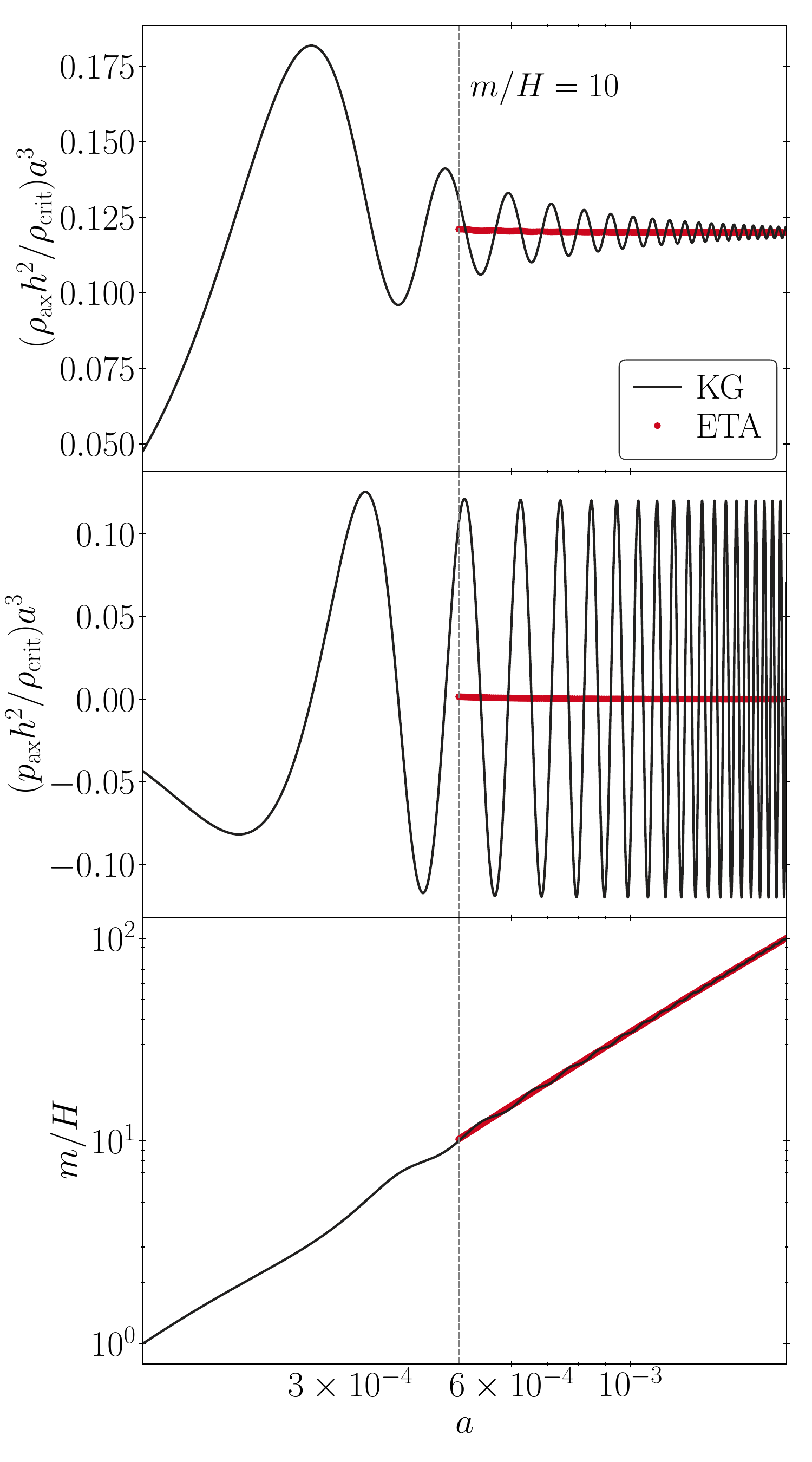}
    \caption{
     Background energy density (top), pressure (middle), and $m/H$ (bottom) for \ULA s with $m=10^{-27}$\,eV and a dark matter fraction $f_\DM = 1$. The \ETA\ values at discrete switch times (dense red points, starting at $m/H_*=10$) effectively average the temporal oscillations of the  KG solution (solid lines).  Here all \ETA\ values take a common initial field value $\phi_{\rm ini}$ determined by matching to the final density for  the $m/H_* = 100$ case and the density is rescaled to reflect this present value $\rho_\ax(a=1) h^2 /\rho_{\rm crit} = \Omega_\ax h^2$. }
    \label{fig:background_ETA}
\end{figure}

As an example in Fig.~\ref{fig:background_ETA}, we take a mass of $10^{-27}$ eV that is comparable to $H$ at matter-radiation equality and $f_\DM = 1$, and plot its  energy density rescaled by $a^3$, specifically $(\rho_\ax h^2/\rho_{\rm crit})a^3$. The normalization chosen such that its final value corresponds to $\Omega_\ax h^2(=0.12)$. The KG solution (thin gray curve) extends to $m/H=100$ at the right edge of the plot. At $m\gg H$, the oscillations in the KG system become much faster than the gravitational dynamical time, here $\Delta \ln a=1$.  The hierarchy of time scales indicates a stiff system that dramatically decreases the efficiency of numerical solvers. On the other hand, this also means that the other cosmological observables effectively only respond to the time average of the \ULA\ oscillations, which motivates our basic approach: 
\begin{enumerate}
\item Directly solve the KG system for $m/H < m/H_*$, where the subscript ``$*$'' labels quantities evaluated at the switch time;
\item At $m/H_*$, compute the effective time average (ETA) over the mass oscillation time scale;
\item Finally, evolve the time-averaged quantities for $m/H>m/H_*$ with an effective fluid approximation (\EFA).
\end{enumerate}

A value of $m/H_* \sim 10$, 
which we will see is sufficient for accuracy given current observational constraints for most masses (cf.\ Sec.\  \ref{sec:CMB_changes}), represents a switch that occurs after a few oscillations of the background density. This corresponds to the left edge of the dashed red line in Fig.~\ref{fig:background_ETA}.  We next proceed to develop the steps and details of this approximation.

\subsection{Effective Time Average}\label{sec:ETA}

When the oscillations have become sufficiently rapid compared to the Hubble rate, we construct quantities that represent the effective time average, or \ETA, over the oscillations, without explicitly performing a cumbersome time average for each $k$-mode and the background. 

Our basic strategy follows Ref.~\cite{Passaglia:2022bcr} and technical details are provided in Appendix \ref{app:ETA_construction}. The idea developed there is to separate the sine and cosine pieces of the field oscillations at  $H_*$ for the \ETA\ [see Eq.~(\ref{eq:BesselApprox})],
\begin{equation}\label{eq:sincos_decomposition}
\begin{aligned}
\phi  &= \varphi_c \cos(\mt -\mt_*) + \varphi_s  \sin(\mt - \mt_* ), \\
\delta \phi  &= \delta \varphi_c \cos(\mt -\mt_* ) +\delta \varphi_s  \sin(\mt - \mt_* ),
\end{aligned}
\end{equation}
where we have introduced $t_*$, the coordinate time at $H_*$, for notational convenience.
The two components still obey the exact KG equation and thus energy-momentum  conservation, but
the redundancy of describing a single field with two components means that we have the freedom to specify additional auxiliary conditions at the switch, which we take to be the same for the background and perturbations:
 \begin{equation}
\begin{aligned}\label{eq:auxiliary_conditions}
\frac{d^2 \varphi_{c,s}/dt^2}{d \varphi_{c,s}/dt}\Bigg|_* & = 
\frac{d^2 \delta\varphi_{c,s}/dt^2}{d \delta\varphi_{c,s}/dt}\Bigg|_* \\
&= 
-\frac{3}{2}  H^\ETA +  \frac{d \ln  H^{\ETA} }{dt } . 
\end{aligned}
\end{equation}
These conditions are consistent with the fact that the ULA number density redshifts specifically as $a^{-3}$ and hence $\varphi_{c,s} \propto a^{-3/2}$ to optimally suppress oscillations and that the perturbation analogues also evolve only on the Hubble time
(see \cite{Passaglia:2022bcr} for a detailed discussion of the performance of the latter). Note, however, that they are implicit equations since they require $H^\ETA$, the \ETA\ of $H$, which must itself be obtained  from constructing $\rho_\ax^\ETA$.

Apart from this subtlety, which we address below, 
$\rho_\ax^\ETA$ and the other corresponding \ETA\ variables are defined from $\varphi_{c,s}$ through Eq.~(\ref{eq:axionstressenergy}) using the usual cycle-average replacement rules for sinusoidal functions ($\cos^2 =\sin^2 \rightarrow 1/2$ and $\cos \cdot \sin \rightarrow 0$) \cite{Passaglia:2022bcr}: for example,
\begin{eqnarray}
 \rho_\ax^\ETA &=& \frac{1}{2} 
 m^2( \varphi_c^2+ \varphi_s^2 )
 + \frac{1}{4 a^2} (
 \dot\varphi_c^2+
 \dot\varphi_s^2) 
 \nonumber\\
 &&+ \frac{m}{2a}(\varphi_s \dot\varphi_c
 - \varphi_c\dot\varphi_s)
\end{eqnarray}
and likewise for
(see Eq.~\ref{eq:rhoPef})
\begin{equation}
\label{eq:ETAstressenergy}
\{   p_\ax^\ETA, 
\delta\rho_\ax^\ETA, \delta p_\ax^\ETA \}, 
\end{equation} 
whereas the same procedure for $(u_\ax \rho_\ax)^\ETA$ 
defines
\begin{equation}
\label{eq:ETAmomentum}
u_\ax^\ETA \equiv \frac{(u_\ax \rho_\ax)^\ETA}{\rho_\ax^\ETA} \equiv \frac{(u \rho)_\ax^\ETA}{\rho_\ax^\ETA}.
\end{equation}
The \ETA\ of the Hubble parameter then becomes
\begin{equation}
\label{eq:ETAHubble}
(H^\ETA)^2 = -\frac{K}{a^2} + \frac{8\pi G}{3}\sum_{i\ne \ax} \rho_i + \rho_\ax^\ETA,
\end{equation}
where $K = -\Omega_K H_0^2$ is the background curvature parameter. Here and below, we use the shorthand notation $(\ldots)_\ax^\ETA$ to refer to \ETA\ evaluation of the \ULA\ quantities.

To address the implicitness of Eq.\,(\ref{eq:auxiliary_conditions}), we iterate the construction of $\varphi_{c, s}$ and $H^\ETA$ from a starting guess of $H=H^\ETA$ until convergence in
Eq.\,(\ref{eq:auxiliary_conditions}) and 
Eq.\,(\ref{eq:ETAHubble}) is achieved (see Appendix~\ref{app:ETA_construction} for details). This iteration, new to our approach, 
is not a significant correction for the case when the \ULA s are themselves gravitationally insignificant compared with other matter components at the switch \cite{Passaglia:2022bcr} but always removes the leading order oscillation in $H$ regardless.  

We show the resulting evolution of $\rho_\ax^\ETA$ and $H^\ETA$
in Fig.~\ref{fig:background_ETA}.
 In this section alone, we take the \ETA\ at multiple but discrete switch times (red dots) to illustrate the dependence on $m/H_*$.  Consequently, the initial field value $\phi_\ini$  is also taken to be the same for all $m/H_*$ switch values and defined so that $m/H_*=100$ gives the correct final density $\Omega_\ax h^2$ even though the other $m/H_*$ choices would then not exactly hit this value.  In the final \ourcode\ method, there is only a single $m/H_*$ corresponding to the switch to \EFA\ and a single methodology afterwards.   Specifically, for \ourcode\ the corresponding value of $\phi_\ini$ is determined by bisecting an initial range until convergence to the desired final $\Omega_\ax h^2$ is obtained 
(see Appendix \ref{sec:constraint} for the impact of this improvement).

\begin{figure*}[htbp]
\includegraphics[width=0.497\linewidth]{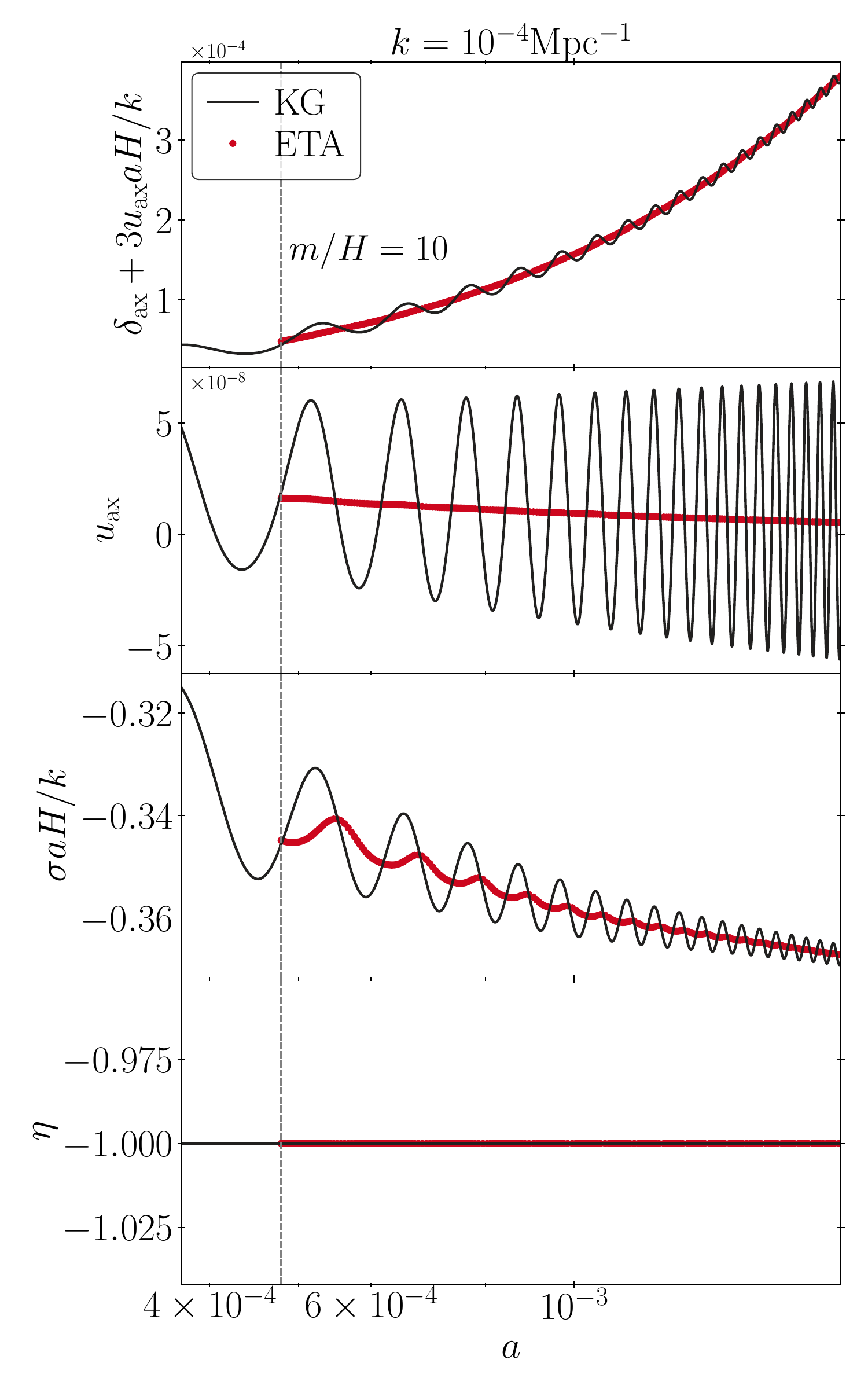}
        \includegraphics[width=0.497\linewidth]{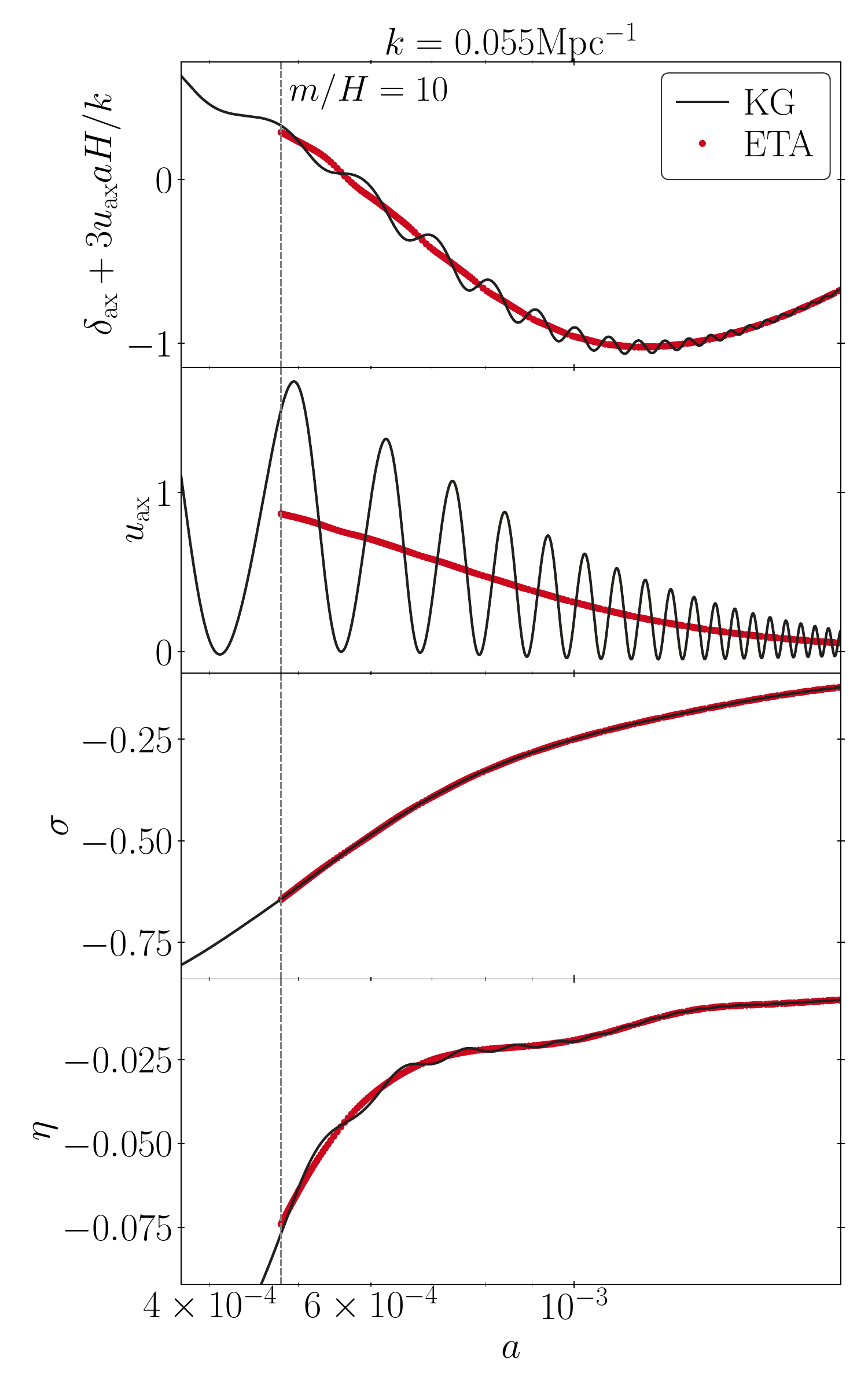}

    \caption{Metric and \ULA\ perturbations (solid lines) and their \ETA\ values (red points) with the same model as  Fig.~\ref{fig:background_ETA} ($m = 10^{-27}$ eV, $f_\DM = 1$) 
    for a superhorizon mode $k = 10^{-4}~$Mpc$^{-1}$ (left) and a subhorizon mode $k = 0.055~$Mpc$^{-1}$ (right).  The temporal range spans  $m/H=7\sim100$, with the start of the \ETA\ at $m/H = 10$ marked with the vertical dashed line. In each case the \ETA\ effectively time averages the leading order KG oscillation (see text for discussion).
    }
    \label{fig:perturbation_ETA}
\end{figure*}

The advantage of this algorithm is that $H^{\rm ETA}$ always gives the effective time-averaged expansion rate regardless of the particular phase at which the switch catches the KG oscillations (see Fig.~\ref{fig:background_ETA}). The design of this procedure introduces a
discontinuity between $H$ and $H^{\rm ETA}$ at the switch to \EFA.   Other approaches demand continuity in $\rho_\ax$ between KG and \EFA\ and therefore $H$ but notice that no condition can provide continuity in both $\rho_\ax$ and $p_\ax$ simultaneously. Pressure discontinuities cause spurious Jeans oscillations in the perturbations as we discuss in Appendix~\ref{app:EFAvars_and_quasistatic}.
Our approach also avoids the problem that the oscillations relative to \ETA\ in $\rho_\ax$ only decay as ${\cal O}(H/m)$, which would cause relatively large errors in the continuous $\rho_\ax$ approach.  

We proceed to define the \ETA\ of metric perturbations in the same way. The metric perturbations themselves are determined from the matter perturbations via the perturbed Einstein equations; in particular, the time-time (00), time-space (0$i$), and their combination with the trace-free space-space ($ij$) Einstein equations are
\begin{align}
\label{eq:perturbedEinstein}
 \frac{1}{2}\frac{\dot{a}}{a}\dot{\hL} -(k^2 - 3K)\etaT  ={}& 4\pi G a^2\sum_i \delta \rho_i,\nonumber\\
     \dot\etaT - \frac{K}{k}\shear ={}& 4\pi G a^2 \sum_i u_i\rho_i /k,\nonumber\\
\dot\shear + \frac{\dot a}{a}\shear 
={}& -4\pi G a^2 \sum_i \Big[ 2 \frac{p_i\pi_i}{k}  \nonumber\\
&{} + \frac{ \delta\rho_i + 3 (\dot a/a)\rho_i u_i/k}{k-3K/k} \Big],
\end{align} 
where $p_i\pi_i$ is the anisotropic stress fluctuation of each component, with $\pi_\ax=0$. Like the background equation for $H$,
these equations transfer the oscillations in $\rho_\ax,\delta\rho_\ax, u_\ax$ to the metric fluctuations.  
Unlike the background equations, these are differential equations, hence those impacts on the metric fluctuations are not solely determined by the KG field variables at the switch, and again require auxiliary conditions to determine fully. 

The auxiliary conditions for the metric fluctuations should uniquely specify the \ETA\ of two independent linear combinations of metric fluctuations, which we can choose without loss of generality to be $\etaT$ and $\shear$. One algebraic combination from the ETA of the matter sources comes from the 00 and 0$i$ equations:
\begin{equation}
\left(\frac{a H}{k} \shear -\etaT \right)^\ETA =  \left( 
4\pi G a^2 \sum_i \frac{ \delta\rho_i + 3 \rho_i u_i a H/k}{k^2-3K} 
\right)^\ETA,
\label{eq:metric_construct}
\end{equation} 
leaving the freedom to define the
partitioning of the \ETA\ of the matter sources between $\etaT^\ETA$ and $\shear^\ETA$.  This choice must obey causality, namely the conservation of curvature for super-horizon adiabatic modes. Given the 0$i$ equation, the deviation induced by free-fall velocities represented by $u_i={\cal O}(k/aH)^3$ (see e.g.~\cite{Hu:2016ssz})
requires that
$\etaT$ is conserved with only ${\cal O}(k/aH)^2$ corrections and hence does not respond to the \ULA\  oscillations. Thus,
\begin{equation}\label{eq:eta_approx}
\etaT^{\ETA} \approx \etaT \qquad (k/aH \ll 1), 
\end{equation}
and $\sigma^\ETA$ follows
from Eq.~(\ref{eq:metric_construct}). 

Conversely, for $k/aH\gg 1$ where $\etaT$  evolves with $u_\ax$ given the $0i$ equation, the 
00 equation implies that $\etaT$ will directly reflect the \ULA\  oscillations from $\delta\rho_\ax$ and evolve $\propto (H/m)$, whereas the $ij$ equation implies that $\shear$ will reflect the time integral of these \ULA\  oscillations, and so be suppressed by an additional $H/m$ factor to be $\propto (H/m)^2$ even without \ETA. Thus, to remove the leading order oscillations, we take
\begin{equation}\label{eq:shear_approx}
\sigma^{\ETA} \approx \sigma \qquad (k/a H \gg 1), 
\end{equation}
and $\etaT^\ETA$ can be obtained in turn
from Eq.~(\ref{eq:metric_construct}) to remove the ${\cal O}(H/m)$ oscillations in $\etaT$.

Thus, $\shear$ and $\etaT$ oscillate in different regimes, motivating an empirical weight
to mediate the transition and fully specify the ETA partitioning
\begin{equation}
\frac{\Delta\etaT}{\Delta\sigma}\equiv
\frac{\etaT-\etaT^\ETA}{\sigma - \sigma^\ETA} = -\frac{a H^\ETA}{k} \frac{W}{1-W},
\label{eq:partition}
\end{equation}
where 
the weight
\begin{equation}\label{eq:metric_weight}
    W = \frac{(k/aH^{\rm ETA})^2}{3 + (k/aH^{\rm ETA})^2}.
\end{equation}
Here the use of $H^\ETA$ rather than $H$ prevents the weight from being sensitive to the phase of the KG oscillation. The scaling is motivated by the
arguments above and the constant $3$ is chosen to roughly optimize the error in evaluating the CMB power spectra and total transfer function, although the error is in fact not very sensitive to the precise value. Nonetheless, this empirical scaling produces excellent agreement with numerical results for the amplitude of the metric oscillations as shown in Appendix~\ref{app:ETA_construction}.

Finally, with the partitioning of Eq.~(\ref{eq:metric_weight}), the KG values of $\shear$ and $\etaT$ at the switch, and Eq.~(\ref{eq:metric_construct}) for the matter \ETA\ source, we solve for $\shear^\ETA$ and $\etaT^\ETA$. In Fig.~\ref{fig:perturbation_ETA}, we illustrate this construction for a super-horizon mode $k = 10^{-4}$\,Mpc$^{-1}$ and a sub-horizon mode $k = 0.055$\,Mpc$^{-1}$, corresponding to the two limiting behaviors of Eqs.\ (\ref{eq:eta_approx}) and (\ref{eq:shear_approx}). In the top panels for both modes, we plot the \ULA\  variables $\delta_\ax + 3u_\ax aH/k$ and $u_\ax$, with $m/H$ varying from $7$ to $100$ across the plot and \ETA\ starting at $m/H = 10$. This verifies that the \ETA\ does effectively average over  the temporal oscillations of the KG system and removes the leading order terms in $H/m$. Note that  $u_\ax^\ETA$ approaches zero as $H/m$ decreases whereas the KG oscillations are unsuppressed by powers of $H/m$ which we will see makes the former a much better match to the \EFA. On the metric side, for the superhorizon mode, $\etaT$ remains constant as expected from
Eq.~(\ref{eq:eta_approx}) and its \ETA\ is accordingly the same value. As dictated by Eq.~(\ref{eq:metric_construct}), the KG oscillations in the \ULA\  are then transferred to $\shear$ through $\shear aH/k$, and its \ETA\ does remove the KG oscillations to leading order in $H/m$, albeit with a noticeable $(H/m)^2$ fractional residual which improves rapidly for later switches.  
 
In the subhorizon panel, we plot $\shear$ rather than $\shear aH/k$ 
and verify the expectations of Eq.~(\ref{eq:shear_approx}) that
$\shear$ in the KG system does not oscillate at ${\cal O}(H/m)$. Instead, $\etaT$  carries these oscillations of the KG system and the \ETA\ effectively removes this leading order term. While $\shear aH/k$ would still oscillate, its oscillations reflect those of $H$ itself. Finally, note that at late times but before the \ULA s fall into baryonic potential wells, the velocity $v_\ax^\ETA \equiv u_\ax^\ETA/(1+w_\ax^{\ETA}) \approx u_\ax^{\ETA} \rightarrow -\shear$, which is the expectation from hydrostatic equilibrium where Newtonian gauge velocities are driven to zero (Appendix~\ref{app:EFAvars_and_quasistatic}). 

In summary, for all matter and metric variables and scales, the \ETA\ procedure effectively removes the KG oscillations at the ${\cal O}(H/m)$ level and attains at least ${\cal O}(H/m)^2$ accuracy. The \ETA\ procedure then removes the sensitivity to the phase of the KG oscillations at the chosen switch time to this order. In principle, this procedure could be refined to capture the next order, but as we shall see in Sec.\ \ref{sec:CMB_changes}, the accuracy of cosmological observables is already sufficiently high for $m\gg 10 H_{\rm eq}$ or for $f_\DM \ll 1$, and when higher accuracy is required, it is easier to simply raise the value of $m/H_*$ accordingly.

\subsection{Effective Fluid Approximation}\label{sec:EFA}

We have shown that the \ETA\ gives us an accurate prescription of the time-average of the KG variables at any desired value of the switch time and is robust to the phase of the KG oscillations at leading order in $H/m$, unlike other techniques. From that switch time, we can then evolve the time average under an effective fluid approximation, i.e.\ \EFA, provided that we can accurately characterize the latter using a fluid description and conservation of energy-momentum.

\begin{figure}
      \includegraphics[width=\linewidth]{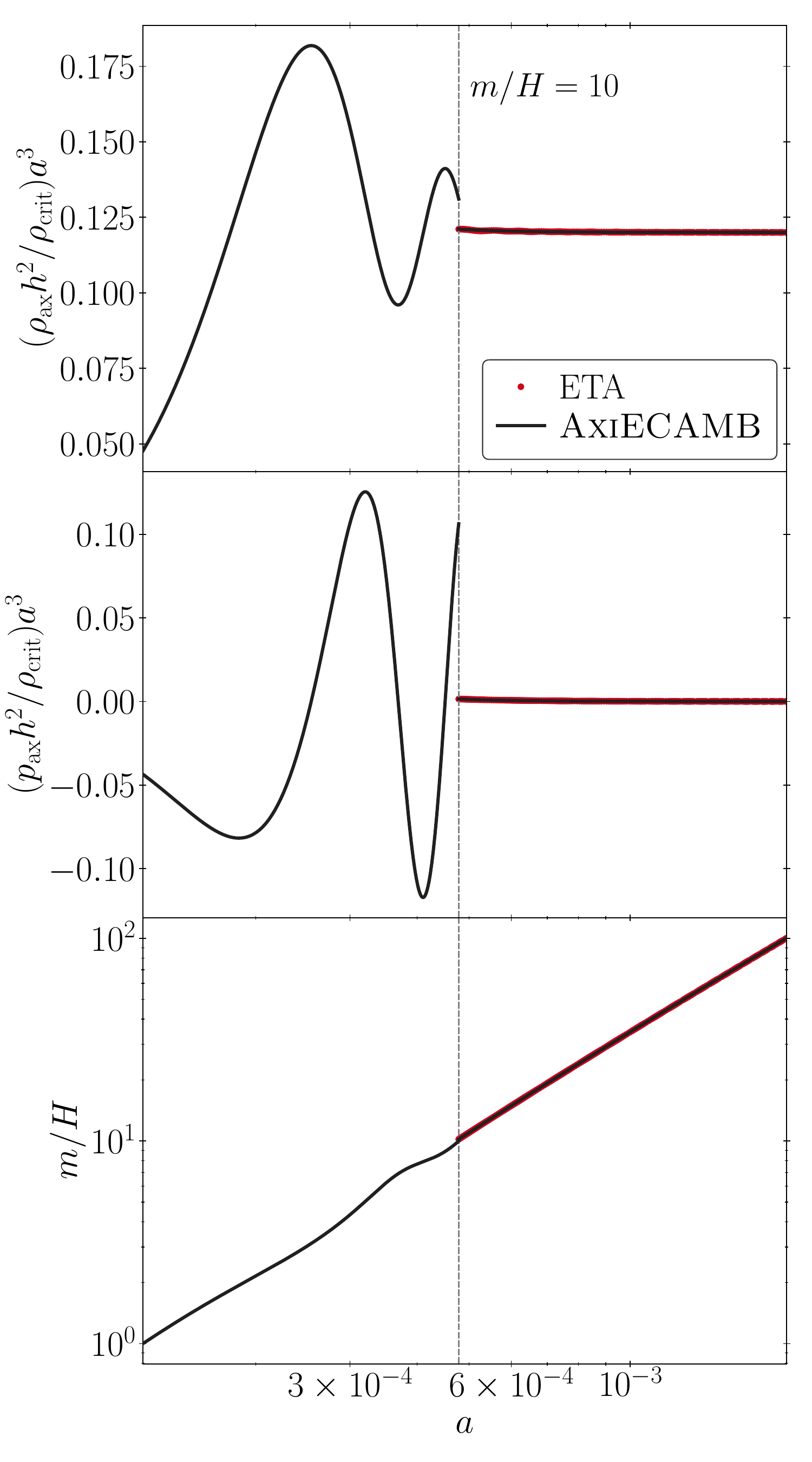}
    \caption{Background evolution of the full \ourcode\ method
    from KG to \EFA\ (solid lines, with the switch to \EFA\ at $m/H_*=10$) compared with \ETA\ (red points after $m/H=10$) for the same case as Fig.~\ref{fig:background_ETA} ($m = 10^{-27}$ eV, $f_\DM = 1$). 
    }
    \label{fig:background_EFA}
\end{figure}

By implementing this as a code switch in \ourcode, no evolution variables are solved across the discontinuity introduced by the switch (see Appendix \ref{app:cmbsources_corr}).  Instead, the \ETA\ provides the initial or switch conditions for the separate evolution of the \EFA. Unlike in the tests of the previous section, there is no need to solve the KG system past the switch. Therefore, in this section, we  compare the \EFA\ evolved system from a fiducial switch time $m/H_*=10$ to the \ETA\ 
evolution established previously, and assess the errors induced by the switch conditions and the \EFA\ modeling.

To model the \EFA, we first need to characterize the effective fluid equation of state $w_\ax^\EFA$.
Following Ref.~\cite{Passaglia:2022bcr}, we parameterize the time evolution of $w_{\rm ax}^{\rm EFA}$ as 
\begin{equation}
\label{eq:wEFA}
w_\ax^\EFA \equiv \frac{p_\ax^\EFA}{\rho_\ax^\EFA} = A_w \left(\frac{H}{m}\right)^2,
\end{equation}
where $A_w$ is a constant to be specified, and the scaling with $H$ is motivated by the behavior of $p^\ETA/\rho^\ETA$ (see Appendix~\ref{app:cmbsources_corr} and Fig.~\ref{fig:wETA_scal}).
This equation of state defines the evolution of the 
background via the continuity equation
\begin{equation}
\dot \rho_\ax^\EFA + 3(1+w_\ax^\EFA) \frac{\dot a}{a} \rho_\ax^\EFA =0, 
\end{equation}
as well as the adiabatic sound speed
\begin{equation}
c_\text{ad}^2 \equiv\frac{\dot{p}_\ax^\EFA}{\dot{\rho}_\ax^\EFA}=
w_\ax^\EFA\left( \frac{1+w_T }{1 + w_\ax^\EFA}+1 \right),
\end{equation}
where $w_T = p_T/\rho_T$ is the total equation of state. 

With the effective method for the background complete,
a final subtlety applies to the definition of $\Omega_\ax h^2$ itself, which we recall determines
the initial field $\phi_{\rm ini}$.  For  masses $m \ge 10 H_0$, we use $\rho_\ax^\ETA(a=1)$ and $H_0=H^\ETA(a=1)$  even when $m/H_*$ is at $a>1$, since cosmological observables reflect the Hubble time averaged effect of \ULA s.\footnote{Technically this is achieved by forcing a switch at  $a = 1 - 10^{-3}$ and using $\rho_\ax^\EFA$ at $a=1$.}  For masses where $m/H_0 < 10$, the oscillations are not rapid at the present and we use the instantaneous $\rho_\ax(a=1)$ and $H_0=H(a=1)$ solved from the KG system.
In this latter regime, the \ULA\ behave as dark energy all the way to the present.

\begin{figure*}[htbp]
        \includegraphics[width=0.49\linewidth]{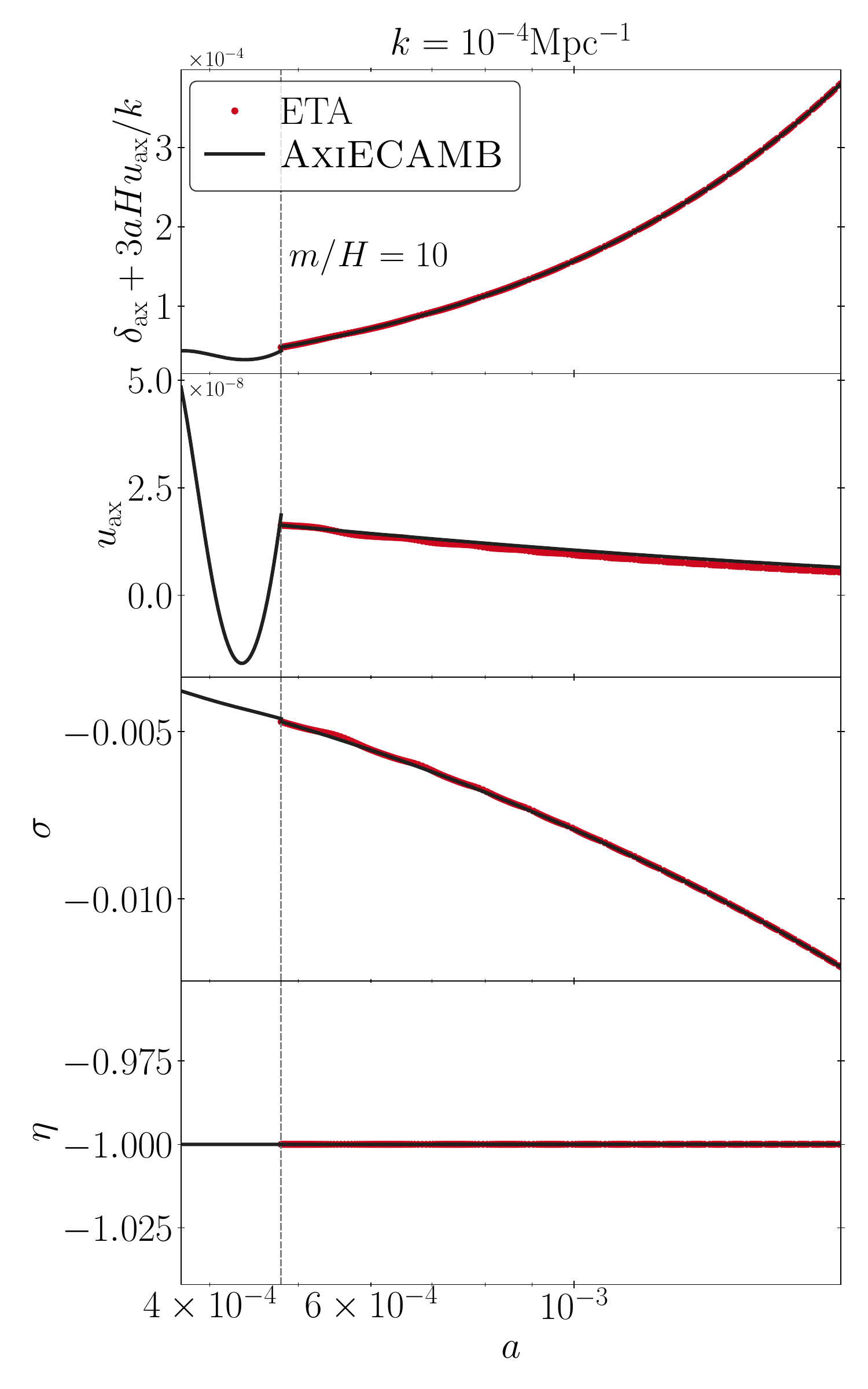}
    \hfill
        \includegraphics[width=0.49\linewidth]{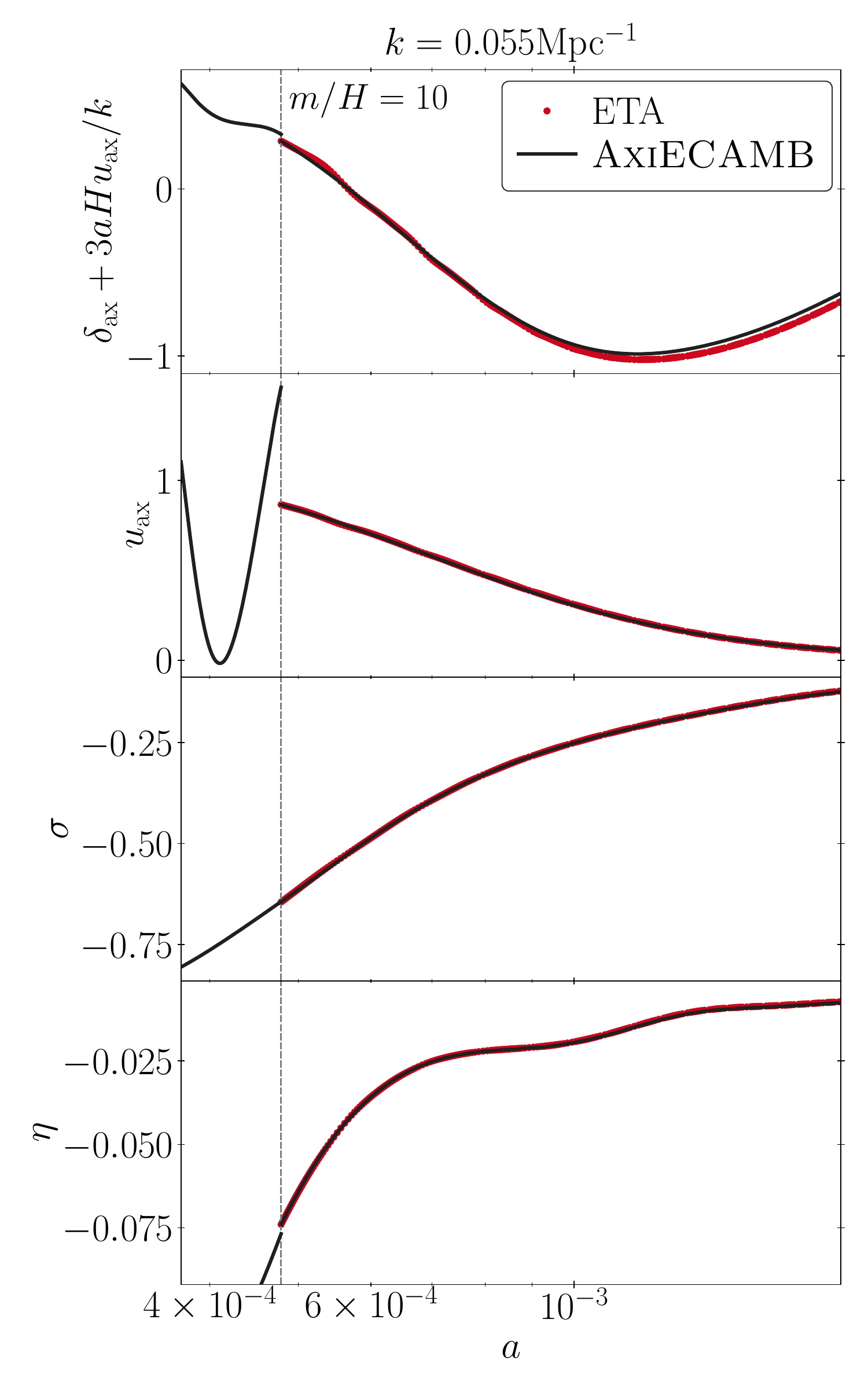}
    \caption{Perturbation evolution of the \ourcode\ method (solid line) compared with \ETA\ (red points) for the same modes as Fig.~\ref{fig:perturbation_ETA}. The switch occurs at $m/H_*=10$ and the temporal range extends to  $m/H = 100$.}
    \label{fig:perturbation_EFA}
\end{figure*}

Proceeding to the perturbations, we define the fluid pressure perturbation by the rest frame sound speed \cite{Hu:1998kj}, which we take as
\begin{align}
\label{eq:cs2}
c_\ax^2 ={}&   \left[ \frac{ \sqrt{1+\left(\frac{k}{a m} \right)^2} -1}{\frac{k}{am}} \right]^2
+ A_c \left(\frac{H}{m}\right)^2,
\end{align}
where $A_c$ is a constant to be specified at the switch by the pressure perturbation as $k\rightarrow 0$ where the first term vanishes.  This form is also motivated by the \ETA\ 
of KG oscillations $\delta p_\ax^\ETA/\delta \rho_\ax^\ETA$ (see Appendix~\ref{app:KG}).
Energy-momentum conservation  dictates that the perturbation fluid variables obey
\begin{align}
\label{eq:fluideom}
\dot{\delta \rho}_\ax^\EFA + 3 {\frac{\dot{a}}{a}} \left(\delta \rho + \delta p\right)_\ax^\EFA ={}& -k (\rho u)_\ax^\EFA 
- (\rho + p)_\ax^\EFA \frac{\dot\hL}{2}, \nonumber\\
\left[\frac{d}{d\ct} + 
4 \frac{\dot a}{a}\right] {(\rho u)_\ax ^\EFA} ={}& k \,\delta p_\ax^\EFA\, ,
\end{align}
where
\begin{equation}
\delta p_\ax^\EFA = c_\ax^2 \delta\rho_\ax^\EFA + 3\left( c_\ax^2- c_\text{ad}^2\right)  \frac{a H}{k} (\rho u)_\ax^\EFA.
\end{equation} 
In the code itself, we evolve $\delta_\ax^\EFA\equiv\delta\rho_\ax^\EFA/\rho_\ax^\EFA$ and the momentum flux $u_\ax^\EFA\equiv(\rho u)_\ax^\EFA/\rho_\ax^\EFA$.

To fully determine the \EFA\ solutions, we need the switch values of the matter and metric variables as well as the constants $A_w$ and $A_c$ which define the equation of state and sound speed after the switch. Errors from both contribute to the final error of our approach. The original approach in Ref.~\cite{Passaglia:2022bcr} was to derive $A_w=3/2$  analytically in radiation domination (i.e.\ appropriate for $m\gg 10 H_{\rm eq}$) to minimize equation of state errors, and to fix $A_{c}=5/4$ by setting $\lim_{k\rightarrow 0} c_\ax^\EFA = c_\ax^\ETA$ to minimize switching errors instead (see Appendix~\ref{app:KG}).
This method has the disadvantage that $p_\ax^\EFA$ and $v_\ax^\EFA $ are discontinuous with their \ETA\ equivalents, which both induces discontinuities in the metric and other observables if $m \lesssim 10 H_{\rm eq}$, and breaks hydrostatic equilibrium below the Jeans scale 
(see Appendix~\ref{app:EFAvars_and_quasistatic}). Furthermore, their procedure would give a result that differs from $A_w=3/2$ when the background is not radiation dominated, removing the advantage of having an analytic prescription.

With these considerations, we modify the approach of Ref.~\cite{Passaglia:2022bcr} to make the \EFA\ to \ETA\ pressure continuous  by instead deriving $A_w$ from its \ETA\ value at the switch. This gives our final choice for the switch conditions to the \EFA\ at $m/H_*$:
\begin{equation}
(p,\rho,\delta \rho, u)_\ax^\EFA= 
(p,\rho,\delta \rho, u)_\ax^\ETA,
\label{eq:fluidmatching}
\end{equation}
which then automatically sets 
\begin{equation}(H,\shear,\etaT)^\EFA=(H,\shear,\etaT)^\ETA
\end{equation}
as well as the value of 
\begin{equation}
A_w= \frac{ p_\ax^\ETA }{\rho_\ax^\ETA}\left( \frac{m}{H_*^\ETA} \right)^2, 
\end{equation} 
since $w^\EFA_\ax=w^\ETA_\ax$.  For simplicity, we keep $A_c=5/4$ since the impact of a refined treatment is minimal in practice (see Appendix~\ref{app:KG} for the comparison to analytic expectations).

For the background, this completes the \ourcode\ method.
In Fig.~\ref{fig:background_EFA} we compare the background evolution obtained using this approach for a switch at $m/H_*=10$ with the \ETA\ after the switch, using the same mass ($10^{-27}$ eV) and \ULA\  fraction in dark matter ($f_\DM = 1$) as in Fig.~\ref{fig:background_ETA}. 
In the \EFA\ regime fractional differences are nearly invisible and substantially below $(H_*/m)^2=0.01$.

The same comparison is carried out for the perturbation variables in Fig.~\ref{fig:perturbation_EFA} after the fiducial switch $m/H_* = 10$ for the same $k$-modes as in Fig.~\ref{fig:perturbation_ETA}. For the superhorizon mode (left panels), small differences in $u_\ax$ come mostly from the modeling of the \EFA\ sound speed $c_\ax^2$ but the overall impact on the dynamically relevant density perturbation is negligible. On these scales, the synchronous gauge $v_\ax$ reflects tiny deviations from free-fall, where $v_\ax=0$,
due to the finite pressure perturbation as $k\rightarrow 0$. For $\shear$, the main error initially is the matching error due to the visible oscillations in $\ETA$ at ${\cal O}(H/m)^2$ which do not track the mean of the KG system. Since $\shear$ is dynamically sourced by density fluctuations according to Eq.~(\ref{eq:perturbedEinstein}), this main error evolves away.  The curvature $\etaT$ remains properly conserved under both \ETA\ and \EFA.

For the subhorizon mode, there are also matching and evolution errors. In the density and momentum perturbations there are small oscillations in the \ETA\ that cause errors that are carried forward as offsets in the \EFA. 
The fractional errors in the density perturbation are in the regime where the perturbation itself is already highly suppressed (see Fig.~\ref{fig:perfTk_allDM}).
Notably, $\etaT$ shows only ${\cal O}(H/m)^2$ differences but carries oscillations on a longer time scale. These are associated with the acoustic oscillations of the photon-baryon system. While the errors that are transferred via the metric 
remain $(8\pi G \rho_\ax/3H^2){\cal O}(H^2/m^2)$,  a later switch time than $m/H_*=10$ is required to achieve better than percent level accuracy for large $f_\DM$, which we will address in detail in the next section.

\subsection{Switch time and CMB sources}\label{sec:CMB_changes}

One major goal of our code \ourcode\ is accurate predictions of CMB observables for all \ULA\ masses. This places additional requirements on the switch time $m/H_*$  when it is near matter-radiation equality or recombination.  In addition, the discontinuities at the switch require that CMB anisotropy source functions be augmented with boundary terms.

The first consideration pertains to cases where the switch is near matter-radiation equality and $f_\DM \sim 1$, which, as we noted in the previous section, needs an $m/H_*>10$ switch to ensure better than 1\% accuracy in the metric and thus also for the CMB.
We expect that at $(8\pi G \rho_\ax/3H^2){\cal O}(H^2/m^2)$,
photons respond to the KG oscillations through the metric perturbations as shown in Appendix~\ref{app:cmbsources_corr}.
The proportionality factor in this scaling can also be $k$-dependent since for $k/aH\gg 1$, the acoustic frequency is larger than the Hubble rate. 

While our \ETA\ procedure in Eq.~(\ref{eq:metric_construct}) removes the leading order effects from the KG oscillations, it does not address these higher-{order} responses. 
Furthermore, the strong sensitivity of CMB anisotropy to the epoch of recombination means that it is always better to avoid placing the switch there regardless of $f_\DM$.

Specifically, starting from a baseline switch choice of $m/H_*$ ($= 10$ for our default accuracy), we raise the $m/H_*$ value of the switch in two ways:

\begin{itemize}
\item For $m<10^{-25}$\,eV, if the baseline switch is after $\rho_\ax/\rho_{r} =0.03$ (estimated by assuming the \ULA\ redshift as matter after the switch),  
but before $z= 1300$, increase $m/H_*$  to around twice the baseline value, specifically corresponding to the best mass-oscillation phase (see below).
\item Then,
if the proposed switch is near recombination $z_*\in(800, 1300]$, increase $m/H_*$  such that the switch is just below $z=800$.
\end{itemize}
In the first step, we have taken advantage of the independence of our \ETA\ technique on the phase of the KG oscillations to further improve accuracy over the naive scaling by choosing the phase that minimizes the response in the photon-baryon acoustic oscillations to the switch phase (see Appendix~\ref{app:cmbsources_corr} for details).  In the second step, we reduce switch artefacts in the CMB anisotropy source function itself  by not allowing the switch to occur near recombination, where it would otherwise be by default for $m \sim  10^{-27.5}$\,eV (see Fig.~\ref{fig:Clerror_recskipcompare}). We discuss this source function next.
The combination of these steps affects masses $10^{-27.5} \lesssim m/{\rm eV}\lesssim 10^{-25}$.  In this regime, the first step only applies to cases where $f_\DM \gtrsim 
(m/10^{-23.5}{\rm eV})^{1/2}$.  
\begin{figure}
    \includegraphics[width=1\linewidth]{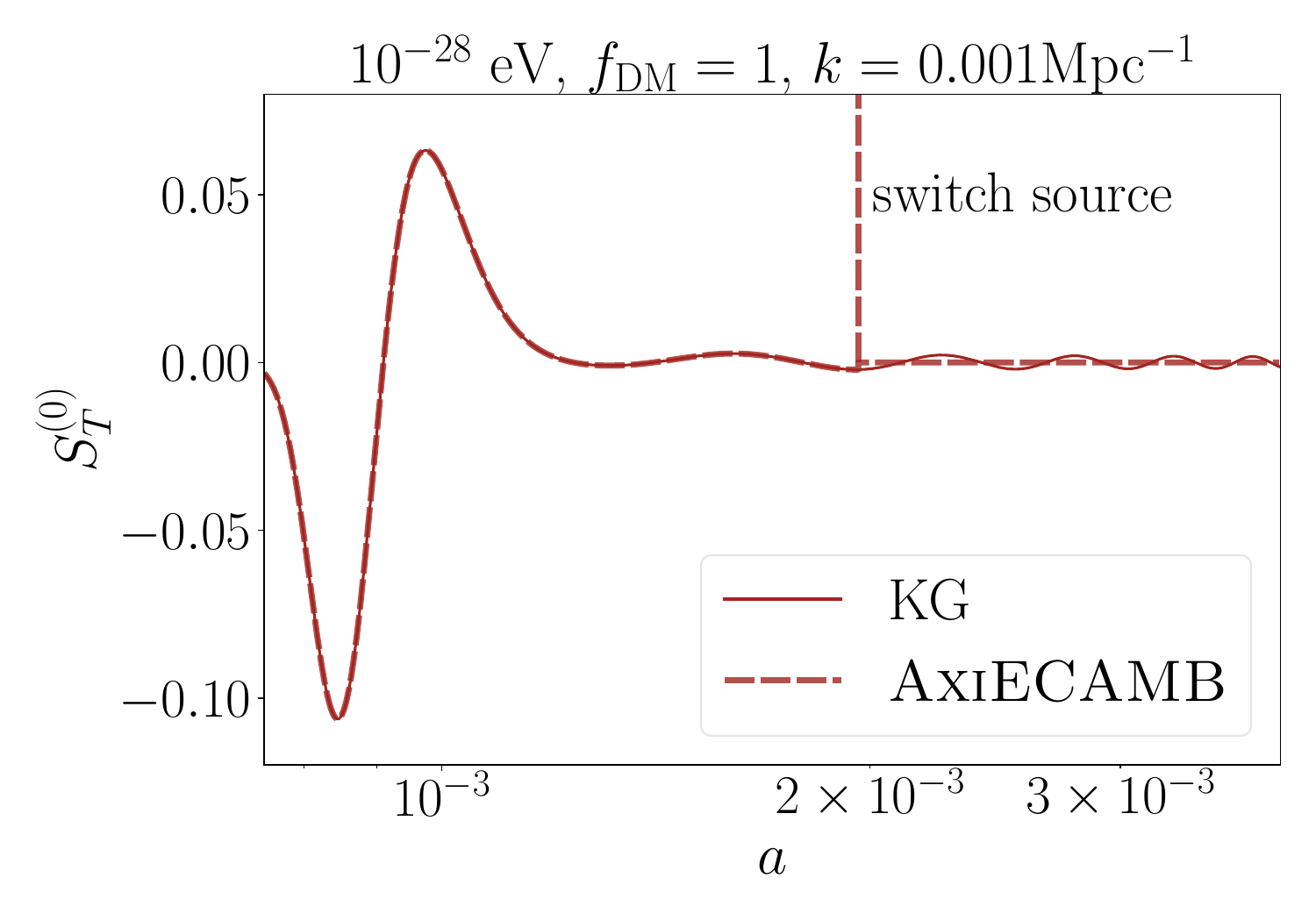}
    \caption{Total CMB source function $S_T^{(0)}$ in  Eq.~(\ref{eq:ST}) for $m = 10^{-28}$ eV, $f_\DM = 1$, $k = 0.001$Mpc$^{-1}$.
    The thin solid black line is the KG solution and the thick red dashed line is our effective theory for a switch at $m/H_* = 10$.  At the switch, the integration by parts adds a boundary term or delta function source represented here by the vertical line.}
    \label{fig:sourcefcn_timedomain}
\end{figure}

With this procedure for setting the switch time $m/H_*$, the methodology of the matter and metric evolution for \ourcode\ is complete. 
On the other hand, CMB Boltzmann codes obtain their anisotropy predictions by a line-of-sight integral of the perturbation sources. 
For example, the CMB temperature power spectrum is given by (see e.g.~Ref.~\cite{Lesgourgues:2013bra}) 
\begin{equation}
C_\ell^{TT} = 4\pi \int d\ln k \Delta_\ell^{T(0)} \Delta_\ell^{T(0)}  {\cal P}_{\cal R}(k),
\end{equation}
where ${\cal P}_{\cal R}(k)$ is the initial curvature power spectrum, and the photon transfer function $\Delta_\ell^{T(0)}$ is the integral over the $S_{T_j}^{(0)}$ source functions
\begin{equation}
\begin{aligned}
\Delta_\ell^{T(0)} &= \int d\tau \sum_{j=0}^{2} 
S_{T_j}^{(0)}\phi_\ell^{j0}[\chi, \nu] \\
&\equiv\int d\tau S_{T}^{(0)} \phi_\ell^{00} [\chi, \nu]
+ B_\ell^{T(0)}.
\end{aligned}
\label{eq:CMBsourcefunction}
\end{equation}
Here $\phi_{\ell}^{jm}$ are the radial eigenfunctions for an $j m$ multipole source and are given explicitly by the 
 hyperspherical Bessel functions \cite{Hu:1997mn} while the radial distance of the source  $\chi = \sqrt{|K|}\cdot(\tau_0 - \tau)$ and wavenumber  $\nu = \sqrt{k^2 + K}/\sqrt{|K|}$ are given in terms of the FRW 
spatial curvature $K$ and $\tau_0=\tau(a=1)$.
The $S_{T_j}$ source functions are given by
\begin{align}
S_{T_0}^{(0)} &= e^{-\mu}\left(\frac{\dot\mu\delta_\gamma}{4} - \frac{\dot\hL}{6}\right),\\
S_{T_1}^{(0)} &= e^{-\mu}\dot\mu v_b ,\\
S_{T_2}^{(0)} &= e^{-\mu}\left(\dot\mu P^{(0)}  + \frac{2}{3}\sqrt{k^2-3K}\,\shear \right),
\label{eq:S2}
\end{align}
where $\delta_\gamma$ is the photon density perturbation, $\dot\mu = n_e \sigma_T a$ is the Thomson opacity, $v_b$ is the baryon velocity, and $P^{(0)}$ is the photon polarization source
\cite{Hu:1997mn}.
 Note that the total source function $S_T^{(0)}$ is then defined via integration by parts using relationships between the hyperspherical Bessel functions,
\begin{equation}\label{eq:ST}
S_T^{(0)} = S_{T_{0}}^{(0)}
+ \frac{\dot S_{T_{1}}^{(0)}}{k}   + \frac{1}{2}\frac{k^2 S_{T_{2}}^{(0)} + 3 \ddot S_{T_{2}}^{(0)}}{k\sqrt{k^2-3K} },
\end{equation}
and $B_\ell^{T(0)}$ is the boundary term associated with the integration by parts.\footnote{Integration by parts is employed in 
codes based on  \textsc{CMBFast} \cite{Seljak:1996is} such as \textsc{CAMB} \href{http://camb.info}{[http://camb.info]} where only the hyperspherical Bessel function itself is constructed, but not in \href{https://lesgourg.github.io/class_public/class.html}{CLASS} \cite{Lesgourgues:2011re}, where the multipole sources are directly integrated using the radial functions.} Usually the boundary term vanishes since the boundaries are at $\tau=0$ and $\tau=\tau_0$ but for our case where there is a new boundary at the switch which implies 
\begin{equation}
\begin{aligned}
B_\ell^{T(0)} =  \frac{3\left( \Delta S_{T_{2}}^{(0)}\dot\phi_\ell^{00} -\Delta \dot S_{T_{2}}^{(0)}\phi_\ell^{00} \right)}{2k \sqrt{k^2-3K} }
-\frac{ \Delta S_{T_{1}}^{(0)}\phi_\ell^{00}}{k} ,
\end{aligned}
\end{equation}
evaluated at $\tau_*$ 
where the boundary change $\Delta$ of any quantity $X$ is defined as
\begin{equation}
\Delta X = X^{\rm KG}- X^\EFA
\end{equation}
and similarly for the derivative.   More explicitly,
given our KG to \EFA\ transition the nonvanishing switch source contributions are\footnote{Technically, the derivatives of $\dot\mu$ carry the difference in the expansion rate ($\Delta H$) but its values are interpolated across the switch and the derivative of the interpolation function is continuous. This also makes the procedure transparent to improvements in the treatment of recombination.}
\begin{eqnarray}
\frac{\Delta S_{T_2}^{(0)}}
{\sqrt{k^2-3K}}
&=& \frac{2}{3} e^{-\mu}
 \Delta\shear,\\
\frac{\Delta \dot S_{T_2}^{(0)}}
{\sqrt{k^2-3K}}
&=& \frac{2}{3} e^{-\mu}\left[k\Delta\etaT +\frac{11}{10}\dot\mu\Delta\shear -2a\Delta(H\shear) \right] ,\nonumber
\end{eqnarray}
where we have used  Eq.~(\ref{eq:perturbedEinstein}) and Ref.~\cite{Hu:1997mn} (their Eqs.\ 38,42)
to obtain
\begin{align}
\Delta \dot \shear &= k\Delta\etaT -2a \Delta(H\shear)  ,\nonumber\\
\frac{\Delta \dot P^{(0)}}{\sqrt{k^2 - 3K}} &= \frac{1}{15} \Delta\shear.
\end{align}

Fig.~\ref{fig:sourcefcn_timedomain} shows the total source function $S_T^{(0)}$ for $m=10^{-28}$\,eV and $k=10^{-3}$\,Mpc$^{-1}$ with a switch after recombination.  In $\Lambda$CDM, the source function after recombination gives the integrated Sachs-Wolfe effect which is negligible for this super-horizon mode at the switch epoch. With \ULA\ dark matter it is suppressed by $(m/H)^2$ in the KG system as well.  Notice however that because the quadrupole source $S_{T_2}^{(0)}$ involves the metric fluctuation $\shear$ in Eq.~(\ref{eq:S2}), the integration by parts enhances these suppressed KG oscillations by a factor of $m/H$ for each derivative in $\ddot S_{T_2}^{(0)}$ in Eq.~(\ref{eq:ST}), leading to prominent oscillations in $S_T^{(0)}$ around a nearly zero mean value.  The second derivative produces an $\dot H \shear$ term which then reflects the ${\cal O}(1)$ amplitude \ULA\ pressure oscillations.   If the KG system were solved to later times (shown to $m/H=100$), the integration of these oscillations would have cancelling contributions but because of the switch, this cancellation is between the leading-order switch source $\Delta (H \shear)$ and the KG integration before the switch (shown schematically as the vertical spike).  The boundary terms can therefore be thought of as a delta function source (henceforth dubbed ``switch source") to the monopole and dipole sources which cancel the incomplete KG integration that is truncated at the switch.  As we show in Appendix \ref{app:cmbsources_corr} and Fig.~\ref{fig:BCsourcecomp_clerror}, omitting the switch sources can produce larger than cosmic variance errors at low multipoles that depend on the exact phase of the KG oscillations at the switch. 

Polarization sources follow a similar form but do not require separate integration by parts and hence do not require additional boundary terms at the switch.

\section{CMB and Matter Power Spectra} \label{sec:Performance}

We evaluate the impact of \ULA s on the CMB and matter power spectra, and assess the accuracy of \ourcode, with the baseline switch set at $m/H_*=10$ (see Sec.~\ref{sec:CMB_changes} for details).
We compare them to a high accuracy calculation which increases code accuracy settings and solves the exact KG system to a much later epoch of $m/H_*=100$, a choice justified by the ${\cal O}(H_*/m)^2$ or smaller errors  seen in Sec.~\ref{sec:EFA_methodology}.
We show the improvement of our technique over previous work through a similar comparison in Appendix \ref{sec:constraint}, where we also test matter power spectra fitting functions that are useful for fuzzy dark matter and more efficient than \ourcode\ for even larger masses.

In Sec.~\ref{sec:accuracymetrics} we first establish the target accuracy for the CMB and matter power spectra based on that of the equivalent $\Lambda$CDM computation.
We then consider the accuracy of the \ULA\ computation in various mass ranges. Since the CMB and large scale structure set the target accuracies, we focus on the mass range where \ULA\ effects are significant, $10^{-33} \lesssim m/{\rm eV} \lesssim 10^{-24}$, and further divide this range into cases where they can comprise of all of the dark matter (Sec.~\ref{sec:totalDM}), part of the dark matter 
(Sec.~\ref{sec:fracDM})
and all of the dark energy (Sec.~\ref{sec:totalDE}), respectively. In Appendix \ref{sec:FDM}, we treat the range $m \gtrsim 10^{-24}$\,eV  where the \ULA s can affect small scale structure and nonlinear objects.

\begin{figure}
    \includegraphics[width=1\linewidth]{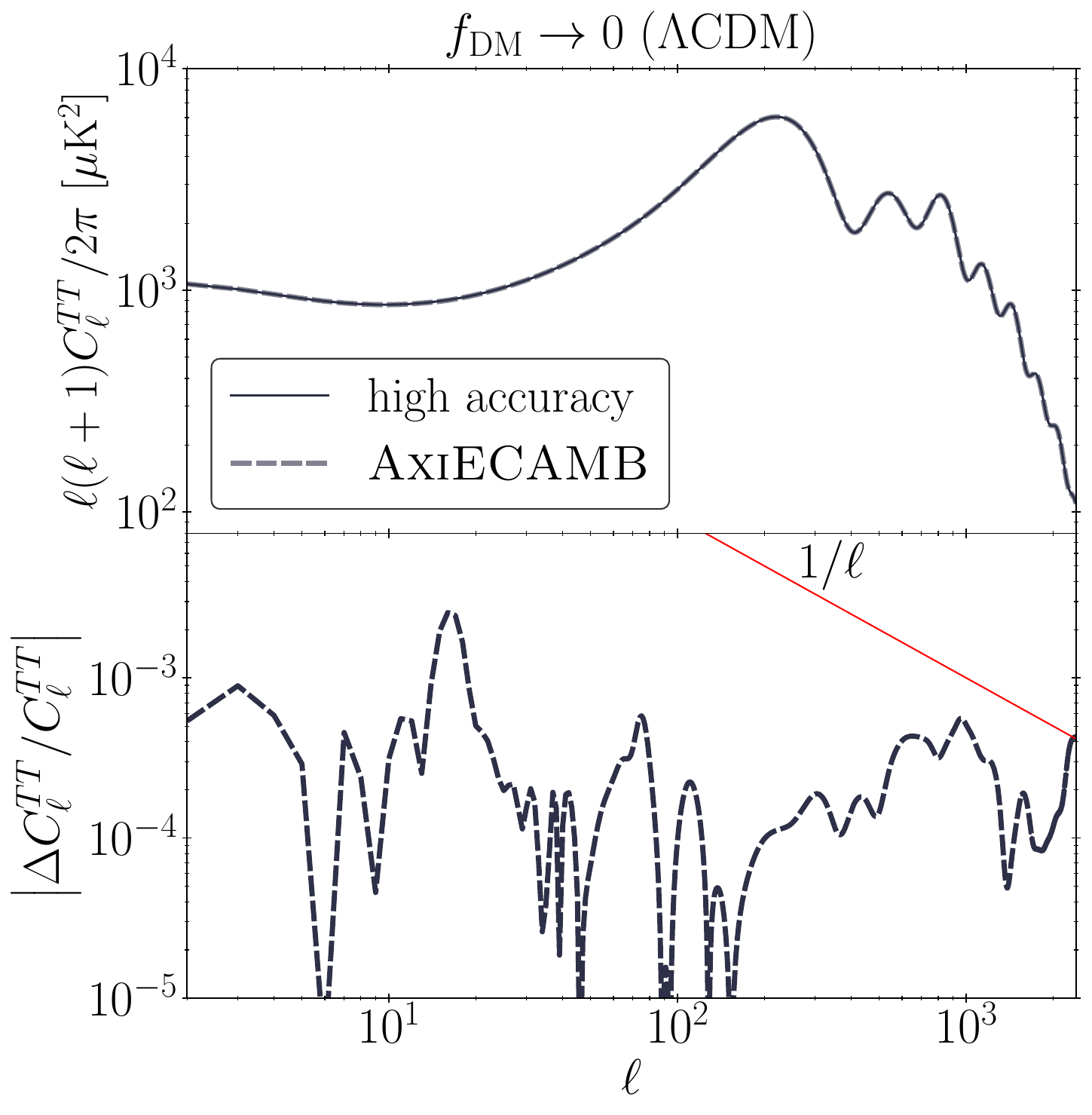}
    \caption{$\Lambda$CDM temperature spectrum $C_\ell^{TT}$ (top) and its fractional error 
    (bottom) between the default \ourcode\ and the high accuracy calculation, benchmarking the error that is not \ULA\ related. 
        }
    \label{fig:ClTT_LCDMbenchmark}
\end{figure}

\begin{figure}
    \includegraphics[width=1\linewidth]{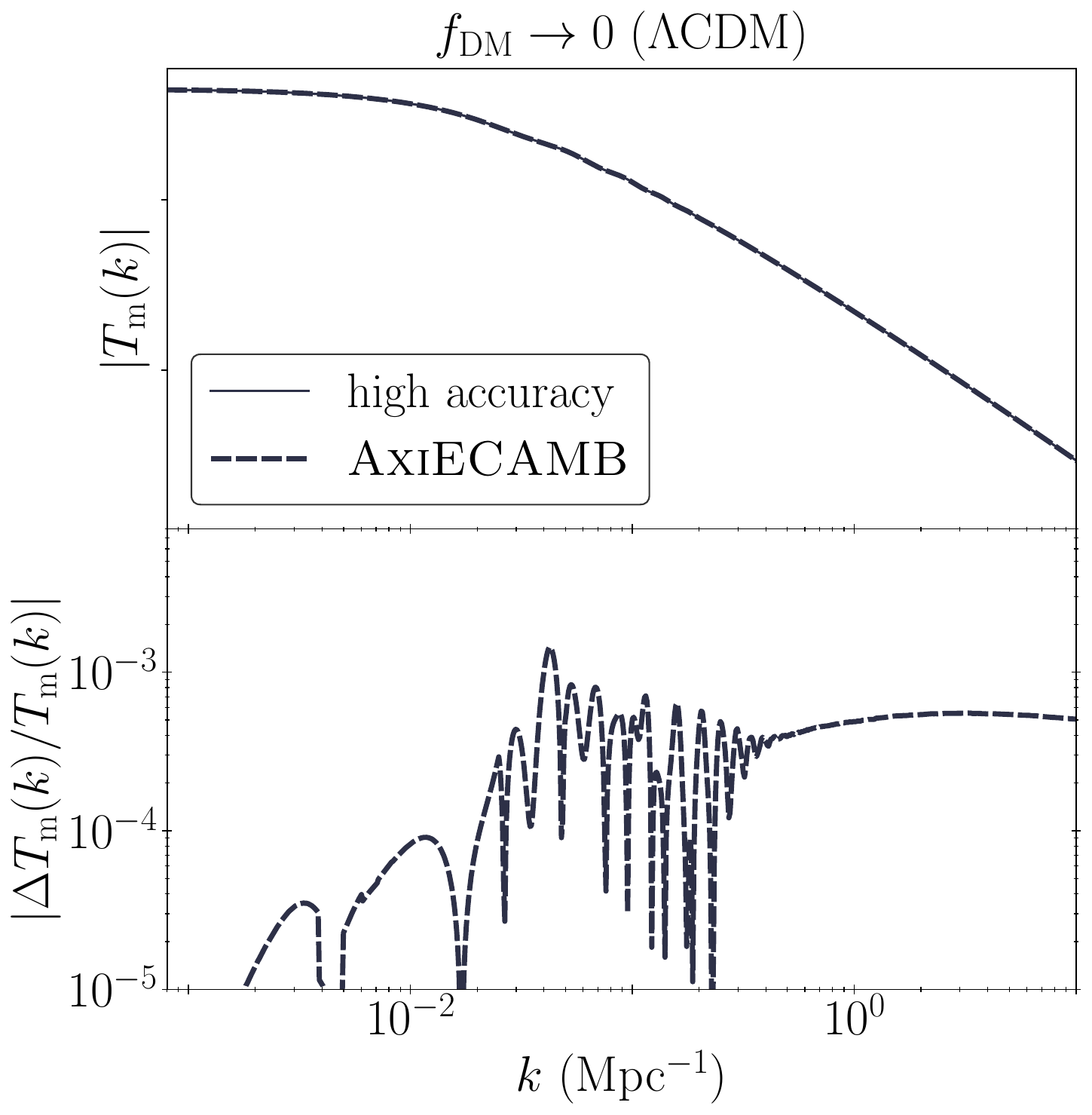}
    \caption{$\Lambda$CDM matter transfer function $|T_{\rm m}(k)|$ (top) 
    and fractional error 
    between the default \ourcode\ and the high accuracy calculation, benchmarking the error $\Delta T_{\rm rel}(k)$ that is not \ULA\ related.}
    \label{fig:Tmk_LCDMbenchmark}
\end{figure}

\subsection{Accuracy metrics}
\label{sec:accuracymetrics}

To assess whether accuracy is sufficient for future CMB experiments, we
compute the $\Delta\chi^2$ between the default and high accuracy calculations:
\begin{equation}
\Delta\chi^2 = \sum\limits_{\ell=2}^{\ell_{\rm max}} \sum_{XY}\Delta C_\ell^{X} (\text{Cov}_\ell)^{-1}_{XY}\Delta C_\ell^{Y} ,
\end{equation}
where ${X,Y}\in {TT,TE,EE}$. 
The covariance matrix is between measurements of
$C_\ell^{X}$ and $C_\ell^{Y}$, which we take to be cosmic variance (CV) limited in all three spectra out to 
$\ell_{\rm max}=2400$ (see e.g.~\cite{Eisenstein:1998hr}). In addition, to establish the connection between the $\ell$-space  errors in $C_\ell^{TT}$ and this type of summary statistic, we consider a cosmic variance limited experiment in ${TT}$ only 
where
\begin{equation}\label{eq:Deltachi2_TT}
\Delta\chi^2_{TT}=\sum_{\ell=2}^{\ell_{\rm max}} \frac{2\ell+1}{2} \left( \frac{ \Delta C_\ell^{TT} } {C_\ell^{TT}}\right)^2.
\end{equation}
Notice that the criterion that $\Delta\chi^2_{TT} \lesssim {\cal O}(1)$ corresponds roughly to $|\Delta C_\ell^{TT}/C_\ell^{TT} |\lesssim 1/\ell$ for all $\ell < \ell_{\rm max}$ and we use this as a rough guide for cosmic variance in figures throughout.

Our choice of $\ell_{\rm max}=2400$ for the CV limit corresponds roughly to the limiting accuracy of the default settings\footnote{
In terms of \textsc{CAMB} numerical accuracy settings, the comparison is specifically between default and high accuracy settings:
\texttt{accuracy\_boost} $= 1, 6$; \texttt{l\_accuracy\_boost} $= 1,8$; \texttt{do\_late\_rad\_truncation} = T,F; \texttt{transfer\_high\_precision} = F,T;
\texttt{do\_nonlinear}=0,0;
\texttt{scalcls}=T,T for $S_8$, \texttt{scalcls}=F,F for $T_{\rm m}(k)$.}
 for the $\Lambda$CDM model where
\begin{align}
\label{eq:CMBbenchmark}
\Delta\chi^2_{\Lambda \rm CDM} \approx{}&0.47.
\end{align}

In Fig.~\ref{fig:ClTT_LCDMbenchmark}, we show the corresponding errors in 
$C_\ell^{TT}$ for $\Lambda$CDM.  
Here and below the $\Lambda$CDM case ($f_\DM=0$) is calculated with the same code using a finite but negligible $f_\DM \rightarrow 0$. 
In the bottom panel we show that the fractional errors that contribute the most to $\Delta\chi^2_{TT}$ are at the highest $\ell$ where they cross $1/\ell$  line. This is mainly due to the default CMB lensing accuracy settings but note that we do not include any other secondary anisotropy sources, recombination accuracy considerations or foregrounds.
Thus $\Delta\chi^2 \approx 1$ is our very conservative criteria both for negligible errors in the \ULA\ effective method compared with the chosen cosmic variance limit and also to match the error floor set by $\Lambda$CDM. 

For the matter power spectra, we treat the masses where the \ULA s behave as dark matter ($m \ge 10 H_0 \approx  1.44\times 10^{-32}$ eV, $0 \le f_\DM \le 1$) and dark energy 
($m < 10 H_0$, $0 \le f_\DE \le 1$)
separately in terms of 
whether the \ULA s are included in the construction of the matter (``m") density field along with the cold dark matter, baryons and massive neutrinos. 
In each regime, the impact of the \ULA s is quantified by the matter transfer function relative to the high accuracy $\Lambda$CDM transfer function
\begin{equation}\label{eq:transfer}
T_{\rm rel}(k; f_{\{\DM, \DE\}})\equiv
\frac{T_{\rm  m}(k;f_{\{\DM, \DE\}})}{T_{\rm m}(k;0)}.
\end{equation}
Note that the matter power spectrum ${\cal P}_{\rm m}(k)\propto (k/H_0)^4 T_{\rm m}^2(k){\cal P_R}(k)$.

The accuracy of the transfer function calculation in $\Lambda$CDM again sets a floor on the target
accuracy of our calculations.
In Fig.~\ref{fig:Tmk_LCDMbenchmark}, we show these  $\Delta T_{\rm m}/T_{\rm m}$ errors in $\Lambda$CDM.
Notice that for $k\lesssim 10^{-1}$\, Mpc$^{-1}$, the errors oscillate such that observable quantities that depend on the integral of the transfer function can have higher accuracy than the transfer function errors.   

One such relevant quantity is $S_8 \equiv \sigma_8\sqrt{\Omega_{\rm m}/0.3}$
where $\sigma_8$ is the rms of the linear matter density field convolved with a spherical top-hat of radius $8h^{-1}$Mpc.  Hence we choose it
as a summary statistic and use its errors in the same way as  $\Delta\chi^2$ for CMB observables. Here the $\Lambda$CDM calculation places a very small floor
\begin{align}\label{eq:S8_floor}
\Delta S_{8, {\Lambda \rm CDM}}  \approx{}& 1.6\times 10^{-4},
\end{align}
with $S_{8,\Lambda{\rm CDM}}=0.85$. 
A more useful benchmark to consider is for $\Delta S_8$ to be much better than the difference between the {\it Planck} $\Lambda$CDM model 
prediction and local measurements, e.g.\ $S_8=0.776\pm 0.017$ with DES-Y3 \cite{DES:2021wwk} under $\Lambda$CDM.
Under this standard, $\Delta S_8$ should be much less than $10^{-2}$.

With these benchmarks in mind, we next consider the predictions and accuracy of \ourcode\ across the range of $m$ that spans the dark matter and dark energy regimes.

 \begin{figure}
    \includegraphics[width=\linewidth]{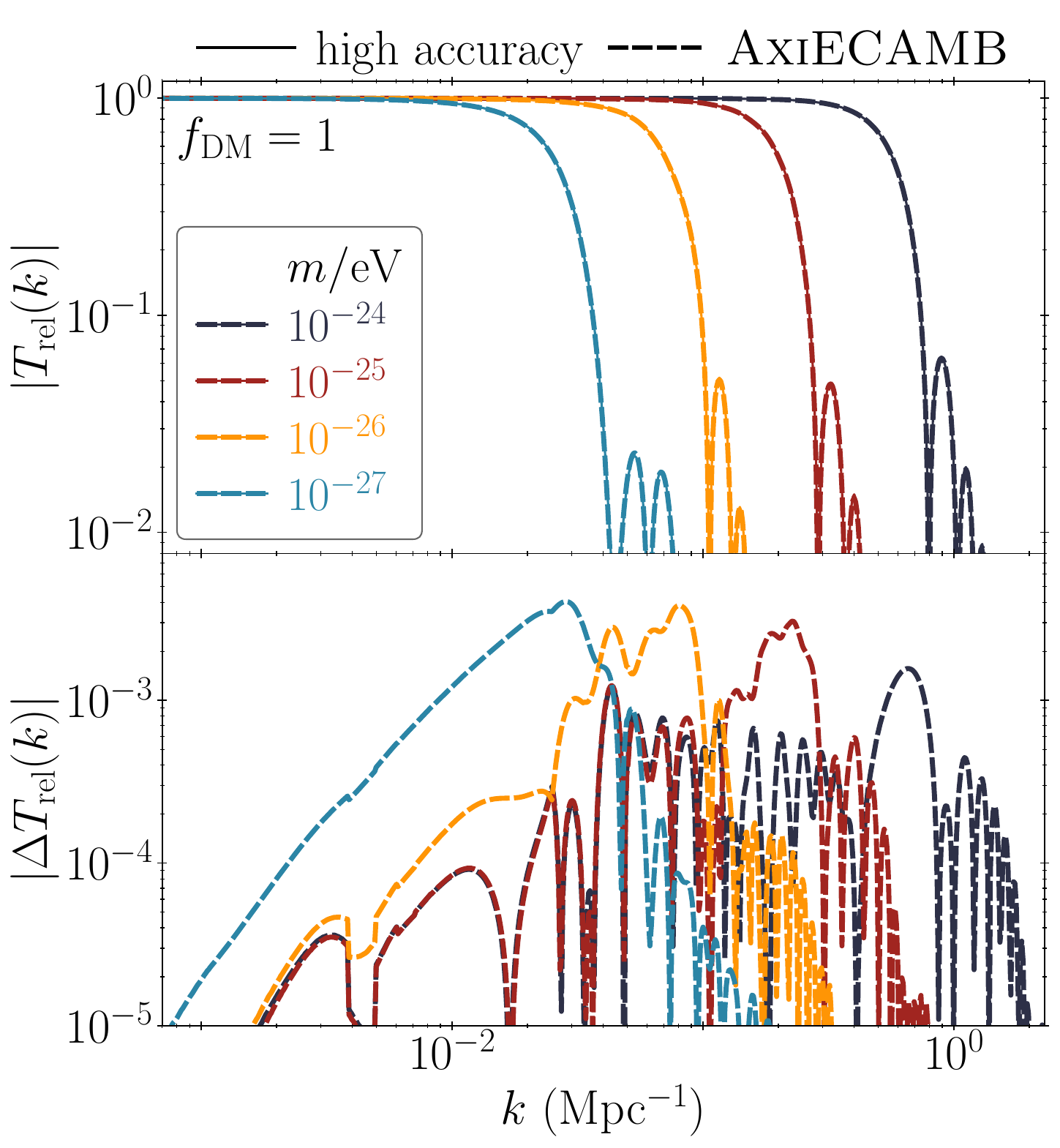}
    \caption{Relative  transfer function (top) $|T_{\rm rel}(k)|$ for total DM cases  ($f_\DM = 1$) and error between the \ourcode\ and high accuracy calculations.   
    } \label{fig:perfTk_allDM}
\end{figure}

\begin{figure}[htbp]
        \includegraphics[width=\linewidth]{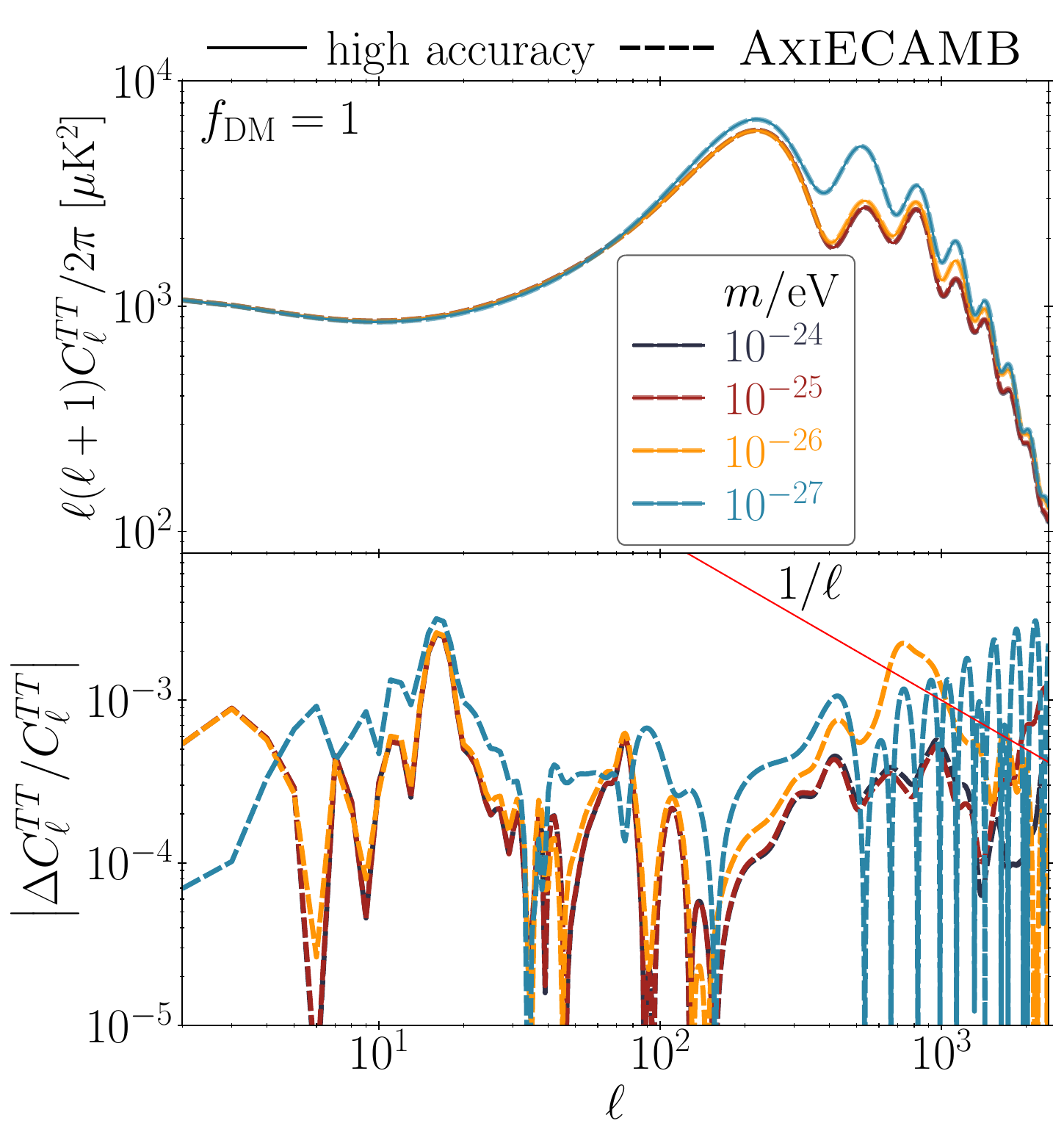}
        \caption{$C_\ell^{TT}$ in the total dark matter cases (top; $f_\DM=1$) and its fractional error between \ourcode\ and high accuracy calculations.  Red line represents the  $1/\ell$ cosmic variance scaling below which errors are undetectable (see text). 
 }
     \label{fig:ClTTperformance_allDM}
\end{figure}

\subsection{Total dark matter}
\label{sec:totalDM}

In the range $m \gtrsim 10^{-27}$\,eV, the CMB and linear large-scale structure allow \ULA s to compose a significant fraction of the dark matter. We take $f_\DM=1$ (well above observational limits in the nonlinear regime, which rule out $f_\DM=1$ for $m\lesssim 10^{-20}$\,eV \cite{Rogers:2020ltq}) to illustrate the most extreme phenomenology and stringent tests of code accuracy.

In the $f_\DM=1$ case, the \ULA\ density fluctuations are sharply cut off below its maximal Jeans scale (see Eq.~\ref{eq:kJ}).  Here we show the $10^{-27}-10^{-24}$\,eV range whereas the fuzzy dark matter regime of 
$10^{-18}-10^{-24}$\,eV  is discussed in Appendix \ref{sec:constraint} and Fig.~\ref{fig:fittingformula_comaprison}.
When $m<10^{-26}$\,eV as shown in
Fig.~\ref{fig:perfTk_allDM} (top panel),  this Jeans scale exceeds the nonlinear scale of $\Lambda$CDM, which corresponds to  $k \sim 10^{-1}$\,Mpc.  To form bound structures at all, \ULA s at lighter masses can no longer comprise most of the dark matter.
Correspondingly as the mass decreases, the \ULA s become more relevant to the primary CMB anisotropy as shown in 
Fig.~\ref{fig:ClTTperformance_allDM}. 
The stabilization of \ULA\ density perturbations then change the metric perturbations which  drive acoustic oscillations (see e.g.~\cite{Lin:2019qug}), much like with neutrinos or dark radiation.  For $m \lesssim 10^{-26}$ eV, this effect or its ISW analogue after recombination is so large that $f_\DM = 1$ is immediately ruled out by CMB data just by inspection of these  modifications to the shape of the spectrum (see Figs.~\ref{fig:ClTTperformance_allDM} and \ref{fig:GT-LCDM} for the significance of the deviation from $\Lambda$CDM.).

\begin{figure}
    \includegraphics[width=0.95\linewidth]{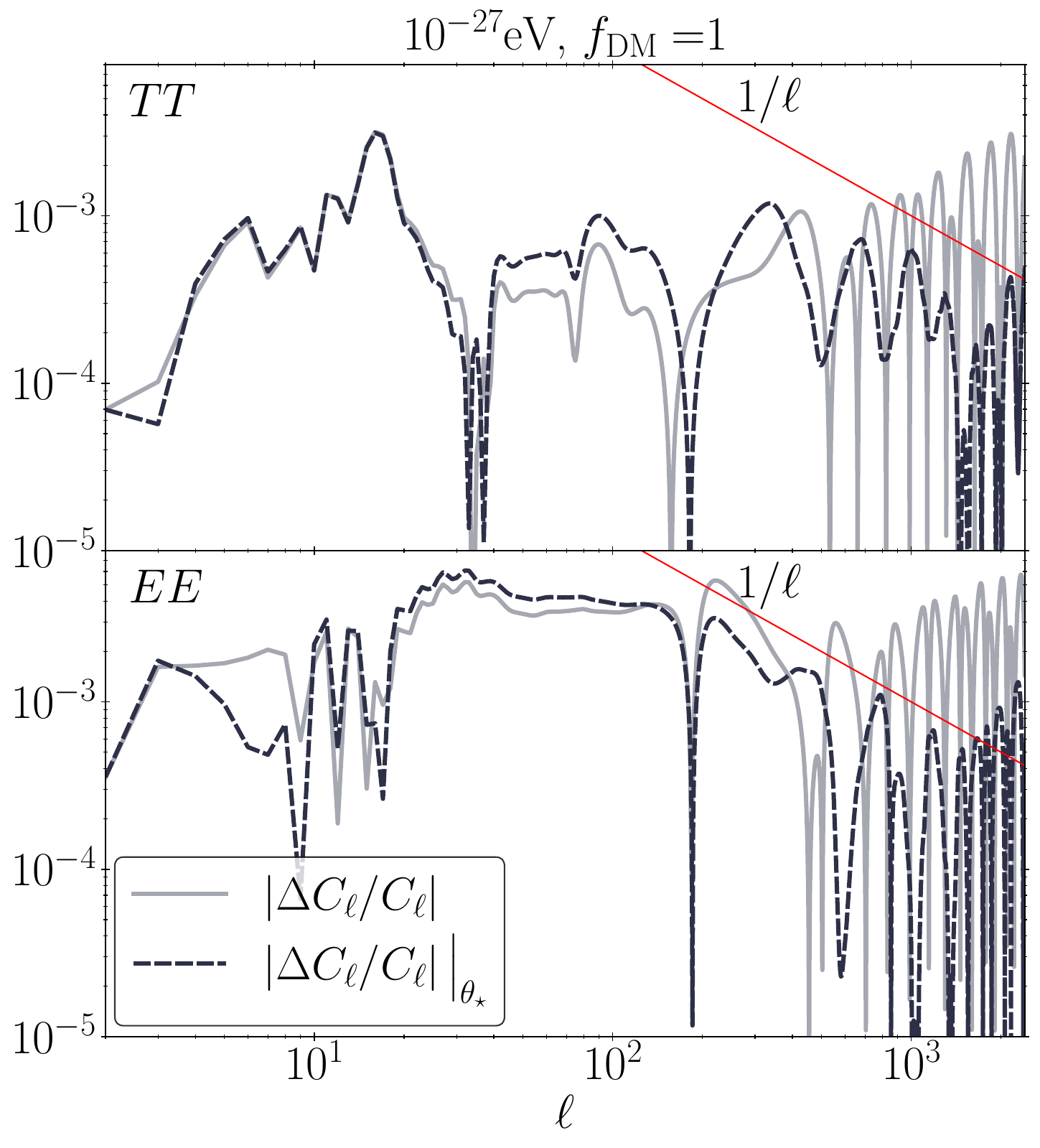}
    \caption{Fractional error in $C_\ell$ for $TT$ and $EE$ before (light solid curve) and after the linearized correction (dashed curve) for the shift in the angular scale of the sound horizon $\Delta\thetas/\thetas=-2.7\times 10^{-4}$ (see Eq.~\ref{eq:thshift})  for
      $m = 10^{-27}$ eV, $f_\DM = 1$. Most of the error at high $\ell$ is due to this tiny shift, which does  not significantly influence other cosmological inferences. 
      }
    \label{fig:clerr_thetacheck}
\end{figure}

Also shown in Figs.~\ref{fig:perfTk_allDM} and \ref{fig:ClTTperformance_allDM} are errors between \ourcode\ vs.~the high accuracy calculation. 
The error in the transfer function $\Delta T_{\rm rel}$ remains well below $10^{-2}$ across this range and is only mildly increasing as the mass decreases. 
This corresponds to $\Delta S_8$ errors below $10^{-3}$ (see Tab.~\ref{tab:ULA_errortable}).   $S_8$ itself drops substantially for masses lower than $\sim 10^{-26}$\,eV.

\begin{figure}
  \includegraphics[width=\linewidth]{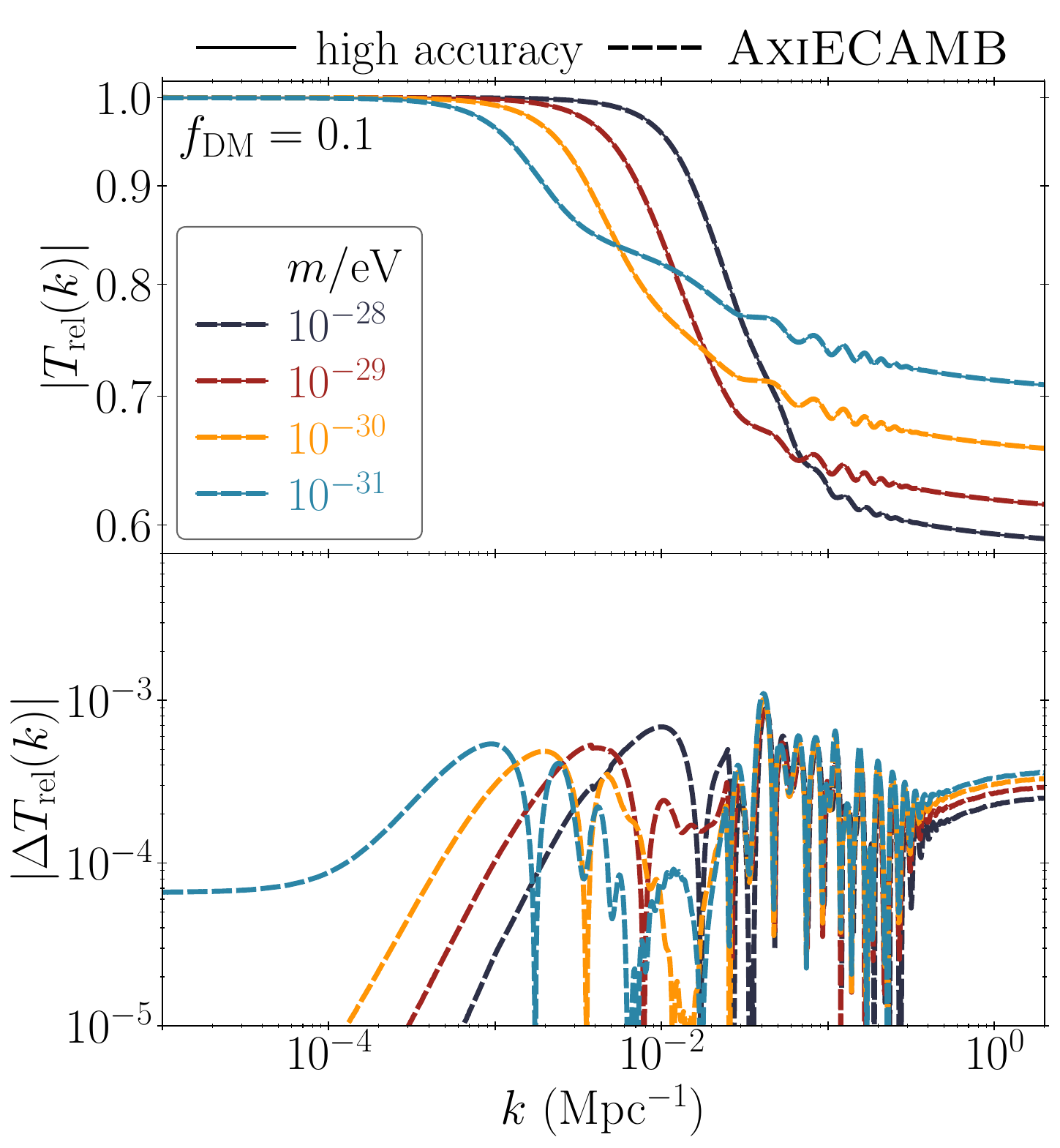}
    \caption{Relative transfer function $|T_{\rm rel}(k)|$ (top) for fractional dark matter cases $(f_\DM=0.1)$ and errors between the \ourcode\ and high accuracy computations (bottom). }
    \label{fig:perfTk_10pDM}
\end{figure}

Likewise for $C_\ell^{TT}$, 
it is only toward this ruled-out lower range of $10^{-26}-10^{-27}$\,eV that the $f_\DM =1$ errors significantly exceed the  $1/\ell$ line and differ noticeably from $\Lambda$CDM errors (see~Fig.~\ref{fig:ClTT_LCDMbenchmark}).  
Except near this end, the corresponding values of $\Delta \chi^2_{TT}$ values are given in Table~\ref{tab:ULA_errortable} and do not significantly exceed unity.
$EE$ and $TE$ polarization power spectra errors are qualitatively similar to temperature but the stronger features in polarization lead to a more significant error in $\Delta \chi^2$ when compared with cosmic variance limits (see also Fig.~\ref{fig:clerr_thetacheck}).

\begin{table*}[t]
\centering
\renewcommand{\arraystretch}{1.2} 
\begin{tabular}{
    >{\centering\arraybackslash}p{1cm} |
    >{\centering\arraybackslash}p{0.6cm} |
    >{\centering\arraybackslash}p{0.9cm} |
    >{\centering\arraybackslash}p{1.5cm} |
    >{\centering\arraybackslash}p{1cm} |
    >{\centering\arraybackslash}p{1cm} |
    >{\centering\arraybackslash}p{1.cm} |
    >{\centering\arraybackslash}p{2.2cm}
}
$m/$eV & $f_\DM$ & $S_8$ & $\Delta S_8/10^{-4}$ & $\Delta\chi^2_{TT}$ & $\Delta \chi^2$ & $\Delta \chi^2|_{\thetas}$ & $(\Delta\thetas/\thetas)/10^{-5}$ \\
\hline
-- & 0 & 
0.85
& 
1.6
& 
0.20
& 
0.47
& 
--
& 
--
\\
\hline
$10^{-24}$ & 1 & 
0.84
& 
1.9
& 
0.20
& 
0.47
& 
0.46
& 
0.07
\\
\hline
$10^{-25}$ & 1 & 
0.72
& 
-4.2
& 
1.0
& 
1.8
& 
1.8
& 
0.30
\\
\hline
$10^{-26}$ & 1 & 
0.36
& 
-9.5
& 
1.5
& 
4.2
& 
4.3
& 
-5.0
\\
\hline
$10^{-27}$ & 1 & 
0.12
& 
-5.9
& 
5.8
& 
53
& 
2.6
& 
-27
\\
\hline
$10^{-28}$ & 0.1 & 
0.55
& 
0.69
& 
0.53 
& 
1.3
& 
0.75
& 
3.2
\\
\hline
$10^{-29}$ & 0.1 & 
0.55
& 
0.85
& 
0.18 
& 
0.56
& 
0.55
& 
0.55
\\
\hline
$10^{-30}$ & 0.1 & 
0.59
 & 
0.95
 & 
0.18 
 & 
0.59
& 
0.55
& 
0.85
\\
\hline
$10^{-31}$ & 0.1 & 
0.63
& 
1.0
& 
0.21
& 
0.66
& 
0.52
& 
1.8
\\
\end{tabular}
\caption{Example \ULA\ models, observables and errors between \ourcode\ and the high accuracy calculation.  The entry for $f_\DM=0$ is the $\Lambda$CDM benchmark model.
}
\label{tab:ULA_errortable}
\end{table*}

\begin{figure}[htbp]
        \includegraphics[width=\linewidth]{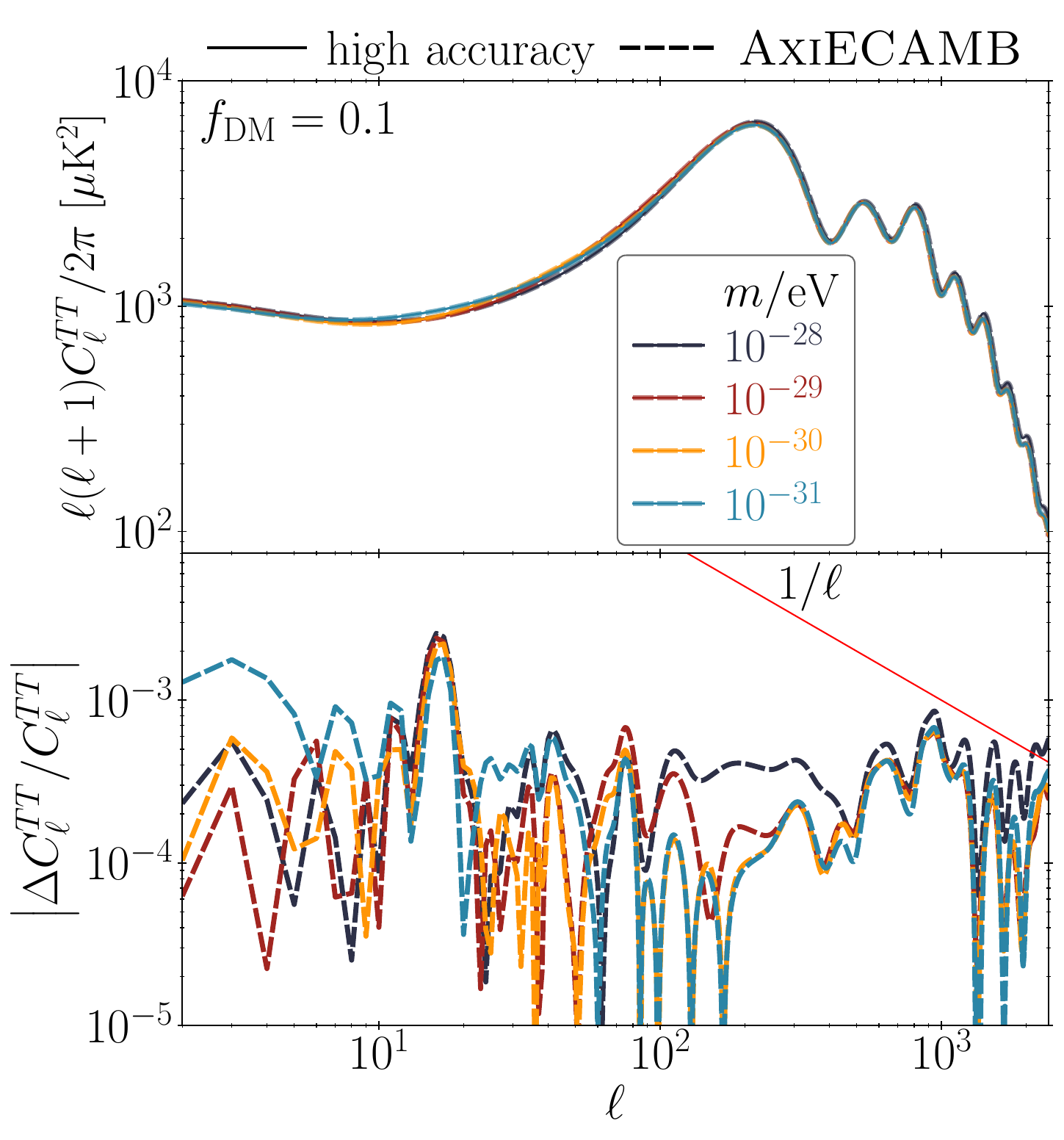}
        \caption{ $C_\ell^{TT}$ in the fractional dark matter cases (
        top; $f_\DM=0.1$) and its fractional error between \ourcode\ and high accuracy calculations as in Fig.~\ref{fig:ClTTperformance_allDM}.}
    \label{fig:ClTTperformance_10percentDM}
\end{figure}

For the most extreme case of $m=10^{-27}$\,eV the large $\Delta\chi^2$ indicates that in principle there could be a $\sqrt{\Delta\chi^2}\approx 7$\,$\sigma$ error in the analysis of such cosmic variance limited data.  In fact, most of this error is associated with a very small shift in the background expansion and hence a very small change in the angular size of the sound horizon $\thetas$.
In this example, the \ourcode\ calculation produces a shift of $\Delta\thetas/\thetas=-2.7\times 10^{-4}$ compared with the high accuracy calculation.  
This small shift is statistically significant compared with the projected Fisher errors on this single parameter $\sigma_{\ln \thetas}=4.1\times 10^{-5}$.  
Compared with the \emph{Planck} 2018 $TT$, $TE$, $EE$ actual errors of $\sigma_{\ln \thetas}\sim 3\times 10^{-4}$ \cite{Aghanim:2018eyx}, this forecast is a factor of $7$ better. For other cases, the improvement is at least a factor of 5 in errors or a factor of $25$ in the implied reduction of $\Delta\chi^2$.
Furthermore, cosmological parameter estimation based on the distance to the recombination surface is already limited by uncertainties in the sound horizon $r_*$, not measurement errors in $\ln \thetas$.  The \emph{Planck} 2018 errors for the former are much larger, $\sigma_{\ln r_\star}=2 \times 10^{-3}$.  Therefore, shifts of $\Delta\thetas/\thetas \sim 10^{-4}$ from our default accuracy calculation would not significantly impact cosmological parameter errors based on distance inferences like $H_0$ at the present or even in the forseeable future.

\begin{figure*}
    \includegraphics[width=1\linewidth]{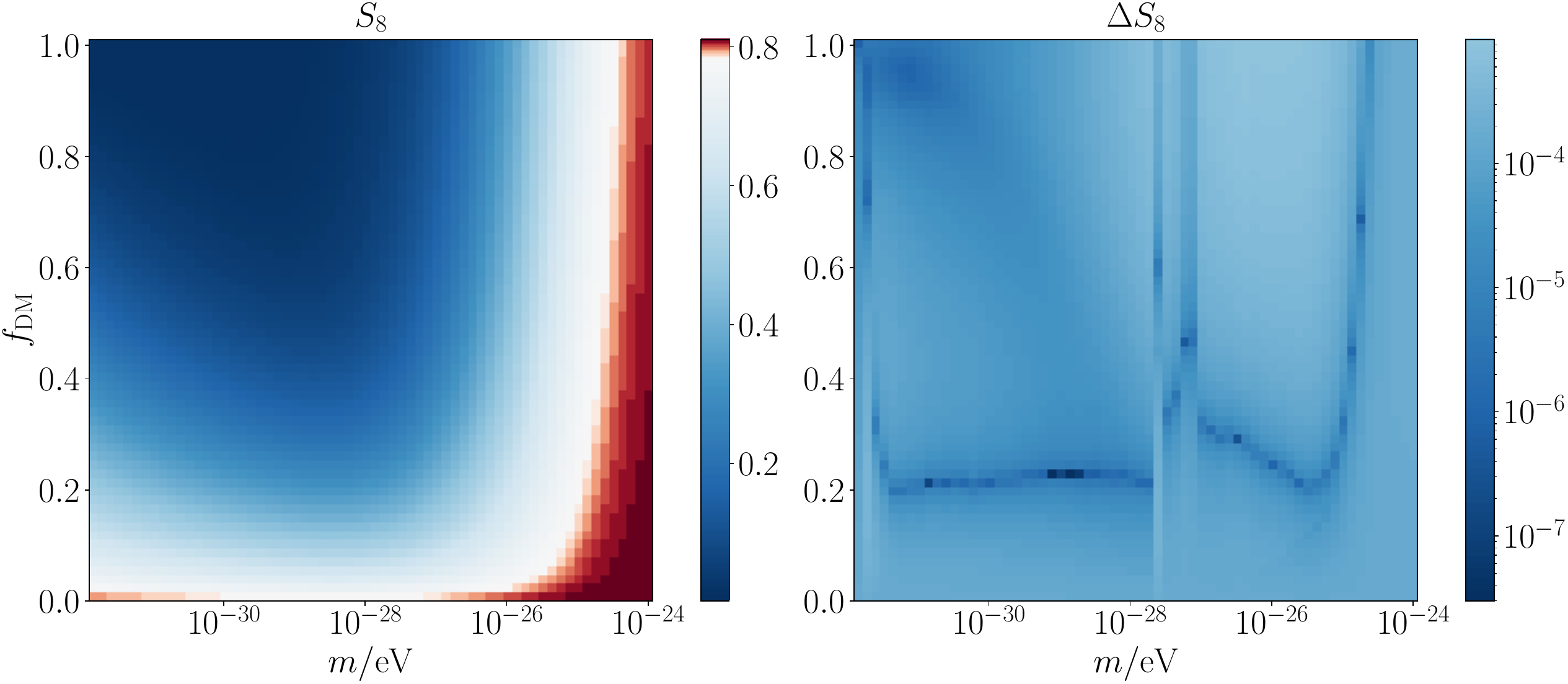}
    \caption{Large scale structure amplitude $S_8$ (left) in the $m,f_\DM$ plane and its error 
    $\Delta S_8$ between \ourcode\ and high accuracy computations. 
    }    \label{fig:performance_S8}
\end{figure*}

To see that this $\thetas$ shift captures the main source of the
$\Delta\chi^2$ error as well, we can approximately correct this error in the acoustic regime of CMB primary anisotropy by an equivalent shift in the multipole spectrum
\begin{equation}
\Delta C_\ell({\rm corrected}) =\Delta C_\ell({\rm original}) - \left[\frac{\Delta\thetas}{\thetas}\frac{d C_\ell}{d\ln\ell}\right].\label{eq:thshift}
\end{equation}
In Fig.~\ref{fig:clerr_thetacheck}, we show that with this correction, the errors for $m=10^{-27}$\,eV and $f_\DM=1$ are reduced below the CV line for $TT$ and comparable to it for $EE$; correspondingly
$\Delta\chi^2|_{\thetas}=2.6$ where $|_{\thetas}$ denotes the remaining errors after fixing $\thetas$ in this manner.  There is no need to correct for this shift in \ourcode\ for constraints on other \ULA\ or cosmological parameters.

\subsection{Fractional dark matter}
\label{sec:fracDM}

For the mass range $10^{-31} < m/{\rm eV} <  10^{-28}$,  the \ULA s can only comprise a small fraction of the dark matter. Current constraints from the linear regime require $f_\DM\lesssim 0.05$ (95\% CL) \cite{Hlozek:2014lca}  and so for illustration purposes we take $f_\DM = 0.1$ to highlight the accuracy of the approach in the most extreme case that is relevant for parameter estimation.

\begin{figure*}
    \includegraphics[width=1\linewidth]{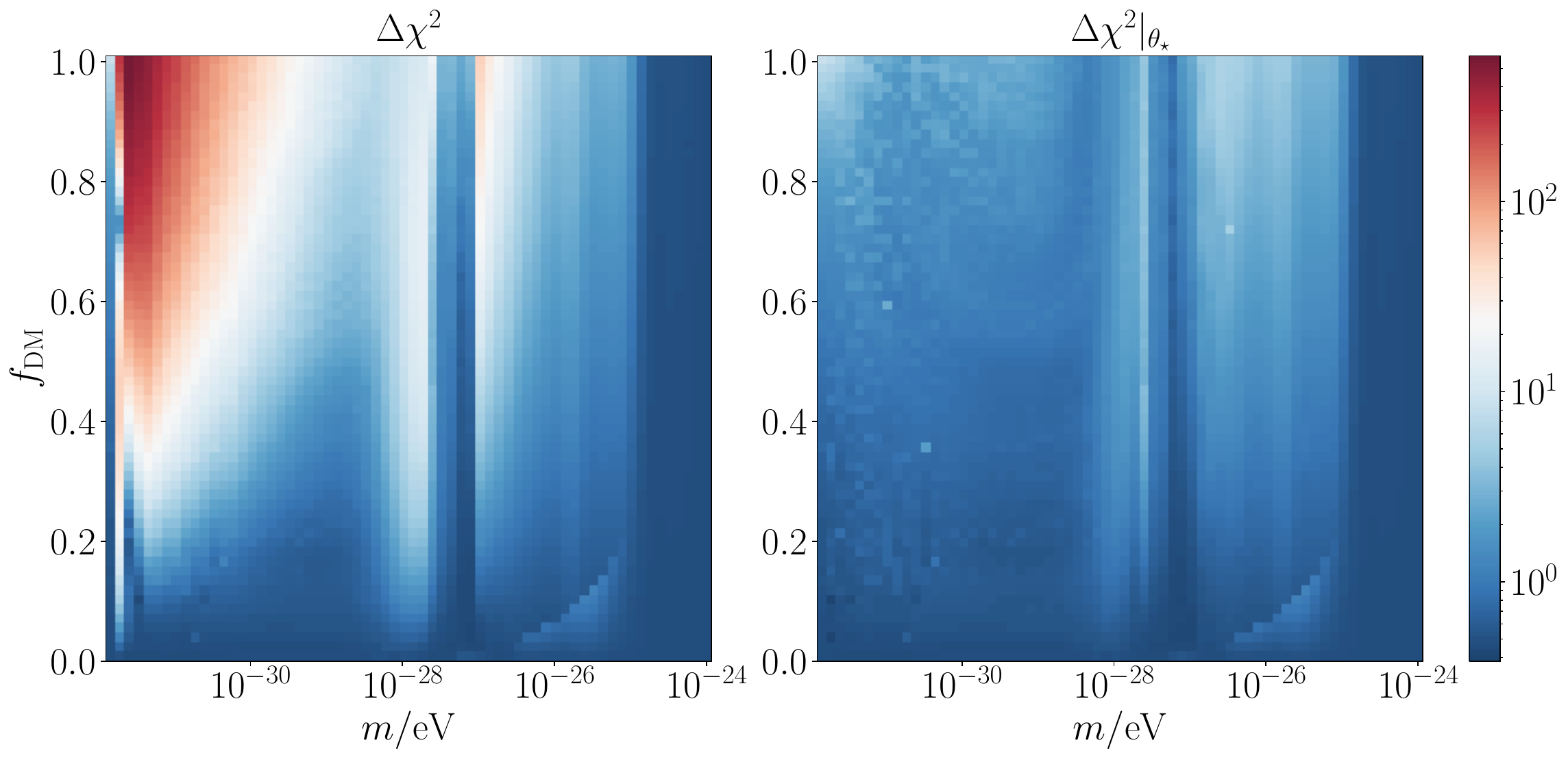}
    \caption{CMB accuracy statistic $\Delta\chi^2$ in the $m-f_\DM$ plane (left panel). The dominant error in \ourcode\ is due to a very small change in the angular scale of the sound horizon $\theta_\star$ that does not substantially impact other parameters, which we demonstrate by its removal (right panel) in the same way as in Fig.\,\ref{fig:clerr_thetacheck}.}   \label{fig:Deltachi2all_CVonly}
\end{figure*}

\begin{figure}
    \includegraphics[width=1\linewidth]{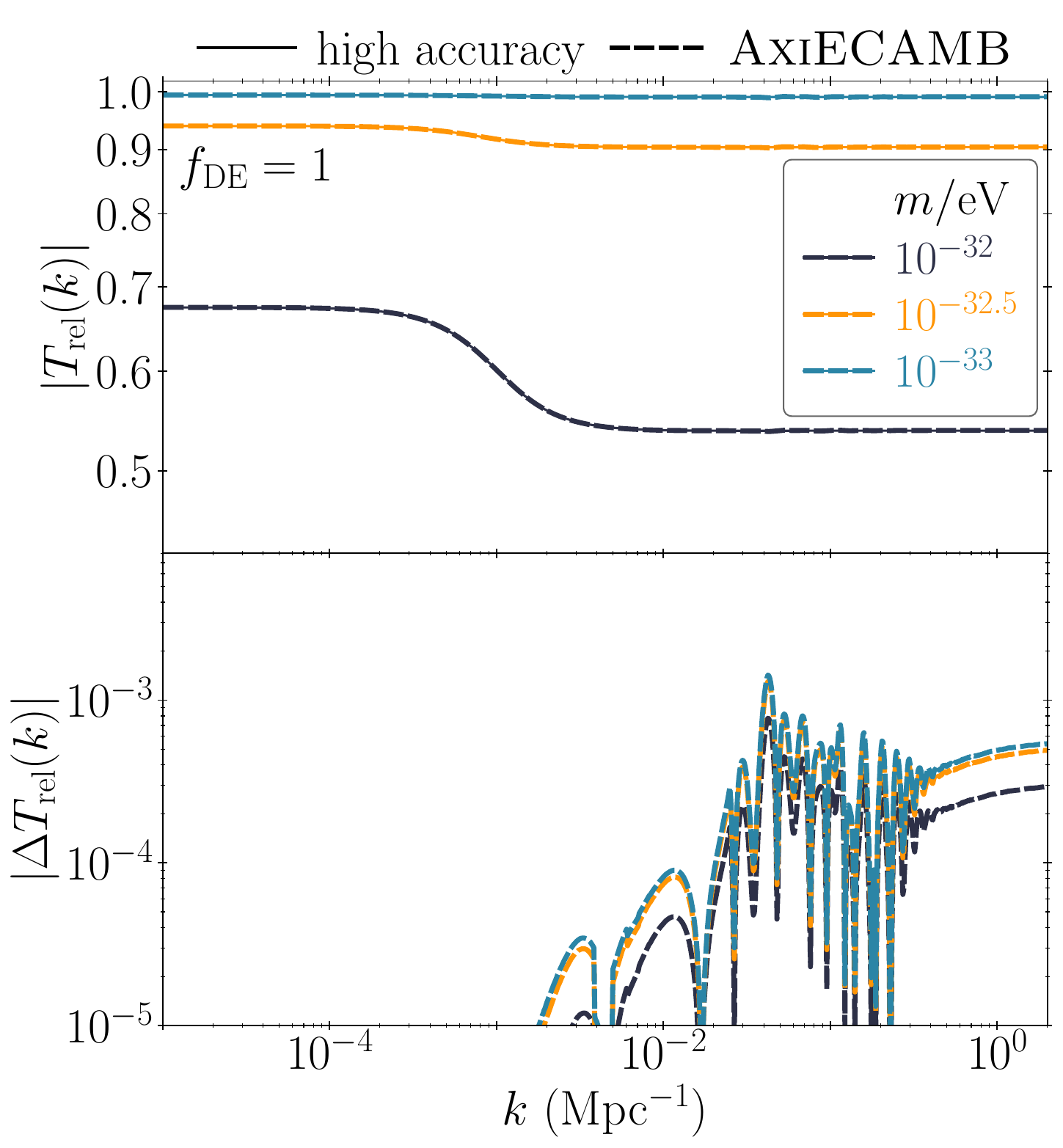}
    \caption{Relative transfer function for total dark energy cases ($m<10 H_0,f_\DE=1$) and errors between \ourcode\ and high accuracy computations.  Here the matter components of the relative transfer function do not include the \ULA.}
    \label{fig:perfTk_allDEtest_dfac10}
\end{figure}

The relative transfer function shown in 
Fig.~\ref{fig:perfTk_10pDM} (top panel) has a sharp decrease at the \ULA\ Jeans scale but then stabilizes at higher $k$ due to the presence of CDM. The suppression at high $k$ results from the slower growth of CDM fluctuations due to the effect of the background \ULA\ density, much like the effect of massive neutrinos (see e.g.~\cite{Amendola:2005ad}).   The errors in $\Delta T_{\rm rel}$ are below the $10^{-3}$ level and are sufficiently small for current and near future constraints.  Correspondingly, the $\Delta S_8$ error in Tab.~\ref{tab:ULA_errortable} is below $10^{-4}$ even though $S_8$ itself is significantly reduced by the \ULA\ suppression of growth to $S_8\sim 0.5-0.6$.

In Fig.~\ref{fig:ClTTperformance_10percentDM} we show the corresponding CMB temperature anisotropy.
Even without the correction for $\thetas$, the error is at or below the CV line for these $f_\DM=0.1$ cases and correspondingly both $\Delta \chi^2_{TT}$ and $\Delta \chi^2$ is order unity or below (see Tab.~\ref{tab:ULA_errortable}).
 For the lighter masses in this range, the main effect of the \ULA\ is to raise the early ISW effect at multipoles before the first peak.  The $\thetas$ correction makes a negligible difference for these models since the shift is only $\sim 10^{-5}$.

For a comprehensive understanding of the errors across the full $m$ and $f_\DM$ space, we evaluate 
$S_8$ and its errors in Fig.~\ref{fig:performance_S8}.
The errors are everywhere $<10^{-3}$ and in fact drop for $m\lesssim 10^{-28}$\,eV.
For lighter masses, the \ULA\ density fluctuations have not caught up by the present with the baryons for $f_\DM=1$ and CDM fluctuations for lower $f_\DM$  at $k \approx 0.1$\,Mpc$^{-1}$ and thus impact $S_8$ as a homogeneous background that suppresses the growth as described above. Thus errors in the \ULA\ density fluctuations themselves become less relevant for $S_8$ (see also the discussion of Figs.~\ref{fig:axionCAMB_S8heatmap}, \ref{fig:quasistaticdemo}).

Likewise, we show 
$\Delta\chi^2$ and $\Delta\chi^2|_{\thetas}$ in Fig.~\ref{fig:Deltachi2all_CVonly}. The largest errors occur near the lightest masses and extreme \ULA\ fractions $f_\DM\rightarrow 1$, which is in the grossly ruled-out regime. Moreover,
correcting for the contribution of $\thetas$ reduces $\Delta\chi^2$
 to low levels compared with CV
even in this region.

Despite being the main source of error in CMB spectra, the maximum error is $\Delta \thetas/\thetas \sim 9\times 10^{-4}$, which occurs around $m = 10^{-31}$\,eV, $f_\DM = 1$. 
  Larger errors occur in this region since the switch takes place at low redshifts $z_* \sim 4$ when the resulting change in the expansion rate has its maximal effect on the distance to the recombination surface, and hence $\thetas$. As $m \rightarrow 10 H_0$ (left edge of Fig.~\ref{fig:Deltachi2all_CVonly}), the switch of both the default and high accuracy calculations is at $z=0$ and the difference between the two diminishes, albeit in an oscillatory fashion that depends on the phase of the KG oscillation at $z=0$.

Other features in the $m \gtrsim 10^{-28}$\,eV range highlight the strategy taken in Sec.~\ref{sec:CMB_changes} for boosting the default $m/H_*$ above the baseline of 10.   
The vertical stripes near $m\sim 10^{-27.5}$\,eV represent the mass range where the baseline switch would have occurred in the recombination window but is raised by our approach to beyond it (see Fig.~\ref{fig:Clerror_recskipcompare}).  
Around these masses and $f_\DM=1$, the main errors are again shifts in $\thetas$, here due to the sound horizon $r_\star$, as the $\thetas$ correction shows. 
In fact, most of the $r_\star$ error when the switch is after recombination comes from our convention that $\Omega_\ax h^2$ is fixed rather than the initial field value $\phi_{\rm ini}$. 
The impact of this shift is at most a $\lesssim 10^{-3}$ change in the value of 
$\Omega_\ax h^2$ corresponding 
to a given CMB prediction.

The curve around $f_\DM = 
(m/10^{-23.5}{\rm eV})^{1/2}$ between $10^{-27.5} < m/{\rm eV} < 10^{-25}$ 
is due to the  boosting of $m/H_*$ in the \ULA-radiation equality window.   The tuning of the phase of this switch is to $m=10^{-26}$\,eV and  improves the errors in this mass range (see Appendix \ref{app:cmbsources_corr}).

\subsection{Total dark energy}
\label{sec:totalDE}

For $m < 10^{-32}$\,eV, $m/H_0 \lesssim 10$ and the KG field is  not highly oscillating  at the present. In this scenario the \ULA\ behave as a quintessence dark energy field and \ourcode\ solves the KG system all the way to the present.
In our examples we take $f_\DE=1$ and consider the \ULA\ as all of the dark energy.

\begin{figure}
    \includegraphics[width=\linewidth]{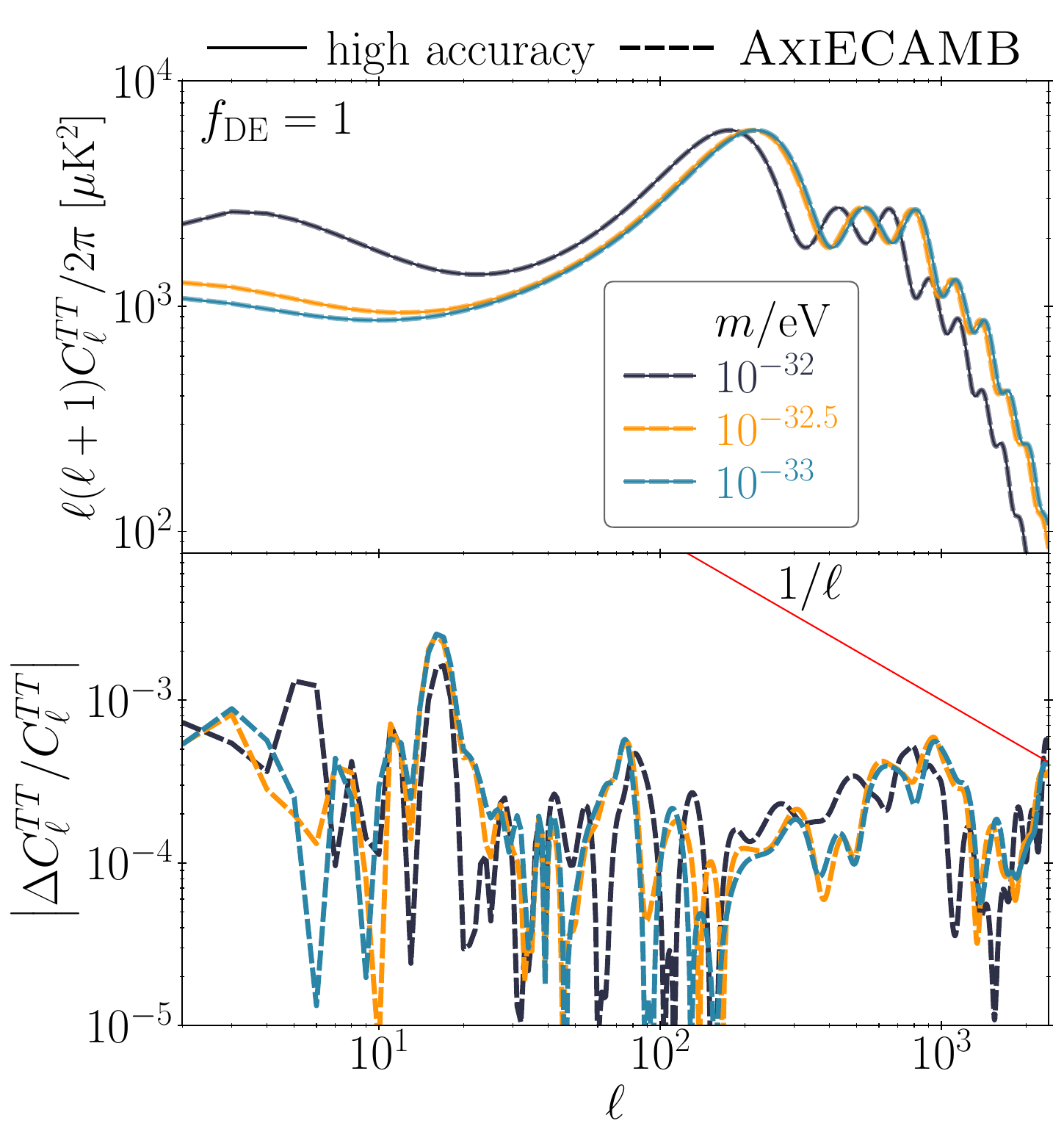}
    \caption{$C_\ell^{TT}$ for total dark energy cases ($m<10H_0,f_\DE = 1$) and fractional errors between \ourcode\ and high accuracy computations.}\label{fig:ClTTperformance_DEs}
\end{figure}

As part of the dark energy, we exclude \ULA\  in $T_{\rm m}$ but the \ULA\ still impacts the growth of matter fluctuations and hence $T_{\rm rel}$ below the Jeans scale if the field $\phi$ rolls substantially before the present.  In Fig.~\ref{fig:perfTk_allDEtest_dfac10} we show the \ULA\ effect on $T_{\rm rel}$, and that the errors are comparable to those of the transfer function in $\Lambda$CDM itself. Here $S_8=0.46$, $0.77$ and $0.84$ for $m/{\rm eV}=10^{-32}, 10^{-32.5}, 10^{-33}$ respectively.

The corresponding CMB $TT$ spectra are plotted in Fig.~\ref{fig:ClTTperformance_DEs}. Here the main effects on the CMB is through the change in the late ISW effect and the low redshift distance redshift relation.  We see once again that our \ourcode\ is accurate to the same level as the $\Lambda$CDM calculation with $\Delta\chi^2=0.96, 0.53$ and $0.47$ for the three cases respectively. There is no analogue to a $\thetas$ correction from the switch in this range since both high and low accuracy calculations solve the KG system to the present.

\section{Discussion}\label{sec:conclusions}

We have developed a method and publicly available code \ourcode\ for the fast and accurate solution of ultralight axion perturbations across the whole mass range from the fuzzy dark matter regime of $10^{-18}$\,eV to the frozen dark energy regime of $10^{-33}$\,eV.   The method switches from solving the Klein-Gordon equation to an effective fluid once the mass oscillations become sufficiently rapid compared with the Hubble rate $m/H \ge m/H_*$. At the switch, the matching conditions are established through an effective time average of the impact of the KG oscillations on the background and perturbations. 

Our baseline choice of switch time $m/H_*=10$  is sufficiently accurate for subpercent accuracy predictions for CMB and matter power spectra, and the accuracy is efficiently improved by increasing this value.   In fact the main source of this error in the CMB is a small fractional shift in the angular size of the sound horizon that is no more than $9 \times 10^{-4}$ even in the most extreme case of models which are not observationally viable.  Aside from this shift, which should not affect cosmological parameter estimation in the forseeable future, the default accuracy approaches the cosmic variance limit at $\ell_{\rm max}=2400$, comparable to $\Lambda$CDM calculations.

Our method improves on previous approaches in a number of ways.  By developing the \ETA\ of both the matter and metric variables, we achieve high accuracy for the low mass regime where the switch occurs near or after matter-radiation equality.  This construction removes the leading order oscillations in these variables and provides ${\cal O}(H_*^2/m^2)$ or better accuracy in the matter and CMB observables even when the \ULA s comprise all of the dark matter. Since the \ETA\ holds for any phase of the KG oscillations, it establishes continuity to leading order with the \EFA\ simultaneously in all of the variables, including the pressure and its fluctuation, unlike methods that match the \ULA\ density field at a specific point.   This method eliminates spurious pressure wave oscillations below the \ULA\ Jeans scale and allows the accuracy of CMB power spectra to be further improved through a choice of the switch phase that reduces higher order effects in the regime around equality, 
$10^{-28} \lesssim m/{\rm eV} \lesssim 10^{-25}$.
We also account for boundary terms in CMB anisotropy source functions that arise from the switch or any background curvature.

As we show in Appendix \ref{sec:constraint} when compared with \textsc{axionCAMB}, in certain regions of parameter space, far from $\Lambda$CDM, the improvements in $\Delta\chi^2$ error can reach 5 orders of magnitude.  In regions closer to $\Lambda$CDM, \textsc{axionCAMB} can still be in error by a significant fraction of the change from $\Lambda$CDM.  While our improvements are unlikely to change cosmological constraints on \ULA s qualitatively, the order unity (and thus possibly relevant for \emph{Planck} constraints) inaccuracy of the \ULA\ change from $\Lambda$CDM in certain regions of parameter space merits further study. Similarly we have improved upon \textsc{axionCAMB} and fitting functions for the transfer function in the fuzzy dark matter regime $10^{-24} < m/{\rm eV}<10^{-18}$ but the changes are quantitative rather than qualitative. Finally, our improvements on both the method and numerical implementation collectively decreased the run time by $\sim 50\%$ compared to \textsc{axionCAMB} in spite of the much higher accuracy.  
Relative to CAMB in $\Lambda$CDM, \textsc{AxiECAMB} is a only factor of $\sim 1.2$ slower for $m \gtrsim 10^{-25}$\,eV and this factor only gradually increases to $\sim 2$ for lighter masses nearly independently of the \ULA\ fraction
and weakly dependent on the maximum $k$.

While we implemented our method specifically for a \ULA\ potential with a purely quadratic form, our technique can be readily extended for cases with self interactions (e.g. \cite{Shapiro:2021hjp,Winch:2023qzl}) by moving the switch time to be when the quadratic approximation holds around the minimum.  Our method can also be applied to \ULA\ isocurvature perturbations.  We leave these and other considerations for future work.

\acknowledgements
We thank T.~L.~Smith, D.~J.~E.~Marsh, R.~Hlo\v{z}ek, S.~Passaglia, H.~Winch, and T.~Cookmeyer for useful discussions.
R.L. and W.H. were supported by U.S. Dept.\ of Energy contract DE-FG02-13ER41958 and the Simons Foundation. D.G. acknowledges support from the Charles Kaufman foundation through grant KA2022-129518, and the provost's office at Haverford College.

\begin{appendix}

\begin{figure*}[t]
    \includegraphics[width=1\linewidth]{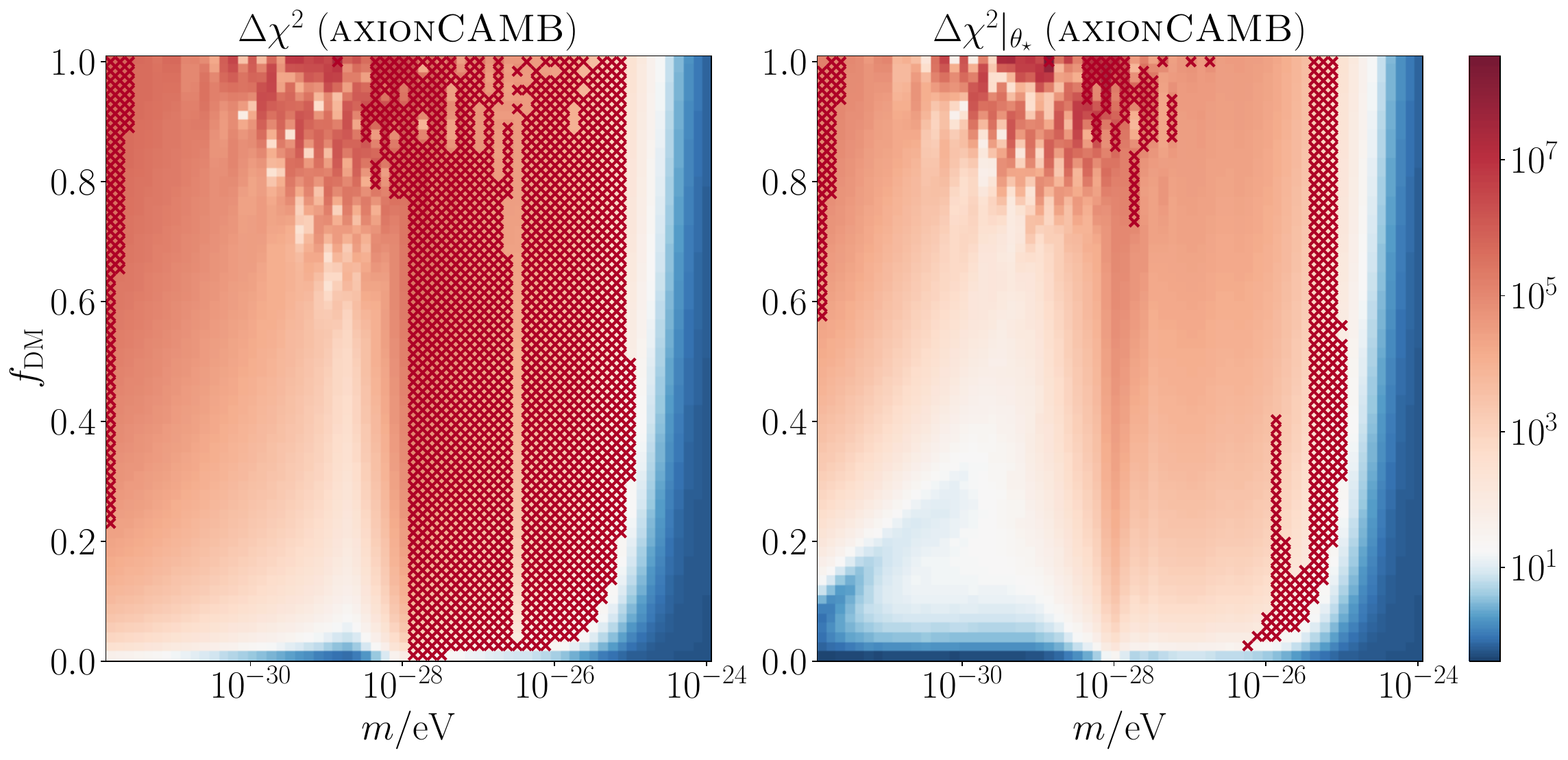}
    \caption{$\Delta \chi^2$ errors for \textsc{axionCAMB} relative to our high accuracy computation before (left) and after (right) the correction for $\thetas$ as in Fig~\ref{fig:Deltachi2all_CVonly}.
    ``$\times$" marks are placed on regions where the error $\Delta\chi^2>25$ and exceeds $\Delta\chi^2(\Lambda$CDM)$/2^2$, the deviation of $\Lambda$CDM from the model.}
    \label{fig:Deltachi2_axionCAMB}
\end{figure*}

\begin{figure*}
    \includegraphics[width=1\linewidth]{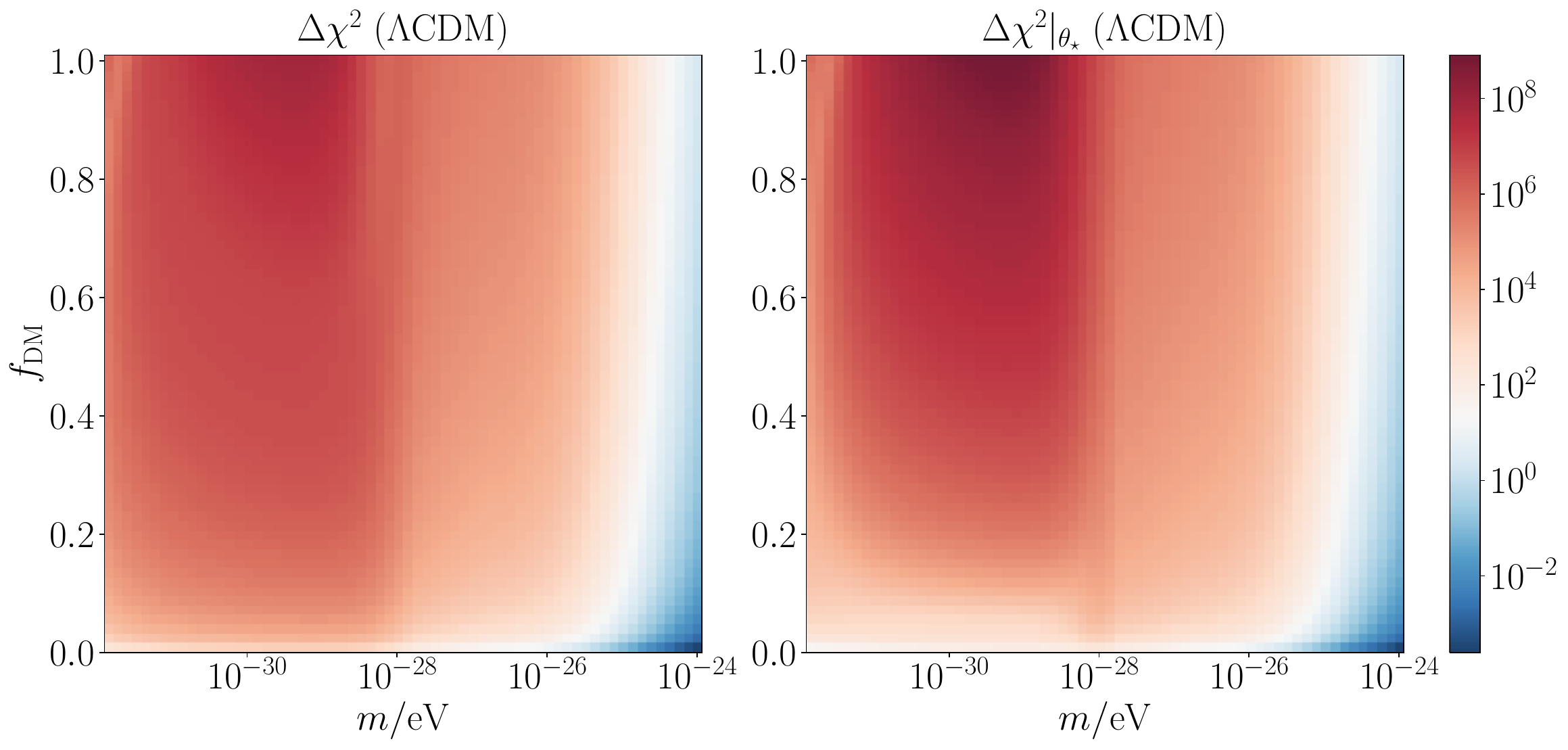}
    \caption{$\Delta \chi^2(\Lambda{\rm CDM})$, the significance of the difference between $\Lambda$CDM and the \ULA\ model, before (left) and after (right) the linearized correction for the different $\thetas$ using Eq.~(\ref{eq:thshift}).
    In regions where the model error in Fig.~\ref{fig:Deltachi2_axionCAMB} are comparable to these differences, the effect of \ULA s are not accurately computed.}
    \label{fig:GT-LCDM}
\end{figure*}

\begin{figure}[t]
    \includegraphics[width=1\linewidth]{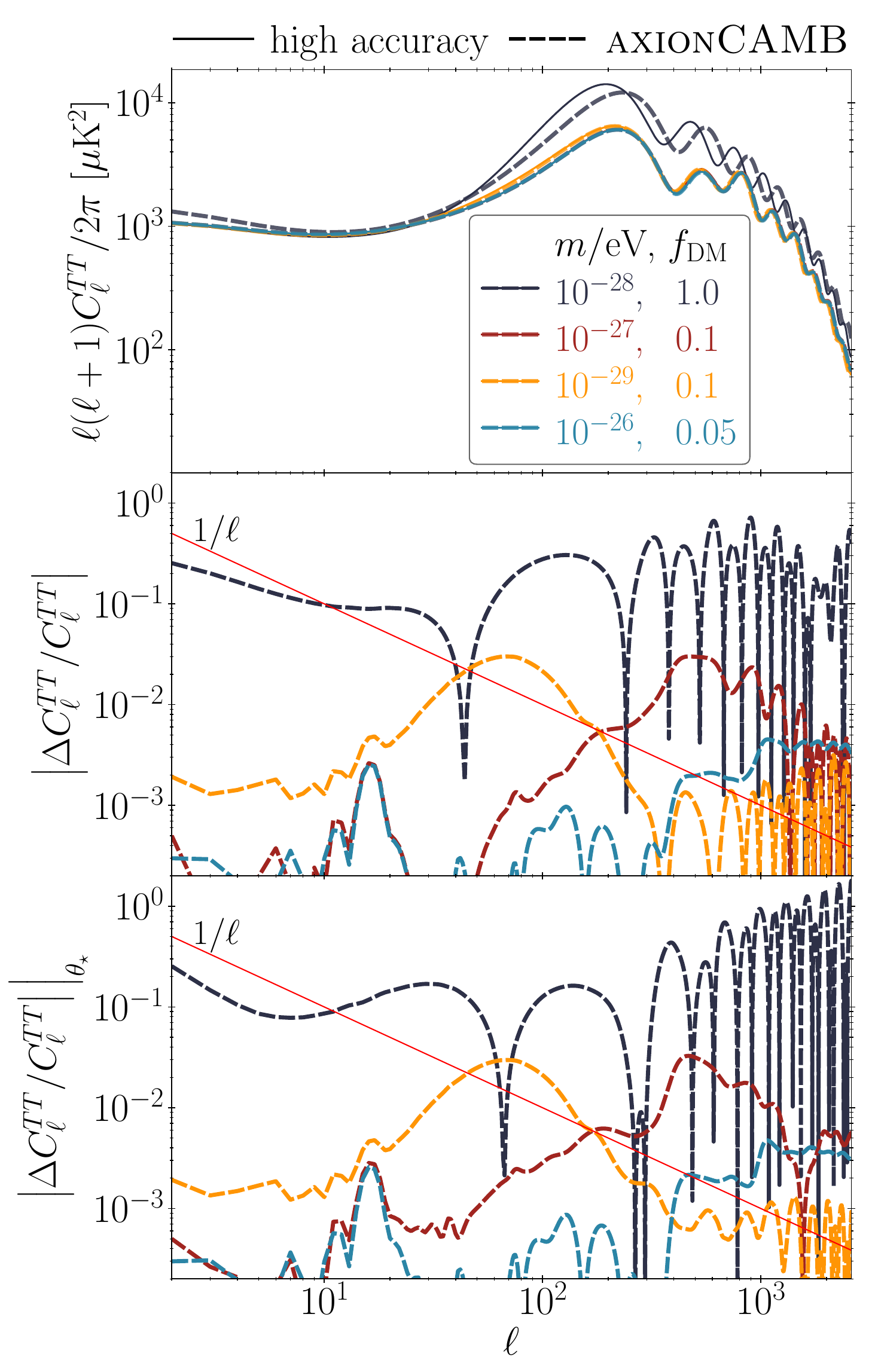}
    \caption{$C_\ell^{TT}$ errors in  \textsc{axionCAMB}  compared with the high accuracy calculation, taken from different representative regions in Fig.~\ref{fig:Deltachi2_axionCAMB} 
    (see text). Errors here are much larger than in \ourcode\ (cf.~Sec.\ \ref{sec:Performance}) and only partially mitigated by the linear correction in the $\thetas$ shift. 
    }
    \label{fig:axionCAMBClerror_chi2rep}
\end{figure}

\section{Comparison with Previous Work}\label{sec:constraint}

We compare the predictions of \ourcode\ with previous codes, primarily \textsc{axionCAMB}. Our code was developed by extensively modifying and rewriting \textsc{axionCAMB} (and consequently shares the same 2013 \textsc{CAMB} code base). It thus provides a fair comparison of errors induced by previous 
fluid approximation techniques. We show regions of parameter space where power spectra are inaccurately computed in \textsc{axionCAMB}, which implies that existing constraints should be re-examined in light of these improvements. We also compare with the transfer function fits from Ref.~\cite{Passaglia:2022bcr} in the fuzzy dark matter regime of 
  $10^{-24} < m/{\rm eV} < 10^{-18}$. In this appendix we always use the high accuracy version of our calculations for comparison as in the main text.

\subsection{\textsc{axionCAMB}}

\textsc{axionCAMB}\,v2.0 solves for the \ULA\ perturbations before a switch at $m/H_*=3
h$ using the generalized dark matter (GDM) fluid approach~\cite{Hu:2000ke} instead of the field perturbations.  Rather than introducing an explicit switch partitioning between the GDM and \EFA, it simply changes the equation of state and sound speed discontinuously at $m/H_*=3h$ and integrates the same fluid-like system across the whole evolution.  While the GDM description is exactly equivalent to the Klein-Gordon equation, it cannot be solved to much higher switch values due to divergences in the adiabatic sound speed $\dot p_\ax/\dot\rho_\ax$ once the KG field starts oscillating whereas the change to the \EFA\ fluid eliminates this problem.

\begin{figure}
    \includegraphics[width=1\linewidth]{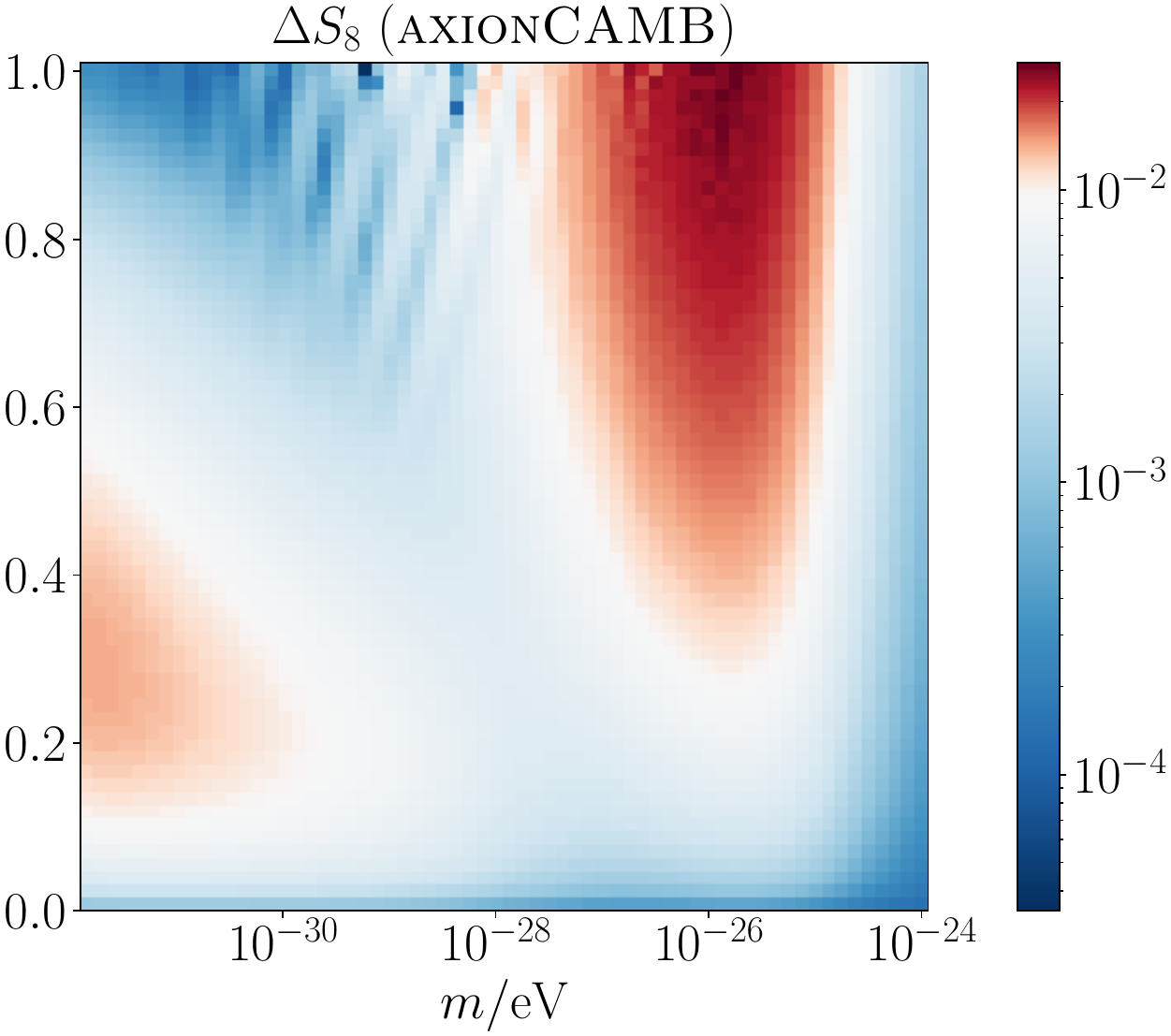}
    \caption{$\Delta S_8$ errors in \textsc{axionCAMB} relative to our high accuracy calculation.  Errors peak at a few $10^{-2}$ and are much larger than those for \ourcode\ (cf.~Fig.~\ref{fig:performance_S8}).
    }
    \label{fig:axionCAMB_S8heatmap}
\end{figure}

We begin by computing $\Delta\chi^2$ between CMB power spectra obtained from \textsc{axionCAMB} and our high-accuracy calculation. The results 
using the same accuracy metrics defined in Sec.~\ref{sec:accuracymetrics}
are shown in Fig.~\ref{fig:Deltachi2_axionCAMB}.  The $\Delta \chi^2$ errors for \textsc{axionCAMB} reach values of $\mathcal{O}(10^7)$ in certain regions of the $m-f_\DM$ parameter space whereas \ourcode\ errors peaks at $\mathcal{O}(10^{2})$. Though correcting for errors in $\thetas$ using the linearized shift of Eq.~(\ref{eq:thshift}) does improve \textsc{axionCAMB} errors in some  regions of the space, they remain much higher than in \ourcode.

\begin{figure}
    \includegraphics[width=1\linewidth]{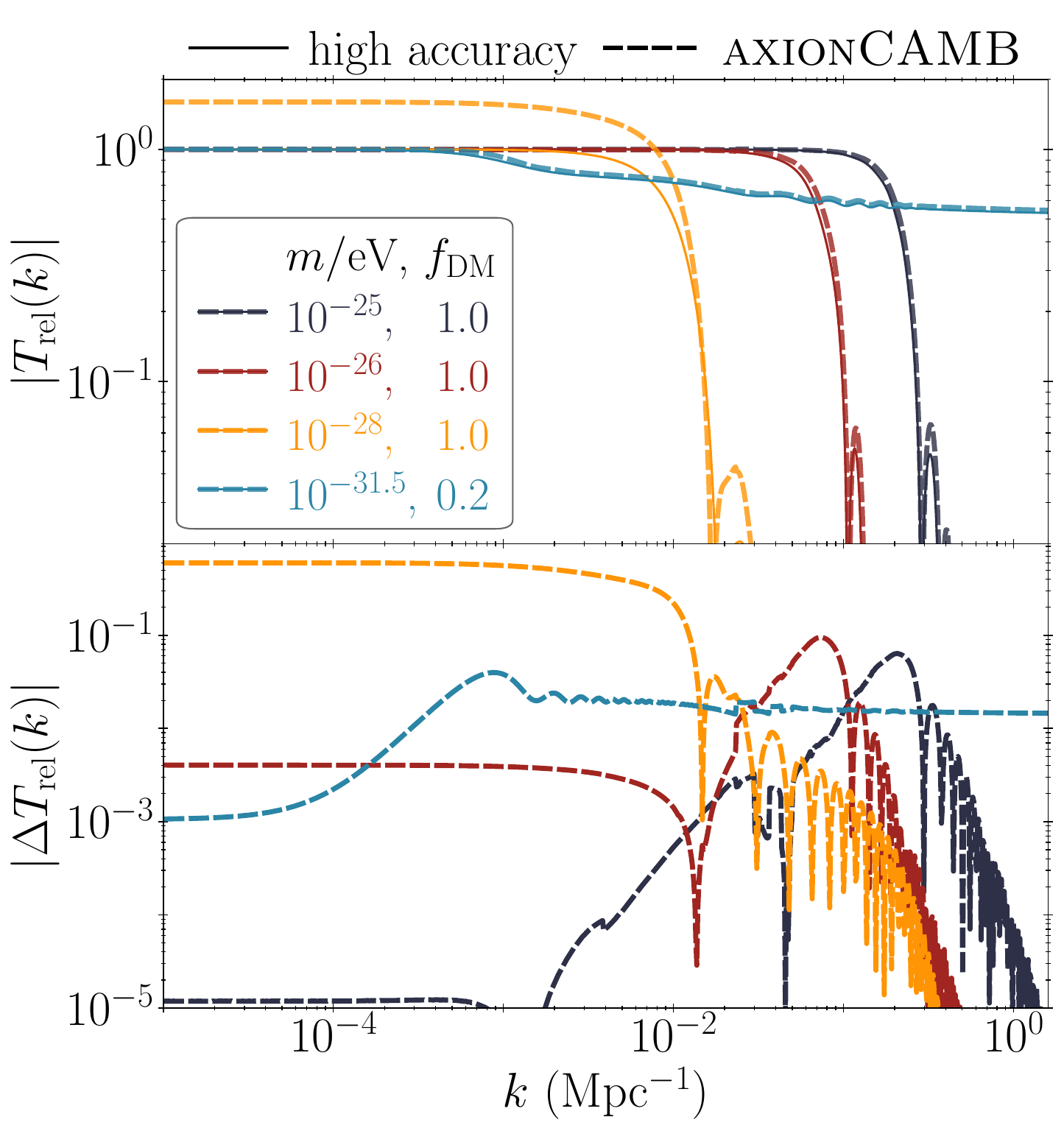}
    \caption{$T_{\rm rel}(k)$ (top) and $\Delta T_{\rm rel}(k)$ (bottom) comparing $\textsc{axionCAMB}$ (dashed) with that run on high accuracy setting (solid), for representative cases in Figs.~\ref{fig:Deltachi2_axionCAMB} and ~\ref{fig:axionCAMB_S8heatmap}. 
    }
\label{fig:Tk_axionCAMB}
\end{figure}

 For reference, we also plot  $\Delta\chi^2(\Lambda{\rm CDM})$, the $\Delta\chi^2$ deviation  between $\Lambda$CDM and our high accuracy calculation in Fig.~\ref{fig:GT-LCDM} before (left panel) and after (right panel) correction for the different $\thetas$ using Eq.~(\ref{eq:thshift}). 
 Note that the linearized correction only reduces the difference when $|\Delta\thetas/\thetas|\ll 0.1$, given the spacing $\Delta \ell\sim 300$ between the acoustic peaks at $\ell_{\rm max}=2400$ sets the criteria for $\Delta\ell/\ell$ to be a perturbative shift of the peaks.
 Instead the correction makes the difference artificially larger near $10^{-29}$\,eV and $f_\DM \sim 1$ but this is not the regime of interest for allowable deviations from $\Lambda$CDM. Many of the largest errors in \textsc{axionCAMB} also occur in such regions.

To identify the more problematic regions for the use of \textsc{axionCAMB} in parameter constraints,
 we place cross marks in Fig.~\ref{fig:Deltachi2_axionCAMB} on parameter points that pass the joint thresholds
 \begin{align}
\Delta \chi^2(\textsc{axionCAMB}) &> 25 \nonumber,\\ \Delta \chi^2(\textsc{axionCAMB}) & > \left( \frac{1}{2} \right)^2 \Delta \chi^2(\Lambda{\rm CDM}),
 \end{align}
where the $1/2$ value is a loose, arbitrarily chosen threshold to reflect an error of half the change in the CMB power spectra $C_\ell$ from $\Lambda$CDM.\footnote{To assess the significance of the error between \textsc{axionCAMB} and our high-accuracy computation, we use the threshold $\Delta \chi^2=25$ rather than $1$ to roughly reflect current rather than future constraints.  In Sec.\,\ref{sec:totalDM}, we found that the future CV limit improves on $\theta_{*}$ errors by at least a factor of 
$\sim 5$ over 
\emph{Planck} errors.   Note however that this scaling only applies to cases where the dominant error is at high $\ell$ since \emph{Planck} is already CV-limited at low $\ell$ in $TT$.}

The source of errors causing both large $\Delta \chi^2$ values and inaccurate depiction of the change from $\Lambda$CDM can be examined on a case-by-case basis. There are at least 4 distinct types of behavior which we illustrate in Fig.~\ref{fig:axionCAMBClerror_chi2rep}. The first is the wildly fluctuating errors near $m =10^{-28}$\,eV and $f_\DM=1$. Here $\Delta C_\ell^{TT}/C_\ell^{TT}$ approaches ${\cal O}(1)$ indicating grossly inaccurate calculations.

We have identified that in this region the \textsc{axionCAMB}
solution for the initial field $\phi_{\rm ini}$ can fail badly as it fits the logarithm of the \ULA\ fraction of matter as a function of $\phi_{\rm ini}$ with spline interpolation to invert the relation. This function saturates as $\phi_{\rm ini} \to \infty$, and so the fit fails. In some cases, due to precision, the implementation in \textsc{axionCAMB} even introduces a zero or nonmonotonic spacing of spline steps resulting in \texttt{NaN} outputs, which we fix to produce Fig.~\ref{fig:Deltachi2_axionCAMB}. \ourcode~avoids this failure by finding the correct $\phi_{\rm ini}$ by bisection, given $\Omega_{a}h^{2}$ instead of the problematic \ULA\ fraction.

While this is mainly a background error, it is so large that it
cannot be corrected with  $\thetas$ as in Eq.~(\ref{eq:thshift}). With $m =10^{-28}$\,eV and $f_\DM=1$ in particular, $\Delta\thetas/\thetas = {-0.16}$, $\Delta \chi^2= 2.8\times 10^6$ 
and $\Delta \chi^2|_{\thetas}= {2.6\times 10^7}$,
indicating a breakdown of the linearized $\thetas$ correction. While a nonlinear correction could potentially improve $\Delta\chi^2$, the error still indicates a substantial misevaluation in distance-related observables in this regime such as $H_0$.

 Secondly, in the lower $f_\DM$ region near the example model with $m =10^{-27}$\,eV and $f_\DM=0.1$, errors of $\Delta C_\ell^{TT}/C_\ell^{TT}$ are around a few percent in the currently well measured acoustic regime. The errors cannot also be fully corrected with $\Delta\thetas/\thetas$, as seen in the example model where $\Delta\thetas/\thetas =0.0011$, $\Delta \chi^2= 1109$ and $\Delta \chi^2|_{\thetas}= 638$. 
 Moreover, the error is a significant fraction of the difference from $\Lambda$CDM: 
 $\sqrt{\Delta \chi^2/\Delta \chi^2(\Lambda\textsc{CDM})}= 0.54$,
and $0.40$
 after the $\thetas$ correction. 
Such an error can make a large fractional change in \ULA\ parameter constraints from the CMB.

Thirdly, near $m =10^{-29}$\,eV and $f_\DM= 0.1$, there remains errors of a few percent in $\Delta C_\ell^{TT}/C_\ell^{TT}$, whose scale has shifted however to lower multipoles and contributes less to $\Delta\chi^2$, although still above cosmic variance in the range where {\it Planck} is CV limited. Larger contributions to $\Delta\chi^2$ at high $\ell$ are correctable
with $\Delta\thetas/\thetas=-3\times 10^{-4}$, with $\Delta\chi^2= 68$ and $\Delta\chi^2|_{\thetas}= 14$
. This case also shows an error that is small compared with the difference from $\Lambda$CDM:  $\sqrt{\Delta \chi^2/\Delta \chi^2(\Lambda\textsc{CDM})} = 0.017$ before and $\approx 0.037$ after correction for $\Delta\thetas/\thetas$.

Finally, a long strip of $\times$'s appears in the right panel of Fig.~\ref{fig:Deltachi2_axionCAMB} between
$10^{-26}-10^{-25}$\,eV.  
For the model at its lowest $f_\DM$ end,  $m=10^{-26}$\,eV and $f_\DM=0.05$, $\Delta\chi^2 = 80$, $\Delta\chi^2|_{\theta_\star}= 71$, and $\sqrt{\Delta\chi^2/\Delta\chi^2(\Lambda\textsc{CDM})} \approx 0.54$ before $\theta_\star$ correction and a negligibly better 0.51 after.
This model is around the mass and $f_\DM$ where the CMB becomes the dominant current constraint \cite{Hlozek:2014lca},
and such an error can shift constraints quantitatively.

In Fig.~\ref{fig:axionCAMB_S8heatmap} we show the errors in $S_8$ as $\Delta S_8$.\footnote{\textsc{axionCAMB} uses a different definition of $\sigma_8$ that is not always the matter density rms defined by our $T_{\rm m}$ in Eq.~(\ref{eq:transfer}).  We calculate the $S_8$ from $\textsc{axionCAMB}$ consistent with our definition here.} 
We highlight specific models where the relationship between $\Delta S_8$ and the
various sources of matter power spectrum error can be better understood  in Fig.~\ref{fig:Tk_axionCAMB} for $T_{\rm rel}(k)$.

The $\Delta S_8$ errors peak  around $m=10^{-26}$\,eV and $f_\DM=1$, where $S_8 = 0.36$
and $\Delta S_8= 0.027$.  Here the \ULA\ Jeans scale is comparable to the scale of $S_8$ and the peak error in $\Delta T_{\rm rel}$ has its maximal impact.

Another local peak in $\Delta S_8$ occurs
at very light masses and intermediate $f_\DM$. As an example we take $m=10^{-31.5}$\,eV and $f_\DM=0.2$, where $S_8=0.5$ and $\Delta S_8=0.013$. This error is attributed to the $\Delta T_{\rm rel}\sim 10^{-2}$ errors at high $k$ below the Jeans scale where the smooth \ULA\ background slows the growth of CDM fluctuations. For much larger $f_\DM$, $T_{\rm rel}$ is so small that even large fractional errors in its value lead to small $\Delta S_8$ despite being large fractional errors in $S_8$ itself. 
There are also spurious oscillations below the Jeans scale between $k=10^{-3} - 10^{-2}$\,Mpc$^{-1}$ which are the consequence of the very large quasistatic equilibrium violation in the \ULA\ that are masked by the much larger CDM and baryon density fluctuations that contribute to $T_{\rm rel}$ (see Fig.~\ref{fig:quasistaticdemo}). 

In the regime around $m=10^{-28}$\,eV and $f_\DM=1$ that we studied for $C_\ell$, the same background errors in $\phi_{\rm init}$ cause ${\cal O}(1)$ errors in the low $k$ amplitude of $T_{\rm rel}(k)$.   On the scales relevant to $S_8$, these problems produce a $\Delta S_8=0.013$
for $S_8=0.031$
and a large fractional error despite 
the model itself being ruled out by its inability to form nonlinear structure regardless of this error.

For a model closer to $\Lambda$CDM, we take $m = 10^{-25}$ eV, $f_\DM = 1$,
where $S_8 = 0.72$  and  $\Delta S_8 = 0.018$, which is a small but non-negligible fraction of the change from the $\Lambda$CDM value of $0.85$.

\begin{figure}
     \includegraphics[width=1\linewidth]{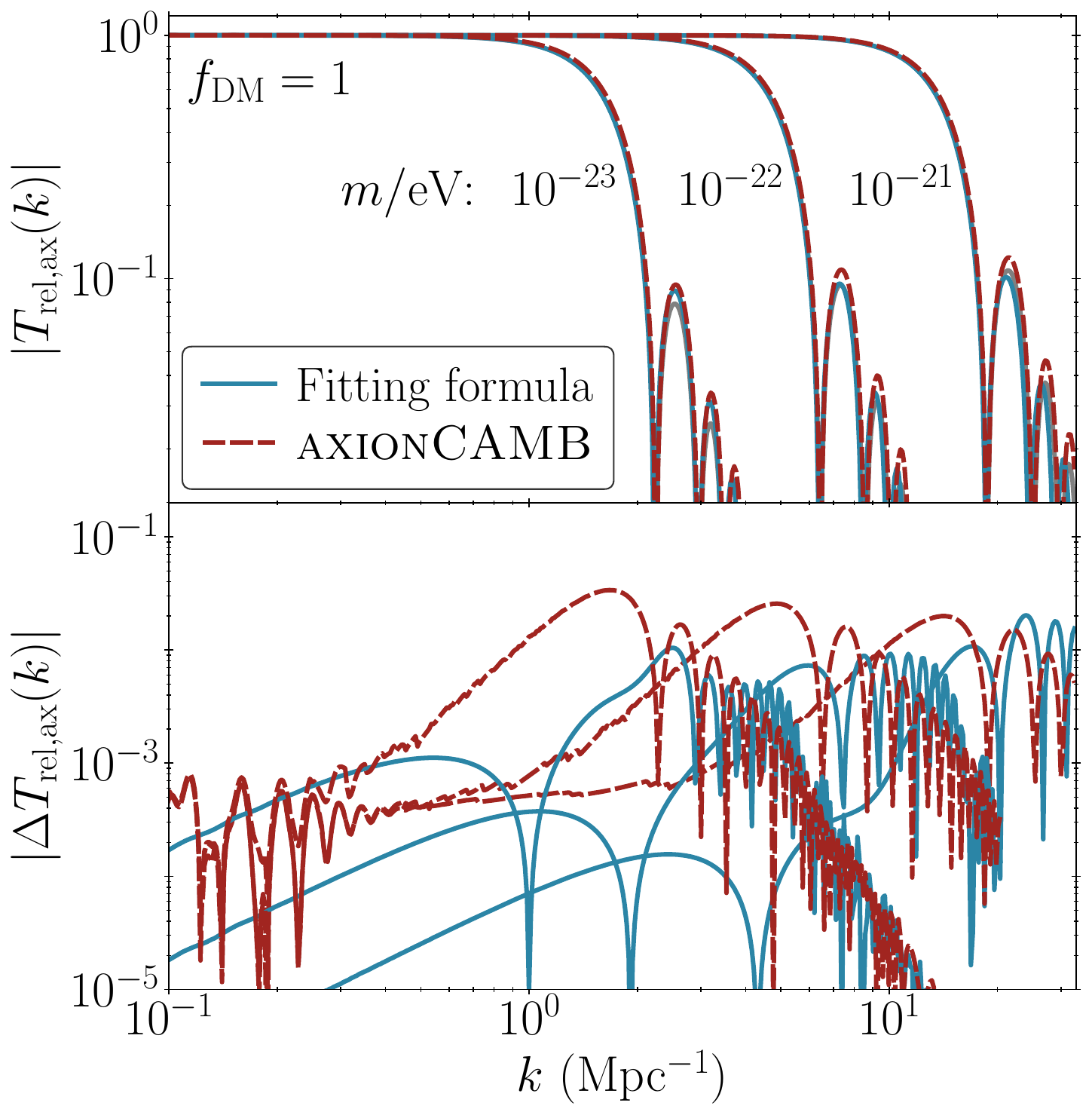}
    \caption{Relative transfer function in the fuzzy dark matter regime $T_{\rm rel,\ax}=T_{\ax}/T_c$ (gray, high accuracy, see text). The errors for \textsc{axionCAMB} and a fitting function from Ref.~\cite{Passaglia:2022bcr} are shown in the bottom panel.}
    \label{fig:fittingformula_comaprison}
\end{figure}

\subsection{Fuzzy Dark Matter}
\label{sec:FDM}

Apart from the above study of masses below $10^{-24}$ eV where the \ULA\ Jeans scale is on cosmological scales, we examine the fuzzy dark matter region where the Jeans scale is instead on astrophysical scales. Here the transfer function is mainly used to set up initial conditions of the \ULA\ for nonlinear simulations.   We therefore exclude the baryons and massive neutrinos from the matter transfer function and take its ratio with the cold dark matter transfer function to define $T_{\rm rel,\ax}= T_{\ax}/T_c(f_\DM=0)$ for $f_\DM=1$.  
In Fig.~\ref{fig:fittingformula_comaprison} we plot this quantity and its errors $\Delta T_{\rm rel,\ax}$ from \textsc{axionCAMB}, which slightly underestimates the Jeans scale leading to an over-estimate of the transfer function in this regime which peak at $\Delta T_{\rm rel,\ax} \sim 10^{-2}$.  
Most of this difference would be absorbed into a small shift in $m$.
\ourcode\ on the other hand performs similarly or better than that shown for $10^{-24}$\,eV in Fig.~\ref{fig:perfTk_allDM} so that these errors are reduced to $10^{-3}$ or below.  While not shown in Fig.~\ref{fig:perfTk_allDM} for clarity reasons, they are entirely negligible for simulation purposes.

We also compare in Fig.~\ref{fig:fittingformula_comaprison} the errors from a fitting function of $T_{\rm rel, ax}$ from Ref.~\cite{Passaglia:2022bcr}:
\begin{equation}
\label{eq:fitting}
T_{\rm rel,\ax}(k) \simeq \frac{\sin(x^{n})}{x^n (1 + B x^{6-n})}\,,
\end{equation}
where \cite{Hu:2000ke}
\begin{equation}
x \equiv A \frac{k}{k_J },\quad k_J = 9 m_{22}^{1/2}\,,
\label{eq:kJ}
\end{equation}
with $m_{22} \equiv m / 10^{-22}\,$eV, $n = 5/2$, and
\begin{eqnarray}
A &=& 2.22 m_{22}^{1/25- 1/1000\ln(m_{22})}\,, \nonumber\\ 
B &=& 0.16 m_{22}^{-1/20}\,. 
\end{eqnarray}
Here most of the amplitude error in $\Delta T_{\rm rel,\ax}$ compared with \textsc{axionCAMB} is removed, leaving mainly a phase error in the Jeans oscillations at  high $k$. 
Since phasing differences would be destroyed in nonlinear evolution, the fitting function should thus also provide sufficiently accurate initial conditions for numerical simulations. 

While the \ourcode\ method works for  $m>10^{-18}$\,eV, the Jeans scale moves into the $k_J \gtrsim 10^3$\,Mpc$^{-1}$ regime where non-\ULA\ perturbations also are difficult to track.   In this regime, the fitting function is a more efficient means of providing initial conditions for simulations.

\section{KG solutions and scalings}\label{app:KG}

Here we discuss field solutions for the KG system that are useful in establishing our effective method for the background and perturbations. 

For the background, the field has an exact solution when the expansion history can be described by a constant total equation of state $w_T$ (e.g.~Ref.~\cite{Marsh:2010wq}): 
\begin{equation}\label{eq:Bessel_KGsol}
\phi \propto a^{-3/2} \sqrt{\mt} J_\nu(\mt), 
\end{equation}
where 
\begin{equation}
\nu = \frac{(1-w_T)}{2(1+w_T)}.
\end{equation}
The Bessel function takes the limiting form
\begin{equation}
\lim_{\mt \ll 1} J_\nu(\mt) \approx 
\frac{2^{-\nu}}{\Gamma(1+\nu)} (\mt)^\nu,
\end{equation}
whereas in the opposite limit, 
\begin{equation}
\lim_{\mt \gg 1} J_\nu(\mt) \approx \sqrt{\frac{2}{\pi \mt}} \left( \cos\beta -\frac{4\nu^2-1}{8 \mt} \sin\beta \right),
\label{eq:BesselApprox}
\end{equation}
with the evolving phase 
\begin{equation}
\beta = mt - \nu\pi/2 -\pi/4.
\label{eq:PhaseApprox}
\end{equation}
Notice that for any $w_T$,
$\phi =$\, constant for $\mt \ll 1$, consistent with a Hubble drag dominated solution.
Conversely for $\mt \gg 1$, i.e.\ $m/H \gg 1$, the amplitude of $\phi \propto a^{-3/2}$ to leading order, consistent with \ULA s redshifting as matter.  

In the radiation-dominated epoch, $w_T=1/3$ and this solution in principle checks all of the steps of our effective method  for $m\gg 10 H_{\rm eq}$.  
First of all here $H t=1/2$ and we can immediately verify the initial conditions for $\dot\phi$ in Eq.~(\ref{eq:KG_IC}) as well as those of the perturbations given that $\dot h\propto \tau$.   This applies to any mass since the initial conditions are always set in radiation domination. Next we can determine the \ETA\ of the oscillations in the $m/H=2\mt \gg 1$ limit 
of Eq.~(\ref{eq:BesselApprox}).   Using the same approach of setting $\cos^2\beta =\sin^2\beta \rightarrow 1/2$ and $\cos\beta\sin\beta \rightarrow 0$, we obtain
\begin{equation}
w^\ETA = \frac{9}{8}(1+w_T) \left( \frac{H}{m} \right)^2,
\label{eq:weta_analytic}
\end{equation}
and hence $w^\ETA = (3/2) (H/m)^2$ in radiation domination, which implies $A_w=3/2$ in Eq.~(\ref{eq:wEFA}).  This value was adopted in Ref.~\cite{Passaglia:2022bcr} where 
$m \gg 10 H_{\rm eq}$ was assumed.

\begin{figure}
    \includegraphics[width=1\linewidth]{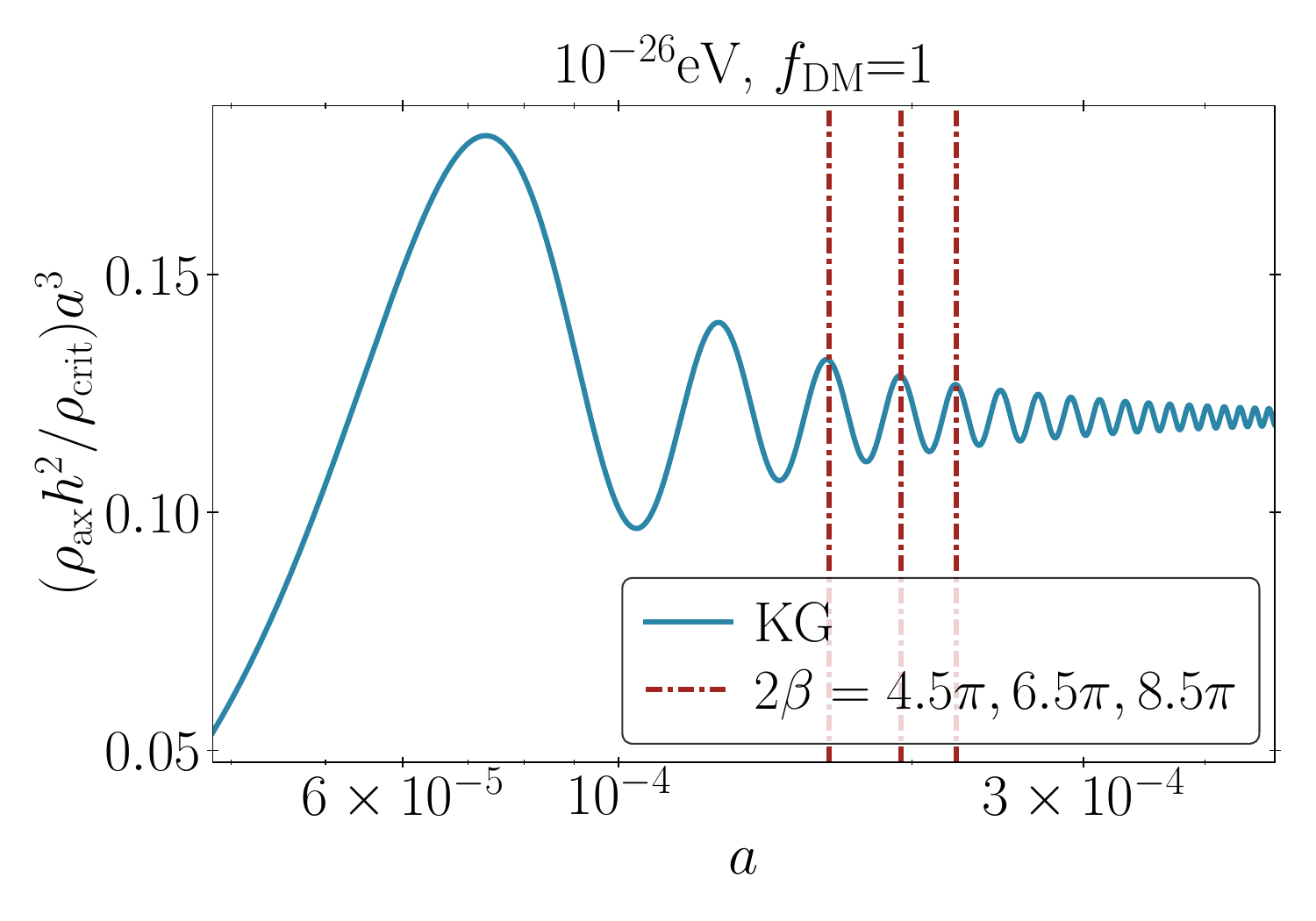}
    \caption{Prediction of temporal location  of peaks the KG oscillations in $\rho_\ax a^3$ for a case where they occur around equality.  Here $m=10^{-26}$\,eV, $f_\DM = 1$ and 
    vertical lines represent the prediction for peaks peaks $n=3,4,5$ where  $2\beta = \pi/2+2\pi(n-1)$ (see text for details).}
    \label{fig:phasefinding_demo}
\end{figure}

  Given that the homogeneous equation of motion for the field perturbation is the same as the background for $k/a m \rightarrow 0$ and in the radiation dominated epoch $\dot\hL \propto \tau$ is an external source, we can construct the inhomogeneous solution using the Green function technique to find
\begin{equation}
\lim_{k\rightarrow 0} \frac{\delta p_\ax^\ETA}{\delta \rho_\ax^\ETA} =
\left[\frac{9}{8} w_T (1+w_T)+1\right] 
\frac{H^2}{m^2}, 
\label{eq:Ac_analytic}
\end{equation}
or $(3/2)(H/m)^2$ in radiation domination,
which in principle specifies $A_c=3/2$ in Eq.~(\ref{eq:cs2}).  As explained in the main text, since we  construct the \ETA\ not from the analytic solution but rather from the auxiliary field construction, we follow Ref.~\cite{Passaglia:2022bcr} and use the empirical value of $A_c =5/4$ determined from this construction for switches deep in radiation domination.  In the opposite subhorizon regime, it is mainly the finite value of $k/am \gg H/m$ that causes the phase of the perturbations to differ from the background rather than the metric fluctuation.  We can derive an \ETA\ for the sound speed by ignoring the metric fluctuation and extracting the natural frequency of oscillation from the dispersion relation 
\begin{equation}
\omega=\sqrt{k^2+a^2 m^2}.
\end{equation}
The phase of the field fluctuation differs from that of the background by
\begin{align}
\delta\beta &= \int d \ct  \omega - mt =
\int d\ct (\omega - a m )
\nonumber\\
&\equiv k\int d\tau c_\phi,
\end{align}
where
\begin{equation}
c_\phi = \left(\frac{k}{am}\right)^{-1} \left[\sqrt{ 1+ \left(\frac{k}{am}\right)^2}-1\right].
\end{equation}
We can then repeat the \ETA\ construction of
${\delta p_\ax^\ETA}/{\delta \rho_\ax^\ETA}$
and identify $c_\ax^2 \rightarrow c_\phi^2$ in this limit (see Ref.~\cite{Passaglia:2022bcr}).  The combination of the two limits produces the full model for the sound speed in
Eq.~(\ref{eq:cs2}).

The specific value for the phase of the background KG oscillations around the \ETA\ as a function of $m/H$ is also of interest.
Since the background density is determined by the square of the field, the relevant quantity is
\begin{equation}
2 \beta = \frac{m}{H} - \frac{3}{4}\pi, \qquad (w_T=1/3). 
\label{eq:RDphase}
\end{equation}
This determines the phase of the switch during radiation domination.

After radiation domination,
the true expansion history cannot be fully characterized by a constant $w_T$, so for switches that occur after radiation domination, these analytic solutions do not directly apply.  
Nonetheless, for $w^\ETA$ and $c_s$, the variation with $w_T$ in Eqs.~(\ref{eq:weta_analytic}) and (\ref{eq:Ac_analytic}) suggest that the equation of state and sound speed at $k\rightarrow 0$ still scale as $(H/m)^2$ but with $A_w$ and $A_c$ that vary by order unity across matter radiation equality.   This motivates the prescription used in our effective method in 
Eqs.~(\ref{eq:wEFA}) and (\ref{eq:cs2}) where they are taken to be adjustable quantities.

Beyond radiation domination we can still obtain a good analytic approximation for the phase by assuming $w_T$ varies very slowly compared with the oscillations. 
For example at high $m/H\gg 1$, the $n$th peak in the energy density $\rho_\ax a^{-3}$ should be located at
\begin{equation}
2\beta = \pi/2 + 2\pi (n-1)
\label{eq:betatarget}
\end{equation}
for some integer $n$ according to Eq.~(\ref{eq:Bessel_KGsol}) and we seek a prescription for the value of the scale factor $a_*$ or $H_*$ at which these maxima occur.
First we take the  $f_\DM\ll1$ limit where
\begin{equation}
Ht = \frac{2}{3}\frac{y^2-y-2+2\sqrt{1+y}}{y^2},
\end{equation}
with
\begin{equation}
y\equiv\frac{a}{a_{\rm eq}}, 
\qquad w_T=\frac{1}{3(1+y)}
\end{equation}
so as to obtain the phase from Eq.~(\ref{eq:PhaseApprox}) as
\begin{equation}
2\beta = \frac{m}{H^\ETA}\left(\frac{4}{3}\frac{y^2-y-2+2\sqrt{1+y}}{y^2}\right) - \frac{1+y}{4+3y} 3\pi. 
\label{eq:betaphase}
\end{equation}
Finally, given a desired phase $2\beta$, e.g.\ from Eq.~(\ref{eq:betatarget})  at the switch, we invert this formula to find
\begin{equation}
\frac{m}{H_*^{\ETA}} = \frac{3}{4}\frac{y^2}{y^2-y-2+2\sqrt{1+y}}\left(2\beta + \frac{1+ y}{4+3 y} 3\pi\right).
\label{eq:switchbeta}
\end{equation}
where the use of $H_*^\ETA$ allows us to generalize this algorithm for finite $f_\DM$.
In this case, we assume that $m/H_*^{\ETA}$ is sufficiently large that the estimation of $y$ using a matter scaling of $\rho_\ax^\ETA \propto a^{-3}$ to the present applies.

 Note that the $y\rightarrow \infty$ solution is
\begin{equation}
\lim_{y\rightarrow \infty} \frac{m}{H_*^{\ETA}} = \frac{3}{2} \beta + \frac{3\pi}{4}
\end{equation}
which, in combination with Eq.~(\ref{eq:RDphase}) for $y\rightarrow 0$, bounds the intermediate range. 
 For a specified value of $\beta$,  \ourcode\ takes an initial zeroth order guess for $m/H_*$ based on an interpolation between the radiation and matter dominated limits:
\begin{equation}
\frac{m}{H_*^{(0)}}
= 2\beta+ \frac{3\pi}{4}- \frac{ \beta^2}{2(\beta + m/H_{\rm eq})}
\end{equation}
and iterates on $m/H_*$ until convergence in $\beta$ is reached with Eq.~(\ref{eq:betaphase}).

In Fig.~\ref{fig:phasefinding_demo} we test this prescription for locating density peaks on the maximally 
difficult case of $f_\DM=1$ and a mass where
the corresponding scale factor is near equality, $m=10^{-26}$\,eV.   We compare the location of the peaks in $\rho_\ax a^{3}$ and their prediction from this procedure for $2\beta = 4.5\pi, 6.5\pi, 8.5\pi$ and show that they closely match.  The resulting values from Eq.~(\ref{eq:switchbeta}) are $m/H_*^\ETA = 15.7, 21.3, 27.0$ respectively.

 This technique applies to any value of $\beta$ and we use this technique in \ourcode\ to choose a specific phase at the switch that minimizes CMB errors around equality (see Appendix \ref{app:cmbsources_corr}).
It may also be useful in other approaches that minimize errors by choosing specific phases in the KG oscillations (e.g.\ \cite{Urena-Lopez:2023ngt}).

\section{\ETA\ construction}\label{app:ETA_construction}
This appendix provides details on the explicit construction for the \ETA\ algorithm corresponding to Secs.~\ref{sec:ETA} in the main text.

In a separate module prior to the usual \textsc{CAMB} Boltzmann-Einstein system for the perturbations we solve for the KG background field by adapting the axion background module from \textsc{axionCAMB}.
 This module supplies 
 KG field solutions at the switch time, parameterized by $H_*/m$
 \begin{align}
\phi_* \equiv& \phi \Big|_{H_*/m}, \nonumber\\
\phi_*' \equiv& \frac{H}{m}  \frac{d \phi}{d\ln a
} \Big|_{H_*/m} .
 \end{align}
 and use the notation $' = d/d\,(\mt)$ here and below. 
 Given the auxiliary field equation (\ref{eq:sincos_decomposition}) and the KG equation (\ref{eq:KG}), the auxiliary variables must satisfy
\begin{equation}
\begin{aligned}
\varphi_c'' + 2\varphi_s' + 3\frac{H}{m}[\varphi_c' + \varphi_s] &= 0, \\
\varphi_s'' - 2\varphi_c' + 3\frac{H}{m}[\varphi_s'-\varphi_c ] &= 0, \\
\varphi_c &= \phi_*,\\
\varphi_s + \varphi_c' &=  \phi'_*,
\end{aligned}    
\label{eq:auxsystem}
\end{equation}
where we implicitly assume the auxiliary variables and their derivatives are evaluated at the switch and drop the $*$ subscript for compactness.
The boundary condition Eq.~(\ref{eq:auxiliary_conditions}) eliminates the second derivatives as $\varphi_{c}'' = \mathcal{A}\varphi_{c}', \varphi_{s}''= \mathcal{A}\varphi_{s}' $, where
\begin{equation}
\mathcal{A} = \frac{1}{m}\left(-\frac{3}{2}  H^\ETA +  \frac{d \ln  H^{\ETA} }{dt }\right),
\end{equation}
and reduces the system (\ref{eq:auxsystem}) 
to a linear algebraic relation which uniquely determines the 4 auxiliary variables
\begin{align}
\varphi_{c} &= \phi_*,\nonumber\\
\varphi_{c}' &= -\frac{3\frac{H_*}{m}\left[2\phi_* + \left(\mathcal{A} + 3\frac{H_*}{m}\right)\phi_*'\right]}{\mathcal{A}^2 + 3\mathcal{A}\frac{H_*}{m} + 4},\nonumber\\
\varphi_{s} &= \phi_*' - \varphi_{c}' ,\nonumber\\
\varphi_{s}' &= \frac{3\frac{H_*}{m}\left(\mathcal{A}\phi_* - 2\phi_*'\right)}{\mathcal{A}^2 + 3\mathcal{A}\frac{H_*}{m} + 4}.
\end{align}   

Note that $H_*$ here is the instantaneous Hubble rate at the switch, and should be distinguished from $H^\ETA$. To evaluate $\mathcal{A}$ requires $H^\ETA$ and \ETA\ equation-of-state at the boundary, since
\begin{equation}
\begin{aligned}
\mathcal{A} &= -\frac{1}{2}\frac{H^\ETA}{m}\left[3 - \frac{H_0^2}{ (H^\ETA)^2}\sum\limits_i -3(1+w_i)\frac{\rho_i}{\rho_\text{c}}\right],\\
\end{aligned}    
\end{equation}
where $\rho_\text{c}$ is the critical density today, and the sum is over all components, including the \ULA, for which we take $w_\ax^\ETA=p_\ax^\ETA/\rho_\ax^\ETA$.

Notice that evaluating $H^\ETA$ and $w_\ax^\ETA$ both determine auxiliary variables through the boundary condition and also require the auxiliary variables for their construction. This is resolved via iteration, where $H^\ETA$ and $w_\ax^\ETA$ are initialized as $H$ and $9(H/m)^2/8$ respectively,\footnote{This starting value is taken from Eq.~(\ref{eq:weta_analytic}) under matter domination $w_T=0$ since $w_\ax^\ETA$ would only affect the construction if the \ULA s contributed substantially to the expansion rate. The choice does not affect our final results here due to our iteration procedure.  Likewise Ref.~\cite{Passaglia:2022bcr} does not need to iterate since they focus on $m \gg 10H_{\rm eq} $ where the  switch is deep in the radiation-dominated epoch and $H^\ETA \approx H$.
} and updated by the resulting $\rho_\ax^\ETA$ in Eq.~(\ref{eq:ETAHubble}) as $H^\ETA$ and $p_\ax^\ETA/\rho_\ax^\ETA$, until convergence is achieved for $w^\ETA$ (which also implies convergence in $H^\ETA$). 

The auxiliary perturbation variables follow a similar logic as the background, only that the $H^\ETA$ and thus $\mathcal{A}$ are already defined by the background. Specifically
\begin{equation}
\begin{aligned}
\delta \varphi_{c} &= \delta \phi_*,\\
\delta \varphi_{c}' &= \frac{-2\mathcal{B}
- \left(\mathcal{A} + 3\frac{H_*}{m}\right)\mathcal{C}
}{2\left(\mathcal{A}^2 + 3\mathcal{A}\frac{H_*}{m} + \frac{2k^2}{a_*^2m^2} + 4\right)},\\
\delta \varphi_{s} &= \delta \phi_*' - \delta \varphi_{c}',\\
\delta \varphi_{s}' &= \frac{\mathcal{A}\mathcal{B}
- \left(2 + \frac{k^2}{a_*^2m^2}\right)
\mathcal{C}
}{2\left(\mathcal{A}^2 + 3\mathcal{A}\frac{H_*}{m} + \frac{2k^2}{a_*^2m^2} + 4\right)},
\end{aligned}
\end{equation}
where
\begin{eqnarray}
\mathcal{B}&=&(\varphi_{c} - \varphi_{s}') \hL_*'   - 2\frac{k^2}{a_*^2m^2}\delta \phi_*'+ 6\frac{H_*}{m}\delta \phi_*,
\nonumber\\
\mathcal{C}&=&
(\varphi_{c}' + \varphi_{s})\hL_*' + 2\frac{k^2}{a_*^2m^2}\delta \phi_*  + 6\frac{H_*}{m}\delta \phi_*'.
\end{eqnarray}

The auxiliary variables are then used to define
 the \ETA\ of the \ULA\ stress energy tensor components \cite{Passaglia:2022bcr}
 \begin{align}
\label{eq:rhoPef}
\rho_\ax^\ETA ={}&  \frac{m^2}{2}  \left(\varphi_c^2 +  \varphi_s^2  + \frac{{\varphi_c'}^2}{2} +\frac{{\varphi_s'}^2}{2}  -  \varphi_c \varphi_s' +  \varphi_s \varphi_c'\right)\!,  \nonumber\\
p_\ax^\ETA ={}& \frac{m^2}{2}  \left(\frac{{\varphi_c'}^2}{2} +\frac{{\varphi_s'}^2}{2}  -  \varphi_c \varphi_s' +  \varphi_s \varphi_c'\right),\nonumber \\
\delta \rho_\ax^\ETA   ={}& \frac{m^2}{2}  \left[\varphi_s \delta\varphi_c' - \varphi_c \delta\varphi_s' + \delta\varphi_c' \varphi_c'+ \delta\varphi_s' \varphi_s' \right.\nonumber\\&+ \left.  \delta\varphi_s (2 \varphi_s + \varphi_c') + \delta\varphi_c  (2 \varphi_c - \varphi_s') \right],\nonumber\\
\delta p_\ax^\ETA  ={}&\ \delta\rho_\ax^\ETA - m^2 \left[\delta \varphi_s \varphi_s + \delta \varphi_c \varphi_c \right],\\
(\rho u)_\ax^\ETA  ={}&\ \frac{  m k}{2 a} \left[\delta \varphi_c \left(\varphi_s+\varphi_c'\right)+ \delta \varphi_s\left(-\varphi_c+\varphi_s'\right)\right].\nonumber
\end{align}

\begin{figure}
    \includegraphics[width=1\linewidth]{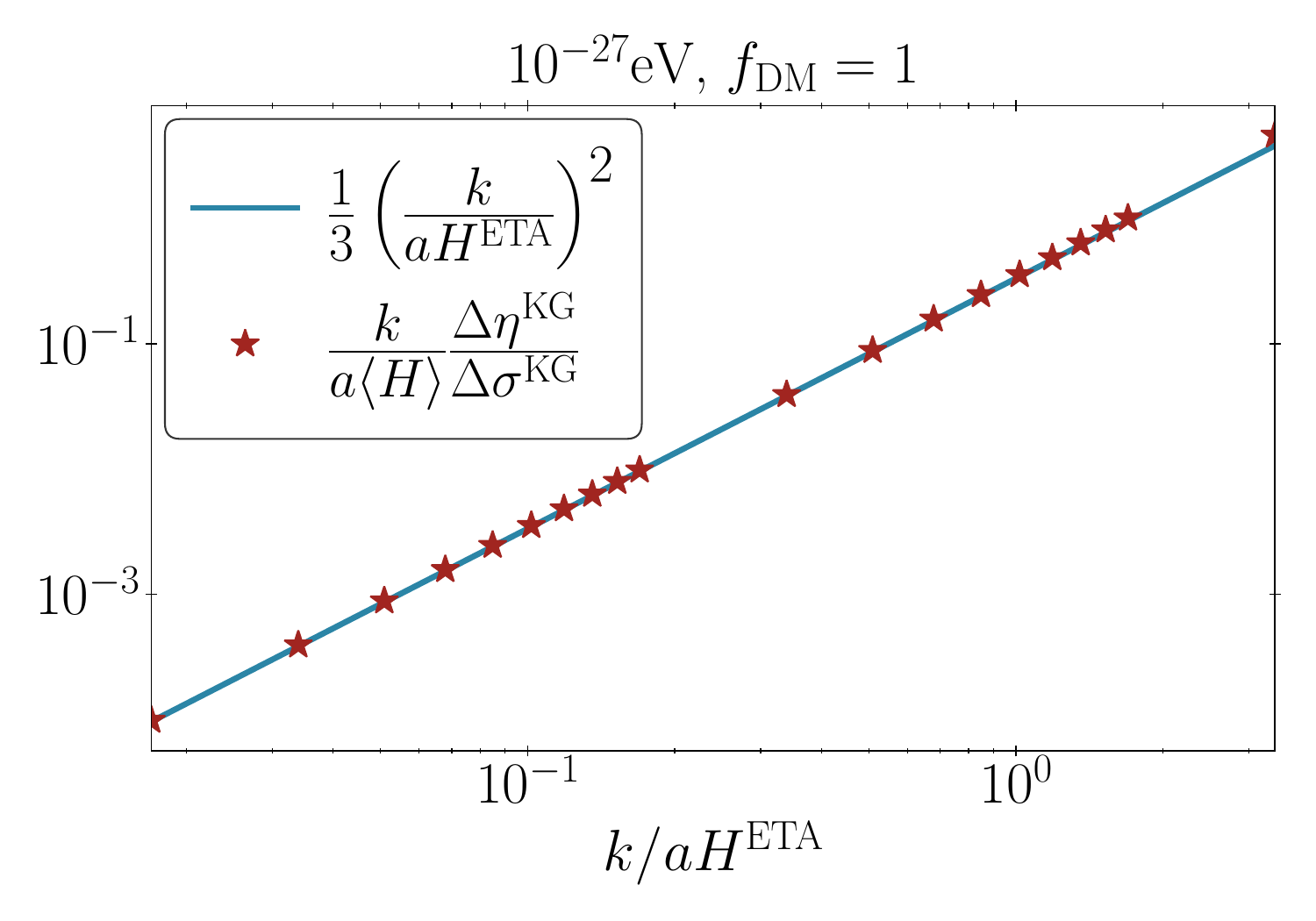}
    \caption{Relative KG oscillation amplitude in the metric perturbations $\Delta\eta^{\rm KG}$ and $\Delta\sigma^{\rm KG}$ (red stars, see discussion around Eq.~(\ref{eq:Deltaln_fit})) compared with the prediction from the partitioning in Eq.~(\ref{eq:partition_Heta}) (blue line) for $f_\DM = 1$, $m = 10^{-27}$\,eV.}
    \label{fig:Wov(1-W)demo}
\end{figure}

The \ETA\ of the metric fluctuations $\etaT$ and $\shear$ follows once the partitioning of  their responses to the KG oscillations are determined.  In the main text, we used causality to specify this division and here we empirically test it against their numerical responses.
To measure the amplitude of oscillations in $\etaT$ and $\shear$,
we examine the range $10 \lesssim m/H \lesssim 20$ in the KG system, and manually take the  difference between a neighboring  peak and trough  of the KG oscillations
\begin{equation}
\begin{aligned}
\Delta \{\shear, \etaT\}^{\rm KG} \equiv \frac{1}{2}\Bigg[\{\shear, \etaT\}&\Big|_{\text{peak} } - \{\shear, \etaT\}\Big|_{\text{trough}}\Bigg],
\end{aligned}
\end{equation} 
with the convention that these are positive definite quantities.
We then construct 
\begin{equation}\label{eq:Deltaln_fit}
\begin{aligned}
\frac{k}{a\langle H\rangle}\frac{\Delta\etaT^{\rm KG}}{\Delta\shear^{\rm KG}},
\end{aligned}
\end{equation}
where $\langle H\rangle$ is the background Hubble taken manually at the ``node'' point between the two and compare it to the model of Eq.~(\ref{eq:partition}) for the oscillation amplitude around the \ETA
\begin{equation}\label{eq:partition_Heta}
\frac{W}{1-W} = \frac{1}{3} \left( \frac{k}{a H^\ETA}\right)^2,
\end{equation}
 The result of this test is plotted in Fig.~\ref{fig:Wov(1-W)demo} for  $m = 10^{-27}$ eV, $f_\DM = 1$, where we find good agreement  across the $k$-modes ranging from $10^{-4}$ to $10^{-2}$ Mpc$^{-1}$, enclosing $k/aH^{\ETA}\sim 1$, the most relevant scale range for the weighting.  We have verified that this agreement is maintained across other masses  in this regime ($m = 10^{-26}\sim10^{-29}$ eV) and  values of  $f_\DM \gtrsim 0.1$ where the oscillations can be accurately measured. 

\section{\EFA\ and quasi-static equilibrium}\label{app:EFAvars_and_quasistatic}

In Sec.~\ref{sec:EFA}, the \EFA\ for the \ULA\ system is obtained by setting the \EFA\ initial conditions by matching to \ETA\ variables at the switch, and also assigning values for the fluid equations of state $w_\ax^\EFA$ and $c_\ax^2$ using Eqs.~(\ref{eq:wEFA}) and~(\ref{eq:cs2}). The choices involved in this method set its accuracy, and differ from the prior approaches of \textsc{axionCAMB} and Ref.~\cite{Passaglia:2022bcr}. 

We now elaborate on these differences and their relationship to
the quasi-static limit (QSL) of the sub-Jeans scale behavior of \ULA\ perturbations. Put simply, a sharp switch from the KG $w_{\ax}$ to $w_\ax^\EFA=0$ in \textsc{axionCAMB}  does not maintain the QSL solution nor does
the analytically motivated $w_{\rm ax}^{\rm EFA}=(3/2)(H/m)^{2}$ chosen in Ref.~\cite{Passaglia:2022bcr}. This becomes a spurious source of  pressure-wave oscillations at the switch.

\begin{figure}
    \includegraphics[width=1\linewidth]{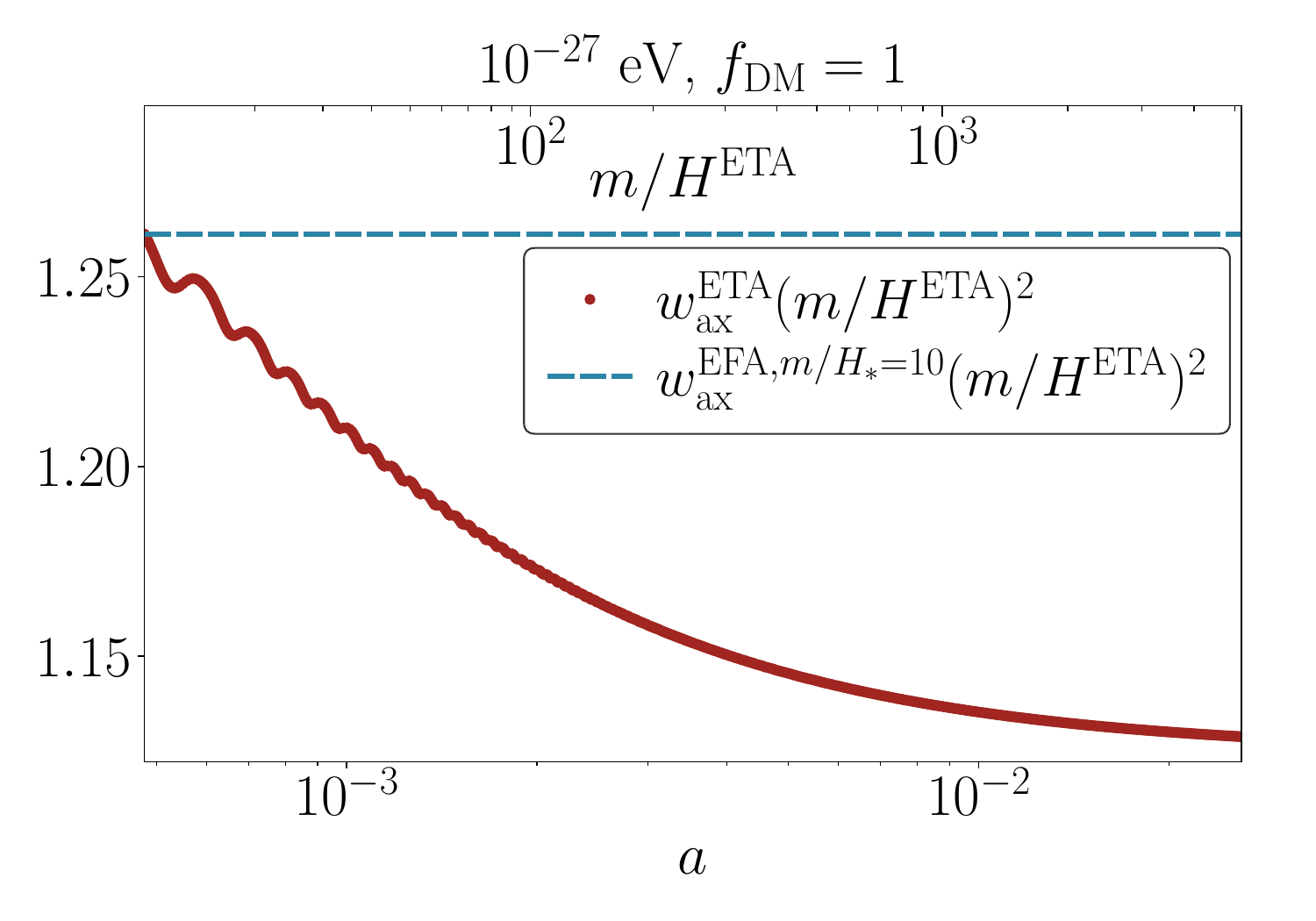}
    \caption{Equation of state in \EFA\ (blue dashed, switched at $m/H_*=10$) and \ETA\ (red dots)  rescaled with $(m/H^\ETA)^2$ (top axis).  
    Here $m=10^{-27}$ eV, $f_\DM = 1$.}
    \label{fig:wETA_scal}
\end{figure}

Consider the small scale limit of  $k/am \gg 1$ and $k/aH\gg 1$. 
Under this condition, Eq.~(\ref{eq:perturbedKG}) yields
\begin{equation}
k^2 \delta \phi \approx -\frac{\dot\hL}{2}\dot{\phi}.
\end{equation}
This closed-form non-dynamical (algebraic rather than differential) equation for $\delta \phi$ is the QSL solution and implies
\begin{equation}
k \rho_\ax  u_\ax \equiv
k ({\rho_\ax + p_\ax}) v_\ax \approx - ({\rho_\ax + p_\ax})\frac{\dot\hL}{2}
\label{eq:vhequilibrium}
\end{equation}
using Eq.~(\ref{eq:axionstressenergy}).
 The source terms of the continuity equation~(\ref{eq:fluideom})
for $\delta\rho_\ax$ cancel in this limit, effectively yielding an equation of hydrostatic equilibrium. 

To understand this result, note that in synchronous gauge, gravitational effects are represented by the changes to the spatial metric through $\dot\hL$. The velocity $v_\ax$ is measured relative to free-fall and  remains finite to counter $\dot\hL$ when in equilibrium. More explicitly, in this limit $\dot\hL$ is much larger than $\dot\etaT$, and so Eq.~(\ref{eq:vhequilibrium}) becomes
\begin{equation}
v_\ax \approx -\shear.
\end{equation}

\begin{figure}
    \includegraphics[width=1\linewidth]{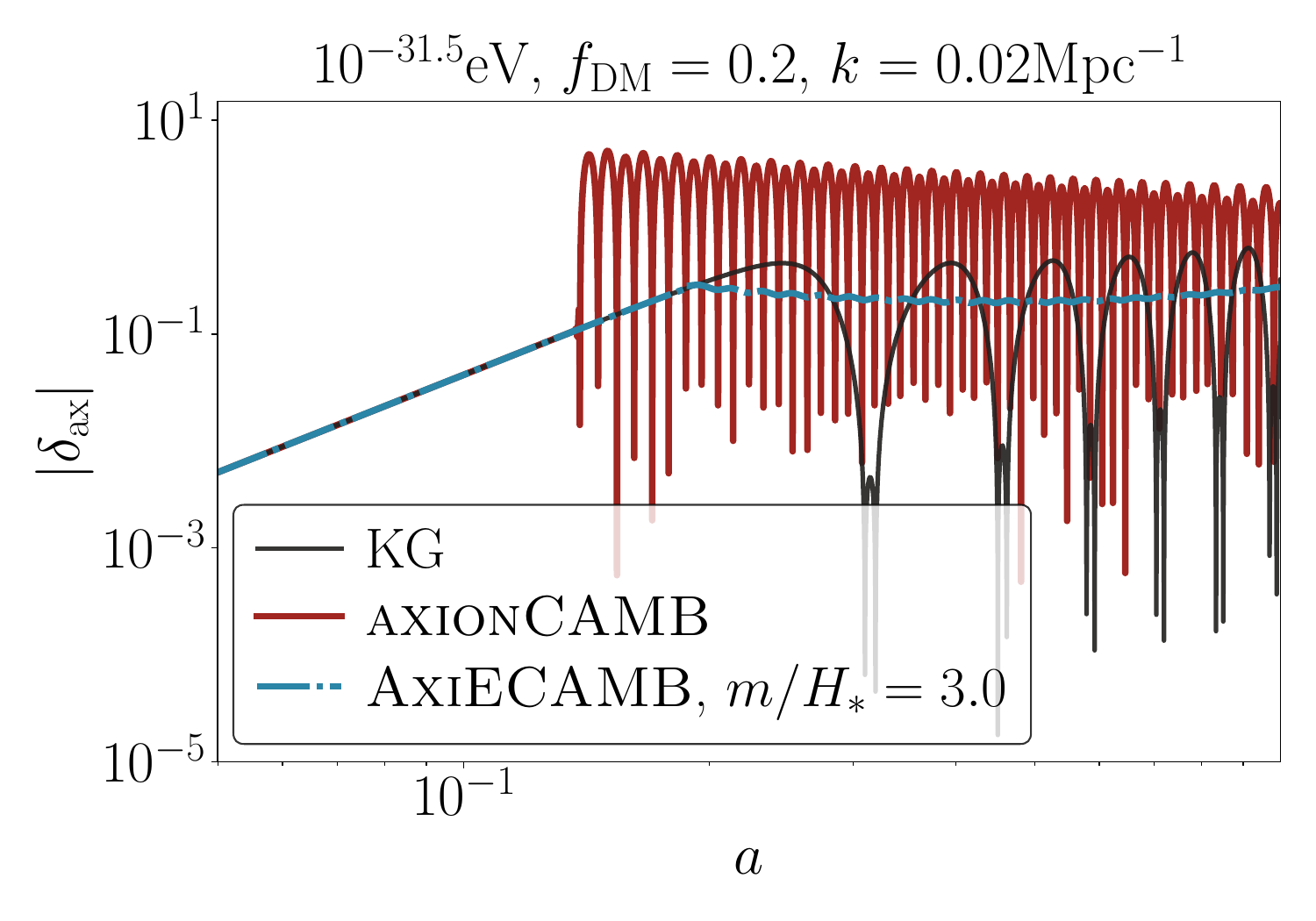}
    \caption{\ULA\  density fluctuation for $10^{-31.5}$ eV, $f_\DM = 0.2$, at $k = 0.02$Mpc$^{-1}$. Plotted together are solutions from KG (gray), \textsc{axionCAMB} (red), and $\textsc{AxiECAMB}$ set at the same $m/H_* = 3$ as \textsc{axionCAMB} (blue). }
\label{fig:quasistaticdemo}
\end{figure}

Our switch conditions should respect this equilibrium, in addition to cycle-averaged momentum conservation (i.e.\ $u_\ax^\ETA=u_\ax^\EFA$).
As $p_\ax$ appears in in the definition of the \ULA\ momentum perturbation,
jointly satisfying these constraints requires continuity in the cycle-averaged value of $p_\ax$. Methods that instantaneously join $p_\ax$ to $w_\ax^\EFA \rho_\ax^\EFA$ break equilibrium even if $\rho_\ax$ is taken to be continuous unless $w_\ax^\EFA=w_\ax^\ETA$.  Since the instantaneous $p_\ax/\rho_\ax$ remains $O(1)$ as $m/H\rightarrow \infty$ this problem remains even for late switches (see Fig.~\ref{fig:background_ETA}). In our technique, the time-averaged velocity and momentum both remain continuous across the switch by choosing the equation of state $w_\ax^\EFA=w_\ax^\ETA$  by matching $A_w$.  This differs from Ref.~\cite{Passaglia:2022bcr}, where $A_w=3/2$ was taken to be the predicted value of Eq.~(\ref{eq:weta_analytic}) in radiation domination. Note that we match the time-averaged or \ETA\ quantities because the we seek to maintain pressure equilibrium rather than instantaneous pressure continuity.

We illustrate this technique in Fig.~\ref{fig:wETA_scal}, where $w_\ax$ is scaled to reflect the value of $A_w$.  At $m/H_*=10$, we equate $w_\ax^\EFA=w_\ax^\ETA$ and compare it to the discrete evaluation of the \ETA\ at later times. Notice that the effective value of $A_w$ decreases with time, as is expected from Eq.~(\ref{eq:weta_analytic}) where $A_w=9/8$ in matter domination.  While some of this fractional evolution could be captured by Eq.~(\ref{eq:weta_analytic}), the correction is small since $w_\ax^\EFA$ itself is small and decreasing as $a^{-3}$ in the matter dominated limit.   Our technique thus has the advantage of maintaining pressure continuity at the expense of a negligible error in the \EFA\ equation of state. 

This advantage is illustrated in 
Fig.~\ref{fig:quasistaticdemo} where we compare the \ULA\ density perturbation $\delta_\ax$ output from \textsc{axionCAMB} to \ourcode\ 
for the same switch time set by the condition $m/H_* = 3$, using the same high-accuracy settings \texttt{accuracy\_boost = l\_accuracy\_boost = 3}, for $m = 10^{-31.5}$ eV, $f_\DM = 0.2$, $k = 0.02$~Mpc$^{-1}$. The KG solution is  plotted for comparison. The sudden large oscillations result from a $p_\ax$ discontinuity in \textsc{axionCAMB} that violates the cancellation of 
$v_\ax^\EFA$ and $\dot\hL$.  The continuity equation then generates a large spurious Jeans oscillation.
This causes a large error in the \ULA\ variables even at the present time.  For the matter power spectrum, its impact is reduced by contributions from the cold dark matter and baryons (see Fig.\ \ref{fig:Tk_axionCAMB}).
Our technique maintains a well-controlled $\delta_\ax^\EFA$ even at the same early switch $m/H_* = 3$ and provides accurate observables for any $f_\DM$.

\begin{figure}
    \includegraphics[width=1\linewidth]{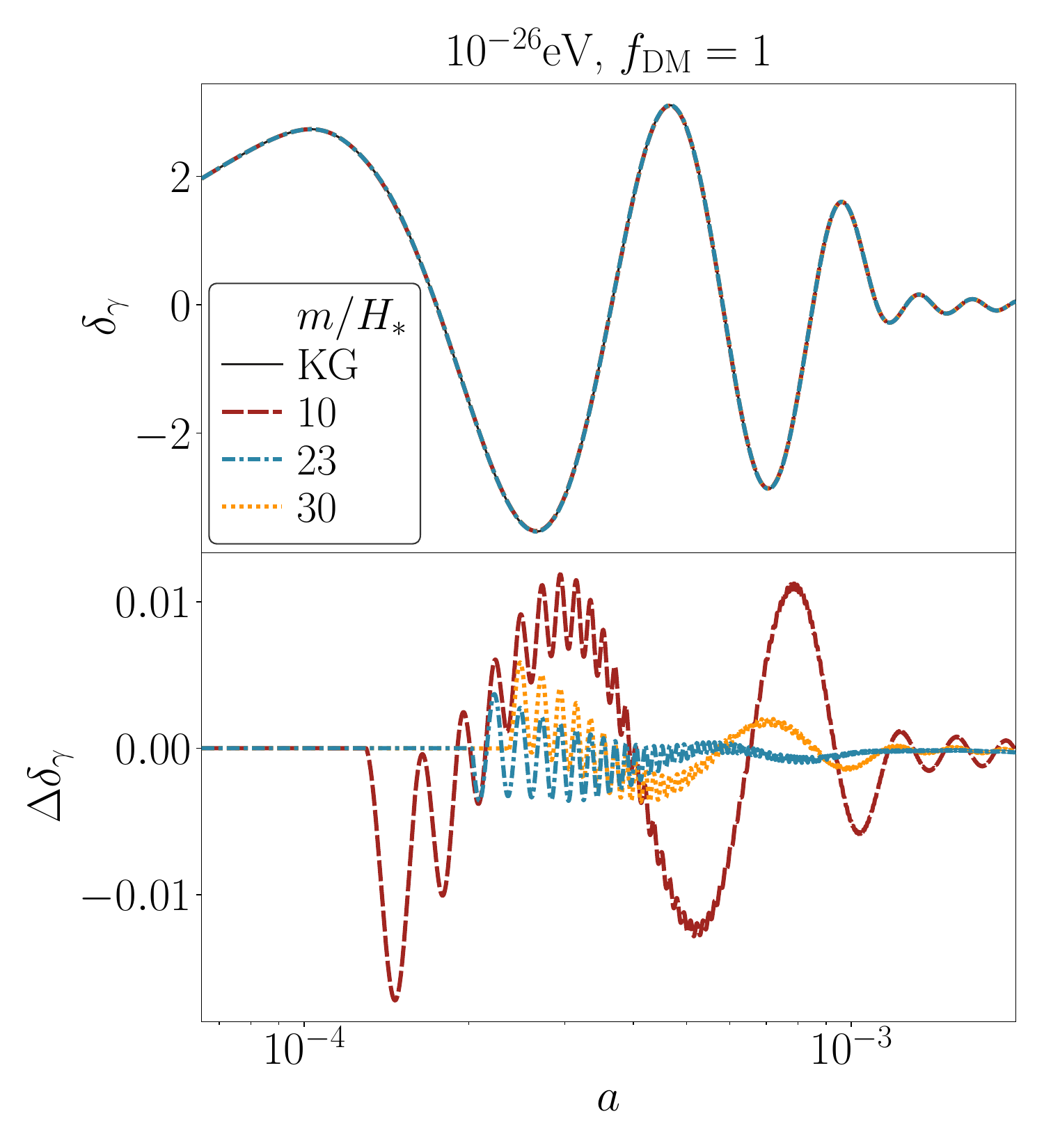}
    \caption{Photon density perturbation (top) and the difference between the \ourcode\ solution switched at various $m/H_*$ values and the KG system solution (bottom) for $m=10^{-26}$\,eV, $f_\DM = 1$ and $k = 0.1$\,Mpc$^{-1}$. 
     At the phase-tuned switch of $m/H_*=23$, the errors at recombination are reduced beyond even that of $m/H_*=30$ and greatly reduced from the baseline of $10$. 
    }
    \label{fig:photonETA_demonstration}
\end{figure}

\begin{figure}
    \includegraphics[width=1\linewidth]{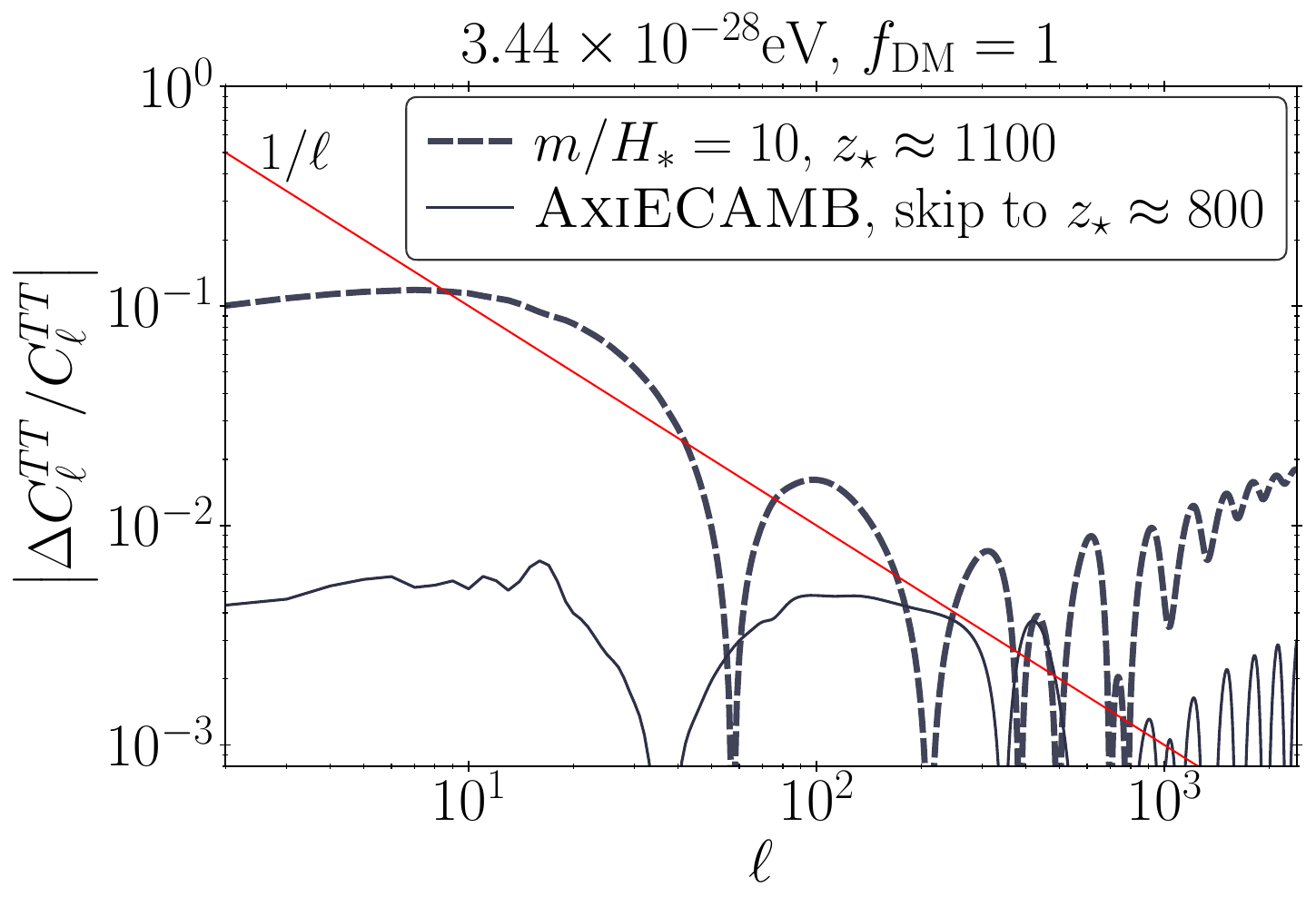}
    \caption{$C_\ell^{TT}$ errors induced by placing the switch around recombination.
    The baseline $m/H_*=10$ switch for $m = 3.44\times 10^{-28}$\,eV, $f_{\DM} = 1$ is at $z_* \approx 1100$ whereas \ourcode\ skips to $z_*\approx 800$.  Errors compared with the high accuracy calculation are much larger than cosmic variance in the former. }
    \label{fig:Clerror_recskipcompare}
\end{figure}

\section{CMB implementation}\label{app:cmbsources_corr}

We now elaborate on the two CMB-related issues discussed in Sec.~\ref{sec:CMB_changes}: the choice of switch time $m/H_*$ for masses $10^{-28} \lesssim m/{\rm eV} \lesssim  10^{-25}$  and large $f_\DM$, and the impact of additional CMB source terms due to discontinuities between KG and \EFA\ at the switch.

The \ETA\ of \ULA\ and metric variables only removes the leading-order effect of the KG oscillations.  
Since the CMB is particularly sensitive to the epoch around equality, the next-to-leading-order effect produces $(8\pi G\rho_\ax/H^2){\cal O}(H_*^2/m^2)$ oscillatory responses 
to the KG oscillations in the photon 
density and velocity perturbations that are not time averaged and remain as artefacts after the switch that change the amplitude and phase of CMB acoustic oscillations.

We show an example in Fig.~\ref{fig:photonETA_demonstration}.  In the top panel, we compare the KG solution (black thin solid line) to ETA/EFA results for a switch at $m/H_*=10$ (red dashed line) for the photon density perturbation $\delta_\gamma$ for
$m = 10^{-26}$\,eV, $f_\DM=1$ and $k=0.1$ Mpc$^{-1}$.
The two agree  to better than percent level accuracy  and both exhibit the dominant photon acoustic oscillations.
We show the error in the photon fluctuations, $\Delta\delta_\gamma$ (bottom panel), where the finer-scale oscillations in $\Delta\delta_\gamma$ after the switch reflect the continued photon response to KG oscillations that are removed in the \EFA, rather than an \EFA\ related error. More specifically, the discrepancy generated by instantaneously matching $\delta_\gamma$ at the switch instead of its time average gives rise to an error in the KG-induced $\delta_\gamma$ response, which then transfers to  a phase and amplitude error in the subsequent baryon-photon acoustic oscillations.
Once generated at the switch, this error does not decay away and its value at recombination is imprinted in $C_\ell^{TT}$ for the \EFA.

\begin{figure}
    \includegraphics[width=1\linewidth]{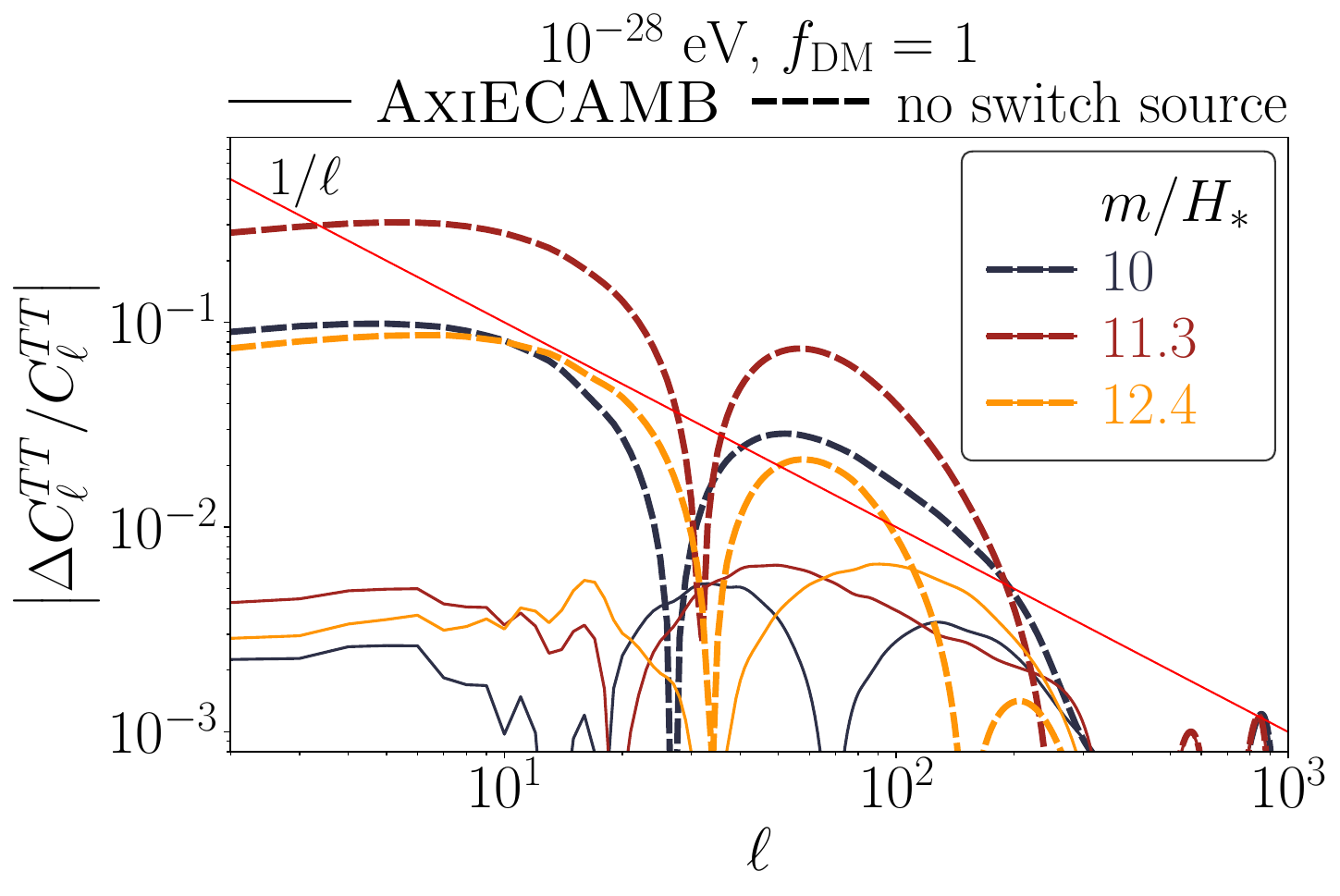}
    \caption{$C_\ell^{TT}$ errors from omitting (dashed lines) or including (thin solid lines) the switch sources (see Sec.\ \ref{sec:CMB_changes}) for $m=10^{-28}$\,eV and $f_\DM=1$ compared with the high accuracy solution.  Omitting them generates larger than cosmic variance errors at low $\ell$ which are strongly dependent on the phase of the KG oscillation at $m/H_*$ with the maximum and minimum around $m/H_*=10$ shown here.}
    \label{fig:BCsourcecomp_clerror}
\end{figure}

This error is reducible with the technique of Appendix \ref{app:KG} since it originates from catching the KG oscillation at a particular phase. In particular, we can choose the best phase as well as raise $m/H_*$ itself in this high sensitivity region. To demonstrate this in Fig.~\ref{fig:photonETA_demonstration}, we also show the $m/H_*=23$ scenario (blue dashdotted line), where the switch has been tuned to approximately
minimize the errors in the acoustic oscillations.  Notice that the errors are substantially less than the naive $(8\pi G\rho_\ax/H^2)(H_*/m)^2$ scaling would imply, which we demonstrate with an example with an even higher value of  $m/H_* = 30$ (orange dotted line) that is not phase matched yet has a higher error.

The technique of Appendix \ref{app:KG} allows us to take this phase tuning  in the $10^{-26}$\,eV, $f_\DM = 1$ case and apply it to other masses, $k$ values, and switch epochs. Using Eq.~(\ref{eq:betaphase}) we find that this best phase around $m/H_*=20$ corresponds to
\begin{equation}
2\beta \approx 7.08\pi .
\end{equation}
  Finally using Eq.~(\ref{eq:switchbeta}), we take this value of $2\beta$ to predict the best $m/H_*^\ETA$ and iteratively solve for the corresponding $m/H_*$.  While the tuning degrades away from $m=10^{-26}$\,eV, it suffices to improve the $10^{-27.5}\lesssim m/{\rm eV}\lesssim 10^{-25}$ regime.   Note that by tuning to $2\beta$ rather than $m/H_*$ in the fiducial model we have automatically allowed for scalings of cosmological parameters such as $\Omega_c h^2$, $\Omega_b h^2$ away from their fiducial values.

Next we illustrate the reason for never allowing the switch to be in the recombination range $z_* \in (800,1300]$.  In Fig.~\ref{fig:Clerror_recskipcompare}, we show the specific mass where the baseline $m/H_*=10$ would place $z_*=1100$ and compare it to the \ourcode\ default of moving it to $z_*\approx 800$. The errors in the former case are larger than cosmic variance even at the intermediate $\ell \sim 10^2$ range and are not of the type that can be corrected by a $\thetas$ shift.   The high sensitivity to recombination of the CMB source function, which carries derivatives of the opacity $\dot\mu$ (see Eq.~\ref{eq:CMBsourcefunction}), makes skipping this redshift window important.
We also increase the CMB source sampling density around recombination to account for the KG oscillation timescale when the switch is after recombination.

Lastly, we consider the impact of the switch sources discussed in Sec.~\ref{sec:CMB_changes}. These  sources account for the boundary terms from integration by parts of the original multipole sources in the construction of the total source.   In Fig.~\ref{fig:BCsourcecomp_clerror} we show the effect of removing these switch sources for the case $m = 10^{-28}$\,eV, $f_\DM = 1$ and switches around our baseline $m/H=10$.  The effect of the switch source can be above the cosmic variance line and depends strongly on the phase of the \ULA\ oscillation at the switch.  In Fig.~\ref{fig:BCsourcecomp_clerror}, we highlight the best and worst cases around $m/H_*=10$.  Even though this $f_\DM=1$ model is far from observationally viable, the errors due omitting the switch sources peak at $\Delta C_\ell/C_\ell \sim 0.3$, which is 
very large by modern CMB code standards.  
In Fig.~\ref{fig:sourcefcn_timedomain}, we show that the switch source is associated with a phase-dependent incomplete cancellation of the KG oscillations before the switch.  With the switch sources included, \ourcode\ is accurate for all phases. The cancellation of the KG oscillations before the switch also requires very dense and symmetric source sampling right around the switch to guarantee accurate line-of-sight integration.

The switch itself is implemented using  the standard \textsc{CAMB} switch mechanism which prevents any equation of motion from being integrated across the discontinuous boundary.
This switch procedure guarantees that the equations on either side of the switch are solved separately and that evolution variables are assigned correctly at the switch epoch.

\end{appendix}
\clearpage
\bibliography{main}

\begin{thebibliography}{70}%
\makeatletter
\providecommand \@ifxundefined [1]{%
 \@ifx{#1\undefined}
}%
\providecommand \@ifnum [1]{%
 \ifnum #1\expandafter \@firstoftwo
 \else \expandafter \@secondoftwo
 \fi
}%
\providecommand \@ifx [1]{%
 \ifx #1\expandafter \@firstoftwo
 \else \expandafter \@secondoftwo
 \fi
}%
\providecommand \natexlab [1]{#1}%
\providecommand \enquote  [1]{``#1''}%
\providecommand \bibnamefont  [1]{#1}%
\providecommand \bibfnamefont [1]{#1}%
\providecommand \citenamefont [1]{#1}%
\providecommand \href@noop [0]{\@secondoftwo}%
\providecommand \href [0]{\begingroup \@sanitize@url \@href}%
\providecommand \@href[1]{\@@startlink{#1}\@@href}%
\providecommand \@@href[1]{\endgroup#1\@@endlink}%
\providecommand \@sanitize@url [0]{\catcode `\\12\catcode `\$12\catcode `\&12\catcode `\#12\catcode `\^12\catcode `\_12\catcode `\%12\relax}%
\providecommand \@@startlink[1]{}%
\providecommand \@@endlink[0]{}%
\providecommand \url  [0]{\begingroup\@sanitize@url \@url }%
\providecommand \@url [1]{\endgroup\@href {#1}{\urlprefix }}%
\providecommand \urlprefix  [0]{URL }%
\providecommand \Eprint [0]{\href }%
\providecommand \doibase [0]{http://dx.doi.org/}%
\providecommand \selectlanguage [0]{\@gobble}%
\providecommand \bibinfo  [0]{\@secondoftwo}%
\providecommand \bibfield  [0]{\@secondoftwo}%
\providecommand \translation [1]{[#1]}%
\providecommand \BibitemOpen [0]{}%
\providecommand \bibitemStop [0]{}%
\providecommand \bibitemNoStop [0]{.\EOS\space}%
\providecommand \EOS [0]{\spacefactor3000\relax}%
\providecommand \BibitemShut  [1]{\csname bibitem#1\endcsname}%
\let\auto@bib@innerbib\@empty
\bibitem [{\citenamefont {Aghanim}\ \emph {et~al.}(2020{\natexlab{a}})\citenamefont {Aghanim} \emph {et~al.}}]{Planck:2018vyg}%
  \BibitemOpen
  \bibfield  {author} {\bibinfo {author} {\bibfnamefont {N.}~\bibnamefont {Aghanim}} \emph {et~al.} (\bibinfo {collaboration} {Planck}),\ }\href {\doibase 10.1051/0004-6361/201833910} {\bibfield  {journal} {\bibinfo  {journal} {Astron. Astrophys.}\ }\textbf {\bibinfo {volume} {641}},\ \bibinfo {pages} {A6} (\bibinfo {year} {2020}{\natexlab{a}})},\ \bibinfo {note} {[Erratum: Astron.Astrophys. 652, C4 (2021)]},\ \Eprint {http://arxiv.org/abs/1807.06209} {arXiv:1807.06209 [astro-ph.CO]} \BibitemShut {NoStop}%
\bibitem [{\citenamefont {Aiola}\ \emph {et~al.}(2020)\citenamefont {Aiola} \emph {et~al.}}]{ACT:2020gnv}%
  \BibitemOpen
  \bibfield  {author} {\bibinfo {author} {\bibfnamefont {S.}~\bibnamefont {Aiola}} \emph {et~al.} (\bibinfo {collaboration} {ACT}),\ }\href {\doibase 10.1088/1475-7516/2020/12/047} {\bibfield  {journal} {\bibinfo  {journal} {JCAP}\ }\textbf {\bibinfo {volume} {12}},\ \bibinfo {pages} {047} (\bibinfo {year} {2020})},\ \Eprint {http://arxiv.org/abs/2007.07288} {arXiv:2007.07288 [astro-ph.CO]} \BibitemShut {NoStop}%
\bibitem [{\citenamefont {Ge}\ \emph {et~al.}(2024)\citenamefont {Ge} \emph {et~al.}}]{SPT-3G:2024atg}%
  \BibitemOpen
  \bibfield  {author} {\bibinfo {author} {\bibfnamefont {F.}~\bibnamefont {Ge}} \emph {et~al.} (\bibinfo {collaboration} {SPT-3G}),\ }\href@noop {} {\  (\bibinfo {year} {2024})},\ \Eprint {http://arxiv.org/abs/2411.06000} {arXiv:2411.06000 [astro-ph.CO]} \BibitemShut {NoStop}%
\bibitem [{\citenamefont {Rubin}\ \emph {et~al.}(2023)\citenamefont {Rubin} \emph {et~al.}}]{Rubin:2023ovl}%
  \BibitemOpen
  \bibfield  {author} {\bibinfo {author} {\bibfnamefont {D.}~\bibnamefont {Rubin}} \emph {et~al.},\ }\href@noop {} {\  (\bibinfo {year} {2023})},\ \Eprint {http://arxiv.org/abs/2311.12098} {arXiv:2311.12098 [astro-ph.CO]} \BibitemShut {NoStop}%
\bibitem [{\citenamefont {Abbott}\ \emph {et~al.}(2024)\citenamefont {Abbott} \emph {et~al.}}]{DES:2024jxu}%
  \BibitemOpen
  \bibfield  {author} {\bibinfo {author} {\bibfnamefont {T.~M.~C.}\ \bibnamefont {Abbott}} \emph {et~al.} (\bibinfo {collaboration} {DES}),\ }\href {\doibase 10.3847/2041-8213/ad6f9f} {\bibfield  {journal} {\bibinfo  {journal} {Astrophys. J. Lett.}\ }\textbf {\bibinfo {volume} {973}},\ \bibinfo {pages} {L14} (\bibinfo {year} {2024})},\ \Eprint {http://arxiv.org/abs/2401.02929} {arXiv:2401.02929 [astro-ph.CO]} \BibitemShut {NoStop}%
\bibitem [{\citenamefont {Adame}\ \emph {et~al.}(2024)\citenamefont {Adame} \emph {et~al.}}]{DESI:2024mwx}%
  \BibitemOpen
  \bibfield  {author} {\bibinfo {author} {\bibfnamefont {A.~G.}\ \bibnamefont {Adame}} \emph {et~al.} (\bibinfo {collaboration} {DESI}),\ }\href@noop {} {\  (\bibinfo {year} {2024})},\ \Eprint {http://arxiv.org/abs/2404.03002} {arXiv:2404.03002 [astro-ph.CO]} \BibitemShut {NoStop}%
\bibitem [{\citenamefont {Abbott}\ \emph {et~al.}(2022)\citenamefont {Abbott} \emph {et~al.}}]{DES:2021wwk}%
  \BibitemOpen
  \bibfield  {author} {\bibinfo {author} {\bibfnamefont {T.~M.~C.}\ \bibnamefont {Abbott}} \emph {et~al.} (\bibinfo {collaboration} {DES}),\ }\href {\doibase 10.1103/PhysRevD.105.023520} {\bibfield  {journal} {\bibinfo  {journal} {Phys. Rev. D}\ }\textbf {\bibinfo {volume} {105}},\ \bibinfo {pages} {023520} (\bibinfo {year} {2022})},\ \Eprint {http://arxiv.org/abs/2105.13549} {arXiv:2105.13549 [astro-ph.CO]} \BibitemShut {NoStop}%
\bibitem [{\citenamefont {Heymans}\ \emph {et~al.}(2021)\citenamefont {Heymans} \emph {et~al.}}]{Heymans:2020gsg}%
  \BibitemOpen
  \bibfield  {author} {\bibinfo {author} {\bibfnamefont {C.}~\bibnamefont {Heymans}} \emph {et~al.},\ }\href {\doibase 10.1051/0004-6361/202039063} {\bibfield  {journal} {\bibinfo  {journal} {Astron. Astrophys.}\ }\textbf {\bibinfo {volume} {646}},\ \bibinfo {pages} {A140} (\bibinfo {year} {2021})},\ \Eprint {http://arxiv.org/abs/2007.15632} {arXiv:2007.15632 [astro-ph.CO]} \BibitemShut {NoStop}%
\bibitem [{\citenamefont {Miyatake}\ \emph {et~al.}(2023)\citenamefont {Miyatake} \emph {et~al.}}]{Miyatake:2023njf}%
  \BibitemOpen
  \bibfield  {author} {\bibinfo {author} {\bibfnamefont {H.}~\bibnamefont {Miyatake}} \emph {et~al.},\ }\href {\doibase 10.1103/PhysRevD.108.123517} {\bibfield  {journal} {\bibinfo  {journal} {Phys. Rev. D}\ }\textbf {\bibinfo {volume} {108}},\ \bibinfo {pages} {123517} (\bibinfo {year} {2023})},\ \Eprint {http://arxiv.org/abs/2304.00704} {arXiv:2304.00704 [astro-ph.CO]} \BibitemShut {NoStop}%
\bibitem [{\citenamefont {Breuval}\ \emph {et~al.}(2024)\citenamefont {Breuval}, \citenamefont {Riess}, \citenamefont {Casertano}, \citenamefont {Yuan}, \citenamefont {Macri}, \citenamefont {Romaniello}, \citenamefont {Murakami}, \citenamefont {Scolnic}, \citenamefont {Anand},\ and\ \citenamefont {Soszy\'nski}}]{Breuval:2024lsv}%
  \BibitemOpen
  \bibfield  {author} {\bibinfo {author} {\bibfnamefont {L.}~\bibnamefont {Breuval}}, \bibinfo {author} {\bibfnamefont {A.~G.}\ \bibnamefont {Riess}}, \bibinfo {author} {\bibfnamefont {S.}~\bibnamefont {Casertano}}, \bibinfo {author} {\bibfnamefont {W.}~\bibnamefont {Yuan}}, \bibinfo {author} {\bibfnamefont {L.~M.}\ \bibnamefont {Macri}}, \bibinfo {author} {\bibfnamefont {M.}~\bibnamefont {Romaniello}}, \bibinfo {author} {\bibfnamefont {Y.~S.}\ \bibnamefont {Murakami}}, \bibinfo {author} {\bibfnamefont {D.}~\bibnamefont {Scolnic}}, \bibinfo {author} {\bibfnamefont {G.~S.}\ \bibnamefont {Anand}}, \ and\ \bibinfo {author} {\bibfnamefont {I.}~\bibnamefont {Soszy\'nski}},\ }\href {\doibase 10.3847/1538-4357/ad630e} {\bibfield  {journal} {\bibinfo  {journal} {Astrophys. J.}\ }\textbf {\bibinfo {volume} {973}},\ \bibinfo {pages} {30} (\bibinfo {year} {2024})},\ \Eprint {http://arxiv.org/abs/2404.08038} {arXiv:2404.08038 [astro-ph.CO]} \BibitemShut {NoStop}%
\bibitem [{\citenamefont {Freedman}\ \emph {et~al.}(2024)\citenamefont {Freedman}, \citenamefont {Madore}, \citenamefont {Jang}, \citenamefont {Hoyt}, \citenamefont {Lee},\ and\ \citenamefont {Owens}}]{Freedman:2024eph}%
  \BibitemOpen
  \bibfield  {author} {\bibinfo {author} {\bibfnamefont {W.~L.}\ \bibnamefont {Freedman}}, \bibinfo {author} {\bibfnamefont {B.~F.}\ \bibnamefont {Madore}}, \bibinfo {author} {\bibfnamefont {I.~S.}\ \bibnamefont {Jang}}, \bibinfo {author} {\bibfnamefont {T.~J.}\ \bibnamefont {Hoyt}}, \bibinfo {author} {\bibfnamefont {A.~J.}\ \bibnamefont {Lee}}, \ and\ \bibinfo {author} {\bibfnamefont {K.~A.}\ \bibnamefont {Owens}},\ }\href@noop {} {\  (\bibinfo {year} {2024})},\ \Eprint {http://arxiv.org/abs/2408.06153} {arXiv:2408.06153 [astro-ph.CO]} \BibitemShut {NoStop}%
\bibitem [{\citenamefont {Sch\"oneberg}\ \emph {et~al.}(2022)\citenamefont {Sch\"oneberg}, \citenamefont {Franco~Abell\'an}, \citenamefont {P\'erez~S\'anchez}, \citenamefont {Witte}, \citenamefont {Poulin},\ and\ \citenamefont {Lesgourgues}}]{Schoneberg:2021qvd}%
  \BibitemOpen
  \bibfield  {author} {\bibinfo {author} {\bibfnamefont {N.}~\bibnamefont {Sch\"oneberg}}, \bibinfo {author} {\bibfnamefont {G.}~\bibnamefont {Franco~Abell\'an}}, \bibinfo {author} {\bibfnamefont {A.}~\bibnamefont {P\'erez~S\'anchez}}, \bibinfo {author} {\bibfnamefont {S.~J.}\ \bibnamefont {Witte}}, \bibinfo {author} {\bibfnamefont {V.}~\bibnamefont {Poulin}}, \ and\ \bibinfo {author} {\bibfnamefont {J.}~\bibnamefont {Lesgourgues}},\ }\href {\doibase 10.1016/j.physrep.2022.07.001} {\bibfield  {journal} {\bibinfo  {journal} {Phys. Rept.}\ }\textbf {\bibinfo {volume} {984}},\ \bibinfo {pages} {1} (\bibinfo {year} {2022})},\ \Eprint {http://arxiv.org/abs/2107.10291} {arXiv:2107.10291 [astro-ph.CO]} \BibitemShut {NoStop}%
\bibitem [{\citenamefont {Di~Valentino}\ \emph {et~al.}(2021)\citenamefont {Di~Valentino}, \citenamefont {Mena}, \citenamefont {Pan}, \citenamefont {Visinelli}, \citenamefont {Yang}, \citenamefont {Melchiorri}, \citenamefont {Mota}, \citenamefont {Riess},\ and\ \citenamefont {Silk}}]{DiValentino:2021izs}%
  \BibitemOpen
  \bibfield  {author} {\bibinfo {author} {\bibfnamefont {E.}~\bibnamefont {Di~Valentino}}, \bibinfo {author} {\bibfnamefont {O.}~\bibnamefont {Mena}}, \bibinfo {author} {\bibfnamefont {S.}~\bibnamefont {Pan}}, \bibinfo {author} {\bibfnamefont {L.}~\bibnamefont {Visinelli}}, \bibinfo {author} {\bibfnamefont {W.}~\bibnamefont {Yang}}, \bibinfo {author} {\bibfnamefont {A.}~\bibnamefont {Melchiorri}}, \bibinfo {author} {\bibfnamefont {D.~F.}\ \bibnamefont {Mota}}, \bibinfo {author} {\bibfnamefont {A.~G.}\ \bibnamefont {Riess}}, \ and\ \bibinfo {author} {\bibfnamefont {J.}~\bibnamefont {Silk}},\ }\href {\doibase 10.1088/1361-6382/ac086d} {\bibfield  {journal} {\bibinfo  {journal} {Class. Quant. Grav.}\ }\textbf {\bibinfo {volume} {38}},\ \bibinfo {pages} {153001} (\bibinfo {year} {2021})},\ \Eprint {http://arxiv.org/abs/2103.01183} {arXiv:2103.01183 [astro-ph.CO]} \BibitemShut {NoStop}%
\bibitem [{\citenamefont {Svrcek}\ and\ \citenamefont {Witten}(2006)}]{Svrcek:2006yi}%
  \BibitemOpen
  \bibfield  {author} {\bibinfo {author} {\bibfnamefont {P.}~\bibnamefont {Svrcek}}\ and\ \bibinfo {author} {\bibfnamefont {E.}~\bibnamefont {Witten}},\ }\href {\doibase 10.1088/1126-6708/2006/06/051} {\bibfield  {journal} {\bibinfo  {journal} {JHEP}\ }\textbf {\bibinfo {volume} {06}},\ \bibinfo {pages} {051} (\bibinfo {year} {2006})},\ \Eprint {http://arxiv.org/abs/hep-th/0605206} {arXiv:hep-th/0605206 [hep-th]} \BibitemShut {NoStop}%
\bibitem [{\citenamefont {Conlon}(2006)}]{Conlon:2006tq}%
  \BibitemOpen
  \bibfield  {author} {\bibinfo {author} {\bibfnamefont {J.~P.}\ \bibnamefont {Conlon}},\ }\href {\doibase 10.1088/1126-6708/2006/05/078} {\bibfield  {journal} {\bibinfo  {journal} {JHEP}\ }\textbf {\bibinfo {volume} {05}},\ \bibinfo {pages} {078} (\bibinfo {year} {2006})},\ \Eprint {http://arxiv.org/abs/hep-th/0602233} {arXiv:hep-th/0602233} \BibitemShut {NoStop}%
\bibitem [{\citenamefont {Arvanitaki}\ \emph {et~al.}(2010)\citenamefont {Arvanitaki}, \citenamefont {Dimopoulos}, \citenamefont {Dubovsky}, \citenamefont {Kaloper},\ and\ \citenamefont {March-Russell}}]{Arvanitaki:2009fg}%
  \BibitemOpen
  \bibfield  {author} {\bibinfo {author} {\bibfnamefont {A.}~\bibnamefont {Arvanitaki}}, \bibinfo {author} {\bibfnamefont {S.}~\bibnamefont {Dimopoulos}}, \bibinfo {author} {\bibfnamefont {S.}~\bibnamefont {Dubovsky}}, \bibinfo {author} {\bibfnamefont {N.}~\bibnamefont {Kaloper}}, \ and\ \bibinfo {author} {\bibfnamefont {J.}~\bibnamefont {March-Russell}},\ }\href {\doibase 10.1103/PhysRevD.81.123530} {\bibfield  {journal} {\bibinfo  {journal} {Phys. Rev.}\ }\textbf {\bibinfo {volume} {D81}},\ \bibinfo {pages} {123530} (\bibinfo {year} {2010})},\ \Eprint {http://arxiv.org/abs/0905.4720} {arXiv:0905.4720 [hep-th]} \BibitemShut {NoStop}%
\bibitem [{\citenamefont {Cicoli}\ \emph {et~al.}(2012)\citenamefont {Cicoli}, \citenamefont {Goodsell},\ and\ \citenamefont {Ringwald}}]{Cicoli:2012sz}%
  \BibitemOpen
  \bibfield  {author} {\bibinfo {author} {\bibfnamefont {M.}~\bibnamefont {Cicoli}}, \bibinfo {author} {\bibfnamefont {M.}~\bibnamefont {Goodsell}}, \ and\ \bibinfo {author} {\bibfnamefont {A.}~\bibnamefont {Ringwald}},\ }\href {\doibase 10.1007/JHEP10(2012)146} {\bibfield  {journal} {\bibinfo  {journal} {JHEP}\ }\textbf {\bibinfo {volume} {10}},\ \bibinfo {pages} {146} (\bibinfo {year} {2012})},\ \Eprint {http://arxiv.org/abs/1206.0819} {arXiv:1206.0819 [hep-th]} \BibitemShut {NoStop}%
\bibitem [{\citenamefont {Stott}\ \emph {et~al.}(2017)\citenamefont {Stott}, \citenamefont {Marsh}, \citenamefont {Pongkitivanichkul}, \citenamefont {Price},\ and\ \citenamefont {Acharya}}]{Stott:2017hvl}%
  \BibitemOpen
  \bibfield  {author} {\bibinfo {author} {\bibfnamefont {M.~J.}\ \bibnamefont {Stott}}, \bibinfo {author} {\bibfnamefont {D.~J.~E.}\ \bibnamefont {Marsh}}, \bibinfo {author} {\bibfnamefont {C.}~\bibnamefont {Pongkitivanichkul}}, \bibinfo {author} {\bibfnamefont {L.~C.}\ \bibnamefont {Price}}, \ and\ \bibinfo {author} {\bibfnamefont {B.~S.}\ \bibnamefont {Acharya}},\ }\href {\doibase 10.1103/PhysRevD.96.083510} {\bibfield  {journal} {\bibinfo  {journal} {Phys. Rev.}\ }\textbf {\bibinfo {volume} {D96}},\ \bibinfo {pages} {083510} (\bibinfo {year} {2017})},\ \Eprint {http://arxiv.org/abs/1706.03236} {arXiv:1706.03236 [astro-ph.CO]} \BibitemShut {NoStop}%
\bibitem [{\citenamefont {Gendler}\ \emph {et~al.}(2023)\citenamefont {Gendler}, \citenamefont {Marsh}, \citenamefont {McAllister},\ and\ \citenamefont {Moritz}}]{Gendler:2023kjt}%
  \BibitemOpen
  \bibfield  {author} {\bibinfo {author} {\bibfnamefont {N.}~\bibnamefont {Gendler}}, \bibinfo {author} {\bibfnamefont {D.~J.~E.}\ \bibnamefont {Marsh}}, \bibinfo {author} {\bibfnamefont {L.}~\bibnamefont {McAllister}}, \ and\ \bibinfo {author} {\bibfnamefont {J.}~\bibnamefont {Moritz}},\ }\href@noop {} {\  (\bibinfo {year} {2023})},\ \Eprint {http://arxiv.org/abs/2309.13145} {arXiv:2309.13145 [hep-th]} \BibitemShut {NoStop}%
\bibitem [{\citenamefont {Graham}\ \emph {et~al.}(2016)\citenamefont {Graham}, \citenamefont {Mardon},\ and\ \citenamefont {Rajendran}}]{Graham:2015rva}%
  \BibitemOpen
  \bibfield  {author} {\bibinfo {author} {\bibfnamefont {P.~W.}\ \bibnamefont {Graham}}, \bibinfo {author} {\bibfnamefont {J.}~\bibnamefont {Mardon}}, \ and\ \bibinfo {author} {\bibfnamefont {S.}~\bibnamefont {Rajendran}},\ }\href {\doibase 10.1103/PhysRevD.93.103520} {\bibfield  {journal} {\bibinfo  {journal} {Phys. Rev. D}\ }\textbf {\bibinfo {volume} {93}},\ \bibinfo {pages} {103520} (\bibinfo {year} {2016})},\ \Eprint {http://arxiv.org/abs/1504.02102} {arXiv:1504.02102 [hep-ph]} \BibitemShut {NoStop}%
\bibitem [{\citenamefont {Amin}\ \emph {et~al.}(2022)\citenamefont {Amin}, \citenamefont {Jain}, \citenamefont {Karur},\ and\ \citenamefont {Mocz}}]{Amin:2022pzv}%
  \BibitemOpen
  \bibfield  {author} {\bibinfo {author} {\bibfnamefont {M.~A.}\ \bibnamefont {Amin}}, \bibinfo {author} {\bibfnamefont {M.}~\bibnamefont {Jain}}, \bibinfo {author} {\bibfnamefont {R.}~\bibnamefont {Karur}}, \ and\ \bibinfo {author} {\bibfnamefont {P.}~\bibnamefont {Mocz}},\ }\href {\doibase 10.1088/1475-7516/2022/08/014} {\bibfield  {journal} {\bibinfo  {journal} {JCAP}\ }\textbf {\bibinfo {volume} {08}},\ \bibinfo {pages} {014} (\bibinfo {year} {2022})},\ \Eprint {http://arxiv.org/abs/2203.11935} {arXiv:2203.11935 [astro-ph.CO]} \BibitemShut {NoStop}%
\bibitem [{\citenamefont {Yang}\ \emph {et~al.}(2025)\citenamefont {Yang}, \citenamefont {Li},\ and\ \citenamefont {Shapiro}}]{Yang:2025vcb}%
  \BibitemOpen
  \bibfield  {author} {\bibinfo {author} {\bibfnamefont {Q.}~\bibnamefont {Yang}}, \bibinfo {author} {\bibfnamefont {B.}~\bibnamefont {Li}}, \ and\ \bibinfo {author} {\bibfnamefont {P.~R.}\ \bibnamefont {Shapiro}},\ }\href@noop {} {\  (\bibinfo {year} {2025})},\ \Eprint {http://arxiv.org/abs/2503.16773} {arXiv:2503.16773 [astro-ph.CO]} \BibitemShut {NoStop}%
\bibitem [{\citenamefont {Su\'arez}\ and\ \citenamefont {Chavanis}(2015)}]{Suarez:2015fga}%
  \BibitemOpen
  \bibfield  {author} {\bibinfo {author} {\bibfnamefont {A.}~\bibnamefont {Su\'arez}}\ and\ \bibinfo {author} {\bibfnamefont {P.-H.}\ \bibnamefont {Chavanis}},\ }\href {\doibase 10.1103/PhysRevD.92.023510} {\bibfield  {journal} {\bibinfo  {journal} {Phys. Rev. D}\ }\textbf {\bibinfo {volume} {92}},\ \bibinfo {pages} {023510} (\bibinfo {year} {2015})},\ \Eprint {http://arxiv.org/abs/1503.07437} {arXiv:1503.07437 [gr-qc]} \BibitemShut {NoStop}%
\bibitem [{\citenamefont {Frieman}\ \emph {et~al.}(1995)\citenamefont {Frieman}, \citenamefont {Hill}, \citenamefont {Stebbins},\ and\ \citenamefont {Waga}}]{Frieman:1995pm}%
  \BibitemOpen
  \bibfield  {author} {\bibinfo {author} {\bibfnamefont {J.~A.}\ \bibnamefont {Frieman}}, \bibinfo {author} {\bibfnamefont {C.~T.}\ \bibnamefont {Hill}}, \bibinfo {author} {\bibfnamefont {A.}~\bibnamefont {Stebbins}}, \ and\ \bibinfo {author} {\bibfnamefont {I.}~\bibnamefont {Waga}},\ }\href {\doibase 10.1103/PhysRevLett.75.2077} {\bibfield  {journal} {\bibinfo  {journal} {Phys. Rev. Lett.}\ }\textbf {\bibinfo {volume} {75}},\ \bibinfo {pages} {2077} (\bibinfo {year} {1995})},\ \Eprint {http://arxiv.org/abs/astro-ph/9505060} {arXiv:astro-ph/9505060 [astro-ph]} \BibitemShut {NoStop}%
\bibitem [{\citenamefont {Choi}(2000)}]{Choi:1999xn}%
  \BibitemOpen
  \bibfield  {author} {\bibinfo {author} {\bibfnamefont {K.}~\bibnamefont {Choi}},\ }\href {\doibase 10.1103/PhysRevD.62.043509} {\bibfield  {journal} {\bibinfo  {journal} {Phys. Rev. D}\ }\textbf {\bibinfo {volume} {62}},\ \bibinfo {pages} {043509} (\bibinfo {year} {2000})},\ \Eprint {http://arxiv.org/abs/hep-ph/9902292} {arXiv:hep-ph/9902292} \BibitemShut {NoStop}%
\bibitem [{\citenamefont {Marsh}\ and\ \citenamefont {Ferreira}(2010)}]{Marsh:2010wq}%
  \BibitemOpen
  \bibfield  {author} {\bibinfo {author} {\bibfnamefont {D.~J.~E.}\ \bibnamefont {Marsh}}\ and\ \bibinfo {author} {\bibfnamefont {P.~G.}\ \bibnamefont {Ferreira}},\ }\href {\doibase 10.1103/PhysRevD.82.103528} {\bibfield  {journal} {\bibinfo  {journal} {Phys. Rev.}\ }\textbf {\bibinfo {volume} {D82}},\ \bibinfo {pages} {103528} (\bibinfo {year} {2010})},\ \Eprint {http://arxiv.org/abs/1009.3501} {arXiv:1009.3501 [hep-ph]} \BibitemShut {NoStop}%
\bibitem [{\citenamefont {Marsh}(2016)}]{Marsh:2015xka}%
  \BibitemOpen
  \bibfield  {author} {\bibinfo {author} {\bibfnamefont {D.~J.~E.}\ \bibnamefont {Marsh}},\ }\href {\doibase 10.1016/j.physrep.2016.06.005} {\bibfield  {journal} {\bibinfo  {journal} {Phys. Rept.}\ }\textbf {\bibinfo {volume} {643}},\ \bibinfo {pages} {1} (\bibinfo {year} {2016})},\ \Eprint {http://arxiv.org/abs/1510.07633} {arXiv:1510.07633 [astro-ph.CO]} \BibitemShut {NoStop}%
\bibitem [{\citenamefont {Klypin}\ \emph {et~al.}(1999)\citenamefont {Klypin}, \citenamefont {Kravtsov}, \citenamefont {Valenzuela},\ and\ \citenamefont {Prada}}]{Klypin:1999uc}%
  \BibitemOpen
  \bibfield  {author} {\bibinfo {author} {\bibfnamefont {A.~A.}\ \bibnamefont {Klypin}}, \bibinfo {author} {\bibfnamefont {A.~V.}\ \bibnamefont {Kravtsov}}, \bibinfo {author} {\bibfnamefont {O.}~\bibnamefont {Valenzuela}}, \ and\ \bibinfo {author} {\bibfnamefont {F.}~\bibnamefont {Prada}},\ }\href {\doibase 10.1086/307643} {\bibfield  {journal} {\bibinfo  {journal} {Astrophys. J.}\ }\textbf {\bibinfo {volume} {522}},\ \bibinfo {pages} {82} (\bibinfo {year} {1999})},\ \Eprint {http://arxiv.org/abs/astro-ph/9901240} {arXiv:astro-ph/9901240} \BibitemShut {NoStop}%
\bibitem [{\citenamefont {de~Blok}(2010)}]{deBlok:2009sp}%
  \BibitemOpen
  \bibfield  {author} {\bibinfo {author} {\bibfnamefont {W.~J.~G.}\ \bibnamefont {de~Blok}},\ }\href {\doibase 10.1155/2010/789293} {\bibfield  {journal} {\bibinfo  {journal} {Adv. Astron.}\ }\textbf {\bibinfo {volume} {2010}},\ \bibinfo {pages} {789293} (\bibinfo {year} {2010})},\ \Eprint {http://arxiv.org/abs/0910.3538} {arXiv:0910.3538 [astro-ph.CO]} \BibitemShut {NoStop}%
\bibitem [{\citenamefont {Boylan-Kolchin}\ \emph {et~al.}(2011)\citenamefont {Boylan-Kolchin}, \citenamefont {Bullock},\ and\ \citenamefont {Kaplinghat}}]{Boylan-Kolchin:2011qkt}%
  \BibitemOpen
  \bibfield  {author} {\bibinfo {author} {\bibfnamefont {M.}~\bibnamefont {Boylan-Kolchin}}, \bibinfo {author} {\bibfnamefont {J.~S.}\ \bibnamefont {Bullock}}, \ and\ \bibinfo {author} {\bibfnamefont {M.}~\bibnamefont {Kaplinghat}},\ }\href {\doibase 10.1111/j.1745-3933.2011.01074.x} {\bibfield  {journal} {\bibinfo  {journal} {Mon. Not. Roy. Astron. Soc.}\ }\textbf {\bibinfo {volume} {415}},\ \bibinfo {pages} {L40} (\bibinfo {year} {2011})},\ \Eprint {http://arxiv.org/abs/1103.0007} {arXiv:1103.0007 [astro-ph.CO]} \BibitemShut {NoStop}%
\bibitem [{\citenamefont {Bullock}\ and\ \citenamefont {Boylan-Kolchin}(2017)}]{Bullock:2017xww}%
  \BibitemOpen
  \bibfield  {author} {\bibinfo {author} {\bibfnamefont {J.~S.}\ \bibnamefont {Bullock}}\ and\ \bibinfo {author} {\bibfnamefont {M.}~\bibnamefont {Boylan-Kolchin}},\ }\href {\doibase 10.1146/annurev-astro-091916-055313} {\bibfield  {journal} {\bibinfo  {journal} {Ann. Rev. Astron. Astrophys.}\ }\textbf {\bibinfo {volume} {55}},\ \bibinfo {pages} {343} (\bibinfo {year} {2017})},\ \Eprint {http://arxiv.org/abs/1707.04256} {arXiv:1707.04256 [astro-ph.CO]} \BibitemShut {NoStop}%
\bibitem [{\citenamefont {Lovell}\ and\ \citenamefont {Zavala}(2023)}]{Lovell:2022vzx}%
  \BibitemOpen
  \bibfield  {author} {\bibinfo {author} {\bibfnamefont {M.~R.}\ \bibnamefont {Lovell}}\ and\ \bibinfo {author} {\bibfnamefont {J.}~\bibnamefont {Zavala}},\ }\href {\doibase 10.1093/mnras/stad216} {\bibfield  {journal} {\bibinfo  {journal} {Mon. Not. Roy. Astron. Soc.}\ }\textbf {\bibinfo {volume} {520}},\ \bibinfo {pages} {1567} (\bibinfo {year} {2023})},\ \Eprint {http://arxiv.org/abs/2209.06834} {arXiv:2209.06834 [astro-ph.CO]} \BibitemShut {NoStop}%
\bibitem [{\citenamefont {{Jahn}}\ \emph {et~al.}(2023)\citenamefont {{Jahn}} \emph {et~al.}}]{2023MNRAS.520..461J}%
  \BibitemOpen
  \bibfield  {author} {\bibinfo {author} {\bibfnamefont {E.~D.}\ \bibnamefont {{Jahn}}} \emph {et~al.},\ }\href {\doibase 10.1093/mnras/stad109} {\bibfield  {journal} {\bibinfo  {journal} {Mon. Not. Roy. Astron. Soc.}\ }\textbf {\bibinfo {volume} {520}},\ \bibinfo {pages} {461} (\bibinfo {year} {2023})},\ \Eprint {http://arxiv.org/abs/2110.00142} {arXiv:2110.00142 [astro-ph.GA]} \BibitemShut {NoStop}%
\bibitem [{\citenamefont {Homma}\ \emph {et~al.}(2024)\citenamefont {Homma} \emph {et~al.}}]{Homma:2023ppu}%
  \BibitemOpen
  \bibfield  {author} {\bibinfo {author} {\bibfnamefont {D.}~\bibnamefont {Homma}} \emph {et~al.},\ }\href {\doibase 10.1093/pasj/psae044} {\bibfield  {journal} {\bibinfo  {journal} {Publ. Astron. Soc. Jap.}\ }\textbf {\bibinfo {volume} {76}},\ \bibinfo {pages} {733} (\bibinfo {year} {2024})},\ \Eprint {http://arxiv.org/abs/2311.05439} {arXiv:2311.05439 [astro-ph.GA]} \BibitemShut {NoStop}%
\bibitem [{\citenamefont {{Mostow}}\ \emph {et~al.}(2024)\citenamefont {{Mostow}} \emph {et~al.}}]{2024arXiv241209566M}%
  \BibitemOpen
  \bibfield  {author} {\bibinfo {author} {\bibfnamefont {O.}~\bibnamefont {{Mostow}}} \emph {et~al.},\ }\href {\doibase 10.48550/arXiv.2412.09566} {\bibfield  {journal} {\bibinfo  {journal} {arXiv e-prints}\ ,\ \bibinfo {eid} {arXiv:2412.09566}} (\bibinfo {year} {2024})},\ \Eprint {http://arxiv.org/abs/2412.09566} {arXiv:2412.09566 [astro-ph.GA]} \BibitemShut {NoStop}%
\bibitem [{\citenamefont {Hu}\ \emph {et~al.}(2000)\citenamefont {Hu}, \citenamefont {Barkana},\ and\ \citenamefont {Gruzinov}}]{Hu:2000ke}%
  \BibitemOpen
  \bibfield  {author} {\bibinfo {author} {\bibfnamefont {W.}~\bibnamefont {Hu}}, \bibinfo {author} {\bibfnamefont {R.}~\bibnamefont {Barkana}}, \ and\ \bibinfo {author} {\bibfnamefont {A.}~\bibnamefont {Gruzinov}},\ }\href {\doibase 10.1103/PhysRevLett.85.1158} {\bibfield  {journal} {\bibinfo  {journal} {Phys. Rev. Lett.}\ }\textbf {\bibinfo {volume} {85}},\ \bibinfo {pages} {1158} (\bibinfo {year} {2000})},\ \Eprint {http://arxiv.org/abs/astro-ph/0003365} {arXiv:astro-ph/0003365 [astro-ph]} \BibitemShut {NoStop}%
\bibitem [{\citenamefont {Hui}\ \emph {et~al.}(2017)\citenamefont {Hui}, \citenamefont {Ostriker}, \citenamefont {Tremaine},\ and\ \citenamefont {Witten}}]{Hui:2016ltb}%
  \BibitemOpen
  \bibfield  {author} {\bibinfo {author} {\bibfnamefont {L.}~\bibnamefont {Hui}}, \bibinfo {author} {\bibfnamefont {J.~P.}\ \bibnamefont {Ostriker}}, \bibinfo {author} {\bibfnamefont {S.}~\bibnamefont {Tremaine}}, \ and\ \bibinfo {author} {\bibfnamefont {E.}~\bibnamefont {Witten}},\ }\href {\doibase 10.1103/PhysRevD.95.043541} {\bibfield  {journal} {\bibinfo  {journal} {Phys. Rev. D}\ }\textbf {\bibinfo {volume} {95}},\ \bibinfo {pages} {043541} (\bibinfo {year} {2017})},\ \Eprint {http://arxiv.org/abs/1610.08297} {arXiv:1610.08297 [astro-ph.CO]} \BibitemShut {NoStop}%
\bibitem [{\citenamefont {Fox}\ \emph {et~al.}(2004)\citenamefont {Fox}, \citenamefont {Pierce},\ and\ \citenamefont {Thomas}}]{Fox:2004kb}%
  \BibitemOpen
  \bibfield  {author} {\bibinfo {author} {\bibfnamefont {P.}~\bibnamefont {Fox}}, \bibinfo {author} {\bibfnamefont {A.}~\bibnamefont {Pierce}}, \ and\ \bibinfo {author} {\bibfnamefont {S.~D.}\ \bibnamefont {Thomas}},\ }\href@noop {} {\  (\bibinfo {year} {2004})},\ \Eprint {http://arxiv.org/abs/hep-th/0409059} {arXiv:hep-th/0409059} \BibitemShut {NoStop}%
\bibitem [{\citenamefont {Poulin}\ \emph {et~al.}(2018)\citenamefont {Poulin}, \citenamefont {Smith}, \citenamefont {Grin}, \citenamefont {Karwal},\ and\ \citenamefont {Kamionkowski}}]{Poulin:2018dzj}%
  \BibitemOpen
  \bibfield  {author} {\bibinfo {author} {\bibfnamefont {V.}~\bibnamefont {Poulin}}, \bibinfo {author} {\bibfnamefont {T.~L.}\ \bibnamefont {Smith}}, \bibinfo {author} {\bibfnamefont {D.}~\bibnamefont {Grin}}, \bibinfo {author} {\bibfnamefont {T.}~\bibnamefont {Karwal}}, \ and\ \bibinfo {author} {\bibfnamefont {M.}~\bibnamefont {Kamionkowski}},\ }\href@noop {} {\  (\bibinfo {year} {2018})},\ \Eprint {http://arxiv.org/abs/1806.10608} {arXiv:1806.10608 [astro-ph.CO]} \BibitemShut {NoStop}%
\bibitem [{\citenamefont {Lin}\ \emph {et~al.}(2019)\citenamefont {Lin}, \citenamefont {Benevento}, \citenamefont {Hu},\ and\ \citenamefont {Raveri}}]{Lin:2019qug}%
  \BibitemOpen
  \bibfield  {author} {\bibinfo {author} {\bibfnamefont {M.-X.}\ \bibnamefont {Lin}}, \bibinfo {author} {\bibfnamefont {G.}~\bibnamefont {Benevento}}, \bibinfo {author} {\bibfnamefont {W.}~\bibnamefont {Hu}}, \ and\ \bibinfo {author} {\bibfnamefont {M.}~\bibnamefont {Raveri}},\ }\href {\doibase 10.1103/PhysRevD.100.063542} {\bibfield  {journal} {\bibinfo  {journal} {Phys. Rev.}\ }\textbf {\bibinfo {volume} {D100}},\ \bibinfo {pages} {063542} (\bibinfo {year} {2019})},\ \Eprint {http://arxiv.org/abs/1905.12618} {arXiv:1905.12618 [astro-ph.CO]} \BibitemShut {NoStop}%
\bibitem [{\citenamefont {Rogers}\ \emph {et~al.}(2023)\citenamefont {Rogers}, \citenamefont {Hlo\v{z}ek}, \citenamefont {Lagu\"e}, \citenamefont {Ivanov}, \citenamefont {Philcox}, \citenamefont {Cabass}, \citenamefont {Akitsu},\ and\ \citenamefont {Marsh}}]{Rogers:2023ezo}%
  \BibitemOpen
  \bibfield  {author} {\bibinfo {author} {\bibfnamefont {K.~K.}\ \bibnamefont {Rogers}}, \bibinfo {author} {\bibfnamefont {R.}~\bibnamefont {Hlo\v{z}ek}}, \bibinfo {author} {\bibfnamefont {A.}~\bibnamefont {Lagu\"e}}, \bibinfo {author} {\bibfnamefont {M.~M.}\ \bibnamefont {Ivanov}}, \bibinfo {author} {\bibfnamefont {O.~H.~E.}\ \bibnamefont {Philcox}}, \bibinfo {author} {\bibfnamefont {G.}~\bibnamefont {Cabass}}, \bibinfo {author} {\bibfnamefont {K.}~\bibnamefont {Akitsu}}, \ and\ \bibinfo {author} {\bibfnamefont {D.~J.~E.}\ \bibnamefont {Marsh}},\ }\href {\doibase 10.1088/1475-7516/2023/06/023} {\bibfield  {journal} {\bibinfo  {journal} {JCAP}\ }\textbf {\bibinfo {volume} {06}},\ \bibinfo {pages} {023} (\bibinfo {year} {2023})},\ \Eprint {http://arxiv.org/abs/2301.08361} {arXiv:2301.08361 [astro-ph.CO]} \BibitemShut {NoStop}%
\bibitem [{\citenamefont {Abazajian}\ \emph {et~al.}(2019)\citenamefont {Abazajian} \emph {et~al.}}]{Abazajian:2019eic}%
  \BibitemOpen
  \bibfield  {author} {\bibinfo {author} {\bibfnamefont {K.}~\bibnamefont {Abazajian}} \emph {et~al.},\ }\href@noop {} {\  (\bibinfo {year} {2019})},\ \Eprint {http://arxiv.org/abs/1907.04473} {arXiv:1907.04473 [astro-ph.IM]} \BibitemShut {NoStop}%
\bibitem [{\citenamefont {Ivezi\'c}\ \emph {et~al.}(2019)\citenamefont {Ivezi\'c} \emph {et~al.}}]{LSST:2008ijt}%
  \BibitemOpen
  \bibfield  {author} {\bibinfo {author} {\bibfnamefont {{\v{Z}}.}~\bibnamefont {Ivezi\'c}} \emph {et~al.} (\bibinfo {collaboration} {LSST}),\ }\href {\doibase 10.3847/1538-4357/ab042c} {\bibfield  {journal} {\bibinfo  {journal} {Astrophys. J.}\ }\textbf {\bibinfo {volume} {873}},\ \bibinfo {pages} {111} (\bibinfo {year} {2019})},\ \Eprint {http://arxiv.org/abs/0805.2366} {arXiv:0805.2366 [astro-ph]} \BibitemShut {NoStop}%
\bibitem [{\citenamefont {{LSST Science Collaboration}}\ \emph {et~al.}(2009)\citenamefont {{LSST Science Collaboration}} \emph {et~al.}}]{2009arXiv0912.0201L}%
  \BibitemOpen
  \bibfield  {author} {\bibinfo {author} {\bibnamefont {{LSST Science Collaboration}}} \emph {et~al.},\ }\href {\doibase 10.48550/arXiv.0912.0201} {\bibfield  {journal} {\bibinfo  {journal} {arXiv e-prints}\ ,\ \bibinfo {eid} {arXiv:0912.0201}} (\bibinfo {year} {2009})},\ \Eprint {http://arxiv.org/abs/0912.0201} {arXiv:0912.0201 [astro-ph.IM]} \BibitemShut {NoStop}%
\bibitem [{\citenamefont {Mandelbaum}\ \emph {et~al.}(2018)\citenamefont {Mandelbaum} \emph {et~al.}}]{LSSTDarkEnergyScience:2018jkl}%
  \BibitemOpen
  \bibfield  {author} {\bibinfo {author} {\bibfnamefont {R.}~\bibnamefont {Mandelbaum}} \emph {et~al.} (\bibinfo {collaboration} {LSST Dark Energy Science}),\ }\href@noop {} {\  (\bibinfo {year} {2018})},\ \Eprint {http://arxiv.org/abs/1809.01669} {arXiv:1809.01669 [astro-ph.CO]} \BibitemShut {NoStop}%
\bibitem [{\citenamefont {Amendola}\ \emph {et~al.}(2018)\citenamefont {Amendola} \emph {et~al.}}]{Amendola:2016saw}%
  \BibitemOpen
  \bibfield  {author} {\bibinfo {author} {\bibfnamefont {L.}~\bibnamefont {Amendola}} \emph {et~al.},\ }\href {\doibase 10.1007/s41114-017-0010-3} {\bibfield  {journal} {\bibinfo  {journal} {Living Rev. Rel.}\ }\textbf {\bibinfo {volume} {21}},\ \bibinfo {pages} {2} (\bibinfo {year} {2018})},\ \Eprint {http://arxiv.org/abs/1606.00180} {arXiv:1606.00180 [astro-ph.CO]} \BibitemShut {NoStop}%
\bibitem [{\citenamefont {Aghamousa}\ \emph {et~al.}(2016)\citenamefont {Aghamousa} \emph {et~al.}}]{DESI:2016fyo}%
  \BibitemOpen
  \bibfield  {author} {\bibinfo {author} {\bibfnamefont {A.}~\bibnamefont {Aghamousa}} \emph {et~al.} (\bibinfo {collaboration} {DESI}),\ }\href@noop {} {\  (\bibinfo {year} {2016})},\ \Eprint {http://arxiv.org/abs/1611.00036} {arXiv:1611.00036 [astro-ph.IM]} \BibitemShut {NoStop}%
\bibitem [{\citenamefont {Hlozek}\ \emph {et~al.}(2015)\citenamefont {Hlozek}, \citenamefont {Grin}, \citenamefont {Marsh},\ and\ \citenamefont {Ferreira}}]{Hlozek:2014lca}%
  \BibitemOpen
  \bibfield  {author} {\bibinfo {author} {\bibfnamefont {R.}~\bibnamefont {Hlozek}}, \bibinfo {author} {\bibfnamefont {D.}~\bibnamefont {Grin}}, \bibinfo {author} {\bibfnamefont {D.~J.~E.}\ \bibnamefont {Marsh}}, \ and\ \bibinfo {author} {\bibfnamefont {P.~G.}\ \bibnamefont {Ferreira}},\ }\href {\doibase 10.1103/PhysRevD.91.103512} {\bibfield  {journal} {\bibinfo  {journal} {Phys. Rev. D}\ }\textbf {\bibinfo {volume} {91}},\ \bibinfo {pages} {103512} (\bibinfo {year} {2015})},\ \Eprint {http://arxiv.org/abs/1410.2896} {arXiv:1410.2896 [astro-ph.CO]} \BibitemShut {NoStop}%
\bibitem [{\citenamefont {Ratra}(1991)}]{Ratra:1990me}%
  \BibitemOpen
  \bibfield  {author} {\bibinfo {author} {\bibfnamefont {B.}~\bibnamefont {Ratra}},\ }\href {\doibase 10.1103/PhysRevD.44.352} {\bibfield  {journal} {\bibinfo  {journal} {Phys. Rev. D}\ }\textbf {\bibinfo {volume} {44}},\ \bibinfo {pages} {352} (\bibinfo {year} {1991})}\BibitemShut {NoStop}%
\bibitem [{\citenamefont {Cookmeyer}\ \emph {et~al.}(2020)\citenamefont {Cookmeyer}, \citenamefont {Grin},\ and\ \citenamefont {Smith}}]{Cookmeyer:2019rna}%
  \BibitemOpen
  \bibfield  {author} {\bibinfo {author} {\bibfnamefont {T.}~\bibnamefont {Cookmeyer}}, \bibinfo {author} {\bibfnamefont {D.}~\bibnamefont {Grin}}, \ and\ \bibinfo {author} {\bibfnamefont {T.~L.}\ \bibnamefont {Smith}},\ }\href {\doibase 10.1103/PhysRevD.101.023501} {\bibfield  {journal} {\bibinfo  {journal} {Phys. Rev.}\ }\textbf {\bibinfo {volume} {D101}},\ \bibinfo {pages} {023501} (\bibinfo {year} {2020})},\ \Eprint {http://arxiv.org/abs/1909.11094} {arXiv:1909.11094 [astro-ph.CO]} \BibitemShut {NoStop}%
\bibitem [{\citenamefont {Passaglia}\ and\ \citenamefont {Hu}(2022)}]{Passaglia:2022bcr}%
  \BibitemOpen
  \bibfield  {author} {\bibinfo {author} {\bibfnamefont {S.}~\bibnamefont {Passaglia}}\ and\ \bibinfo {author} {\bibfnamefont {W.}~\bibnamefont {Hu}},\ }\href {\doibase 10.1103/PhysRevD.105.123529} {\bibfield  {journal} {\bibinfo  {journal} {Phys. Rev. D}\ }\textbf {\bibinfo {volume} {105}},\ \bibinfo {pages} {123529} (\bibinfo {year} {2022})},\ \Eprint {http://arxiv.org/abs/2201.10238} {arXiv:2201.10238 [astro-ph.CO]} \BibitemShut {NoStop}%
\bibitem [{\citenamefont {Chen}\ and\ \citenamefont {Soda}(2023)}]{Chen:2023unc}%
  \BibitemOpen
  \bibfield  {author} {\bibinfo {author} {\bibfnamefont {C.-B.}\ \bibnamefont {Chen}}\ and\ \bibinfo {author} {\bibfnamefont {J.}~\bibnamefont {Soda}},\ }\href {\doibase 10.1088/1475-7516/2023/06/049} {\bibfield  {journal} {\bibinfo  {journal} {JCAP}\ }\textbf {\bibinfo {volume} {06}},\ \bibinfo {pages} {049} (\bibinfo {year} {2023})},\ \Eprint {http://arxiv.org/abs/2303.00999} {arXiv:2303.00999 [astro-ph.CO]} \BibitemShut {NoStop}%
\bibitem [{\citenamefont {Blas}\ \emph {et~al.}(2011)\citenamefont {Blas}, \citenamefont {Lesgourgues},\ and\ \citenamefont {Tram}}]{Blas:2011rf}%
  \BibitemOpen
  \bibfield  {author} {\bibinfo {author} {\bibfnamefont {D.}~\bibnamefont {Blas}}, \bibinfo {author} {\bibfnamefont {J.}~\bibnamefont {Lesgourgues}}, \ and\ \bibinfo {author} {\bibfnamefont {T.}~\bibnamefont {Tram}},\ }\href {\doibase 10.1088/1475-7516/2011/07/034} {\bibfield  {journal} {\bibinfo  {journal} {JCAP}\ }\textbf {\bibinfo {volume} {1107}},\ \bibinfo {pages} {034} (\bibinfo {year} {2011})},\ \Eprint {http://arxiv.org/abs/1104.2933} {arXiv:1104.2933 [astro-ph.CO]} \BibitemShut {NoStop}%
\bibitem [{\citenamefont {Baryakhtar}\ \emph {et~al.}(2024)\citenamefont {Baryakhtar}, \citenamefont {Simon},\ and\ \citenamefont {Weiner}}]{Baryakhtar:2024rky}%
  \BibitemOpen
  \bibfield  {author} {\bibinfo {author} {\bibfnamefont {M.}~\bibnamefont {Baryakhtar}}, \bibinfo {author} {\bibfnamefont {O.}~\bibnamefont {Simon}}, \ and\ \bibinfo {author} {\bibfnamefont {Z.~J.}\ \bibnamefont {Weiner}},\ }\href {\doibase 10.1103/PhysRevD.110.083505} {\bibfield  {journal} {\bibinfo  {journal} {Phys. Rev. D}\ }\textbf {\bibinfo {volume} {110}},\ \bibinfo {pages} {083505} (\bibinfo {year} {2024})},\ \Eprint {http://arxiv.org/abs/2405.10358} {arXiv:2405.10358 [astro-ph.CO]} \BibitemShut {NoStop}%
\bibitem [{\citenamefont {Ure\~{n}a L\'{o}pez}\ and\ \citenamefont {Gonzalez-Morales}(2016)}]{Urena-Lopez:2015gur}%
  \BibitemOpen
  \bibfield  {author} {\bibinfo {author} {\bibfnamefont {L.~A.}\ \bibnamefont {Ure\~{n}a L\'{o}pez}}\ and\ \bibinfo {author} {\bibfnamefont {A.~X.}\ \bibnamefont {Gonzalez-Morales}},\ }\href {\doibase 10.1088/1475-7516/2016/07/048} {\bibfield  {journal} {\bibinfo  {journal} {JCAP}\ }\textbf {\bibinfo {volume} {1607}},\ \bibinfo {pages} {048} (\bibinfo {year} {2016})},\ \Eprint {http://arxiv.org/abs/1511.08195} {arXiv:1511.08195 [astro-ph.CO]} \BibitemShut {NoStop}%
\bibitem [{\citenamefont {Cede\~no}\ \emph {et~al.}(2017)\citenamefont {Cede\~no}, \citenamefont {Gonz\'alez-Morales},\ and\ \citenamefont {Ure\~na L\'opez}}]{Cedeno:2017sou}%
  \BibitemOpen
  \bibfield  {author} {\bibinfo {author} {\bibfnamefont {F.~X.~L.}\ \bibnamefont {Cede\~no}}, \bibinfo {author} {\bibfnamefont {A.~X.}\ \bibnamefont {Gonz\'alez-Morales}}, \ and\ \bibinfo {author} {\bibfnamefont {L.~A.}\ \bibnamefont {Ure\~na L\'opez}},\ }\href {\doibase 10.1103/PhysRevD.96.061301} {\bibfield  {journal} {\bibinfo  {journal} {Phys. Rev. D}\ }\textbf {\bibinfo {volume} {96}},\ \bibinfo {pages} {061301} (\bibinfo {year} {2017})},\ \Eprint {http://arxiv.org/abs/1703.10180} {arXiv:1703.10180 [gr-qc]} \BibitemShut {NoStop}%
\bibitem [{\citenamefont {Ure\~na L\'opez}\ and\ \citenamefont {Linares Cede\~no}(2023)}]{Urena-Lopez:2023ngt}%
  \BibitemOpen
  \bibfield  {author} {\bibinfo {author} {\bibfnamefont {L.~A.}\ \bibnamefont {Ure\~na L\'opez}}\ and\ \bibinfo {author} {\bibfnamefont {F.~X.}\ \bibnamefont {Linares Cede\~no}},\ }\href@noop {} {\  (\bibinfo {year} {2023})},\ \Eprint {http://arxiv.org/abs/2307.05600} {arXiv:2307.05600 [astro-ph.CO]} \BibitemShut {NoStop}%
\bibitem [{\citenamefont {Aghanim}\ \emph {et~al.}(2020{\natexlab{b}})\citenamefont {Aghanim} \emph {et~al.}}]{Aghanim:2018eyx}%
  \BibitemOpen
  \bibfield  {author} {\bibinfo {author} {\bibfnamefont {N.}~\bibnamefont {Aghanim}} \emph {et~al.} (\bibinfo {collaboration} {Planck}),\ }\href {\doibase 10.1051/0004-6361/201833910} {\bibfield  {journal} {\bibinfo  {journal} {Astron. Astrophys.}\ }\textbf {\bibinfo {volume} {641}},\ \bibinfo {pages} {A6} (\bibinfo {year} {2020}{\natexlab{b}})},\ \Eprint {http://arxiv.org/abs/1807.06209} {arXiv:1807.06209 [astro-ph.CO]} \BibitemShut {NoStop}%
\bibitem [{\citenamefont {Ma}\ and\ \citenamefont {Bertschinger}(1995)}]{Ma:1995ey}%
  \BibitemOpen
  \bibfield  {author} {\bibinfo {author} {\bibfnamefont {C.-P.}\ \bibnamefont {Ma}}\ and\ \bibinfo {author} {\bibfnamefont {E.}~\bibnamefont {Bertschinger}},\ }\href {\doibase 10.1086/176550} {\bibfield  {journal} {\bibinfo  {journal} {Astrophys. J.}\ }\textbf {\bibinfo {volume} {455}},\ \bibinfo {pages} {7} (\bibinfo {year} {1995})},\ \Eprint {http://arxiv.org/abs/astro-ph/9506072} {arXiv:astro-ph/9506072 [astro-ph]} \BibitemShut {NoStop}%
\bibitem [{\citenamefont {Hu}\ \emph {et~al.}(2016)\citenamefont {Hu}, \citenamefont {Chiang}, \citenamefont {Li},\ and\ \citenamefont {LoVerde}}]{Hu:2016ssz}%
  \BibitemOpen
  \bibfield  {author} {\bibinfo {author} {\bibfnamefont {W.}~\bibnamefont {Hu}}, \bibinfo {author} {\bibfnamefont {C.-T.}\ \bibnamefont {Chiang}}, \bibinfo {author} {\bibfnamefont {Y.}~\bibnamefont {Li}}, \ and\ \bibinfo {author} {\bibfnamefont {M.}~\bibnamefont {LoVerde}},\ }\href {\doibase 10.1103/PhysRevD.94.023002} {\bibfield  {journal} {\bibinfo  {journal} {Phys. Rev. D}\ }\textbf {\bibinfo {volume} {94}},\ \bibinfo {pages} {023002} (\bibinfo {year} {2016})},\ \Eprint {http://arxiv.org/abs/1605.01412} {arXiv:1605.01412 [astro-ph.CO]} \BibitemShut {NoStop}%
\bibitem [{\citenamefont {Hu}(1998)}]{Hu:1998kj}%
  \BibitemOpen
  \bibfield  {author} {\bibinfo {author} {\bibfnamefont {W.}~\bibnamefont {Hu}},\ }\href {\doibase 10.1086/306274} {\bibfield  {journal} {\bibinfo  {journal} {Astrophys. J.}\ }\textbf {\bibinfo {volume} {506}},\ \bibinfo {pages} {485} (\bibinfo {year} {1998})},\ \Eprint {http://arxiv.org/abs/astro-ph/9801234} {arXiv:astro-ph/9801234 [astro-ph]} \BibitemShut {NoStop}%
\bibitem [{\citenamefont {Lesgourgues}\ and\ \citenamefont {Tram}(2014)}]{Lesgourgues:2013bra}%
  \BibitemOpen
  \bibfield  {author} {\bibinfo {author} {\bibfnamefont {J.}~\bibnamefont {Lesgourgues}}\ and\ \bibinfo {author} {\bibfnamefont {T.}~\bibnamefont {Tram}},\ }\href {\doibase 10.1088/1475-7516/2014/09/032} {\bibfield  {journal} {\bibinfo  {journal} {JCAP}\ }\textbf {\bibinfo {volume} {09}},\ \bibinfo {pages} {032} (\bibinfo {year} {2014})},\ \Eprint {http://arxiv.org/abs/1312.2697} {arXiv:1312.2697 [astro-ph.CO]} \BibitemShut {NoStop}%
\bibitem [{\citenamefont {Hu}\ \emph {et~al.}(1998)\citenamefont {Hu}, \citenamefont {Seljak}, \citenamefont {White},\ and\ \citenamefont {Zaldarriaga}}]{Hu:1997mn}%
  \BibitemOpen
  \bibfield  {author} {\bibinfo {author} {\bibfnamefont {W.}~\bibnamefont {Hu}}, \bibinfo {author} {\bibfnamefont {U.}~\bibnamefont {Seljak}}, \bibinfo {author} {\bibfnamefont {M.~J.}\ \bibnamefont {White}}, \ and\ \bibinfo {author} {\bibfnamefont {M.}~\bibnamefont {Zaldarriaga}},\ }\href {\doibase 10.1103/PhysRevD.57.3290} {\bibfield  {journal} {\bibinfo  {journal} {Phys. Rev. D}\ }\textbf {\bibinfo {volume} {57}},\ \bibinfo {pages} {3290} (\bibinfo {year} {1998})},\ \Eprint {http://arxiv.org/abs/astro-ph/9709066} {arXiv:astro-ph/9709066} \BibitemShut {NoStop}%
\bibitem [{\citenamefont {Seljak}\ and\ \citenamefont {Zaldarriaga}(1996)}]{Seljak:1996is}%
  \BibitemOpen
  \bibfield  {author} {\bibinfo {author} {\bibfnamefont {U.}~\bibnamefont {Seljak}}\ and\ \bibinfo {author} {\bibfnamefont {M.}~\bibnamefont {Zaldarriaga}},\ }\href {\doibase 10.1086/177793} {\bibfield  {journal} {\bibinfo  {journal} {Astrophys. J.}\ }\textbf {\bibinfo {volume} {469}},\ \bibinfo {pages} {437} (\bibinfo {year} {1996})},\ \Eprint {http://arxiv.org/abs/astro-ph/9603033} {arXiv:astro-ph/9603033 [astro-ph]} \BibitemShut {NoStop}%
\bibitem [{\citenamefont {Lesgourgues}(2011)}]{Lesgourgues:2011re}%
  \BibitemOpen
  \bibfield  {author} {\bibinfo {author} {\bibfnamefont {J.}~\bibnamefont {Lesgourgues}},\ }\href@noop {} {\  (\bibinfo {year} {2011})},\ \Eprint {http://arxiv.org/abs/1104.2932} {arXiv:1104.2932 [astro-ph.IM]} \BibitemShut {NoStop}%
\bibitem [{\citenamefont {Eisenstein}\ \emph {et~al.}(1999)\citenamefont {Eisenstein}, \citenamefont {Hu},\ and\ \citenamefont {Tegmark}}]{Eisenstein:1998hr}%
  \BibitemOpen
  \bibfield  {author} {\bibinfo {author} {\bibfnamefont {D.~J.}\ \bibnamefont {Eisenstein}}, \bibinfo {author} {\bibfnamefont {W.}~\bibnamefont {Hu}}, \ and\ \bibinfo {author} {\bibfnamefont {M.}~\bibnamefont {Tegmark}},\ }\href {\doibase 10.1086/307261} {\bibfield  {journal} {\bibinfo  {journal} {Astrophys. J.}\ }\textbf {\bibinfo {volume} {518}},\ \bibinfo {pages} {2} (\bibinfo {year} {1999})},\ \Eprint {http://arxiv.org/abs/astro-ph/9807130} {arXiv:astro-ph/9807130 [astro-ph]} \BibitemShut {NoStop}%
\bibitem [{\citenamefont {Rogers}\ and\ \citenamefont {Peiris}(2021)}]{Rogers:2020ltq}%
  \BibitemOpen
  \bibfield  {author} {\bibinfo {author} {\bibfnamefont {K.~K.}\ \bibnamefont {Rogers}}\ and\ \bibinfo {author} {\bibfnamefont {H.~V.}\ \bibnamefont {Peiris}},\ }\href {\doibase 10.1103/PhysRevLett.126.071302} {\bibfield  {journal} {\bibinfo  {journal} {Phys. Rev. Lett.}\ }\textbf {\bibinfo {volume} {126}},\ \bibinfo {pages} {071302} (\bibinfo {year} {2021})},\ \Eprint {http://arxiv.org/abs/2007.12705} {arXiv:2007.12705 [astro-ph.CO]} \BibitemShut {NoStop}%
\bibitem [{\citenamefont {Amendola}\ and\ \citenamefont {Barbieri}(2006)}]{Amendola:2005ad}%
  \BibitemOpen
  \bibfield  {author} {\bibinfo {author} {\bibfnamefont {L.}~\bibnamefont {Amendola}}\ and\ \bibinfo {author} {\bibfnamefont {R.}~\bibnamefont {Barbieri}},\ }\href {\doibase 10.1016/j.physletb.2006.08.069} {\bibfield  {journal} {\bibinfo  {journal} {Phys. Lett. B}\ }\textbf {\bibinfo {volume} {642}},\ \bibinfo {pages} {192} (\bibinfo {year} {2006})},\ \Eprint {http://arxiv.org/abs/hep-ph/0509257} {arXiv:hep-ph/0509257} \BibitemShut {NoStop}%
\bibitem [{\citenamefont {Shapiro}\ \emph {et~al.}(2021)\citenamefont {Shapiro}, \citenamefont {Dawoodbhoy},\ and\ \citenamefont {Rindler-Daller}}]{Shapiro:2021hjp}%
  \BibitemOpen
  \bibfield  {author} {\bibinfo {author} {\bibfnamefont {P.~R.}\ \bibnamefont {Shapiro}}, \bibinfo {author} {\bibfnamefont {T.}~\bibnamefont {Dawoodbhoy}}, \ and\ \bibinfo {author} {\bibfnamefont {T.}~\bibnamefont {Rindler-Daller}},\ }\href {\doibase 10.1093/mnras/stab2884} {\bibfield  {journal} {\bibinfo  {journal} {Mon. Not. Roy. Astron. Soc.}\ }\textbf {\bibinfo {volume} {509}},\ \bibinfo {pages} {145} (\bibinfo {year} {2021})},\ \Eprint {http://arxiv.org/abs/2106.13244} {arXiv:2106.13244 [astro-ph.CO]} \BibitemShut {NoStop}%
\bibitem [{\citenamefont {Winch}\ \emph {et~al.}(2024)\citenamefont {Winch}, \citenamefont {Hlozek}, \citenamefont {Marsh}, \citenamefont {Grin},\ and\ \citenamefont {Rogers}}]{Winch:2023qzl}%
  \BibitemOpen
  \bibfield  {author} {\bibinfo {author} {\bibfnamefont {H.}~\bibnamefont {Winch}}, \bibinfo {author} {\bibfnamefont {R.}~\bibnamefont {Hlozek}}, \bibinfo {author} {\bibfnamefont {D.~J.~E.}\ \bibnamefont {Marsh}}, \bibinfo {author} {\bibfnamefont {D.}~\bibnamefont {Grin}}, \ and\ \bibinfo {author} {\bibfnamefont {K.~K.}\ \bibnamefont {Rogers}},\ }\href {\doibase 10.1103/PhysRevD.110.043517} {\bibfield  {journal} {\bibinfo  {journal} {Phys. Rev. D}\ }\textbf {\bibinfo {volume} {110}},\ \bibinfo {pages} {043517} (\bibinfo {year} {2024})},\ \Eprint {http://arxiv.org/abs/2311.02052} {arXiv:2311.02052 [astro-ph.CO]} \BibitemShut {NoStop}%
\end{thebibliography}%

\end{document}